\documentclass[11pt,letterpaper]{article}
\usepackage{times}
\usepackage{epsfig}
\usepackage{graphicx}
\usepackage{amsmath}
\usepackage{amssymb}
\usepackage[utf8]{inputenc} 
\usepackage[T1]{fontenc}    
\usepackage{url}            
\usepackage{breakurl}

\usepackage{booktabs}       
\usepackage{microtype}      
\usepackage{amsthm,amsfonts,subfigure,bm}
\usepackage{multirow}
\usepackage{float}
\restylefloat{table}
\theoremstyle{plain}

\theoremstyle{remark}     
\theoremstyle{definition} 

\usepackage[margin=0.8in]{geometry}

\newcommand{\ind}{\stackrel{\mathrm{ind}}{\sim}}
\newcommand\om{\Omega}

\newcommand\A{\mathcal{A}}

\newcommand\T{\mathcal{T}}

\newcommand\bx{\bm{x}}
\newcommand\by{\bm{y}}

\newcommand\tlam{\tilde{\lambda}}

\newcommand\blam{\bm\lambda}

\def\independenT#1#2{\mathrel{\setbox0\hbox{$#1#2$}%
		\copy0\kern-\wd0\mkern4mu\box0}}

\usepackage[export]{adjustbox}
\usepackage[pagebackref=true,breaklinks=true,letterpaper=true,colorlinks,bookmarks=false]{hyperref}

\usepackage{booktabs}
\usepackage{multirow}
\usepackage{colortbl}
\usepackage{xcolor}
\usepackage{xfrac}

\begin{document}

\title{Efficient in-situ image and video compression through probabilistic image representation} 

\author{Rongjie Liu \\
  Department of Statistics\\
  Rice University\\
  Houston, TX 77005\\
{\tt\small rongjie.liu@rice.edu}
\and
Meng Li \\
   Department of Statistics\\
  Rice University\\
  Houston, TX 77005\\
{\tt\small meng@rice.edu}
\and
   Li Ma\\
   Department of Statistical Science\\
  Duke University\\
  Durham, NC 27708\\
  {\tt\small li.ma@duke.edu}
}

\maketitle
\begin{abstract}
Fast and effective image compression for multi-dimensional images has become increasingly important for efficient storage and transfer of massive amounts of high-resolution images and videos. Desirable properties in compression methods include (1) high reconstruction quality at a wide range of compression rates while preserving key local details, (2) computational scalability, (3) applicability to a variety of different image/video types and of different dimensions, (4) progressive transmission, and (5) ease of tuning. We present such a method for multi-dimensional image compression called Compression via Adaptive Recursive Partitioning (CARP). CARP uses an optimal permutation of the image pixels inferred from a Bayesian probabilistic model on recursive partitions of the image to reduce its effective dimensionality, achieving a parsimonious representation that preserves information. CARP uses a multi-layer Bayesian hierarchical model to achieve in-situ compression along with self-tuning and regularization, with just one single parameter to be specified by the user to achieve the desired compression rate. Extensive numerical experiments using a variety of datasets including 2D 
still images, real-life YouTube videos, and surveillance videos show that CARP dominates the state-of-the-art image/video compression approaches---including JPEG, JPEG2000, BPG, MPEG4, HEVC and a neural network-based method---for all of these different image types and on nearly all of the individual images and videos over some methods.
 \end{abstract}

\section{Introduction}

Image compression is a long-standing, fundamental problem in computer vision and image processing, and is key to efficient storage and transfer of the vast amount of high-resolution images and videos that are routinely collected in a variety of applications. Efficient compression relies on parsimonious representations of images that preserve important spatial and contextual features. Standards such as JPEG \cite{wallace1992jpeg} and JPEG2000 \cite{skodras2001jpeg} utilize fixed, deterministic linear function transforms, such as wavelets followed by optimized encoding under such transforms. BPG \cite{bpg2017}, short for Better Portable Graphics, is based on the intra-frame encoding of the High Efficiency Video Coding (HEVC) video compression standard and has also been widely used in image compression as an alternative to JPEG.
These approaches give excellent stability and scalability in practical implementation, and require little training and tuning. However, they lack adaptivity to image-specific features and as such achieve only suboptimal compression efficiency. The more recent convolutional neural network (CNN) based approaches \cite{Liu2018DeepIC,Balle17a,NIPS2017_6714,li2018learning,baig2017learning,minnen2018joint} utilize much more flexible, nonlinear transformations of the original image. This additional flexibility often leads to improved compression efficiency, but at the same time leads to substantially more extensive training and tuning of the methods. In this paper, we aim to develop a compression method that is able to mitigate the restrictiveness of deterministic transforms and attain state-of-the-art performance without separate training, achieving both computational and compression efficiency.

To this end, we strike a middle ground between these two approaches by introducing a Bayesian probabilistic modeling strategy for incorporating adaptivity to image features into the wavelet transform based image processing framework, while maintaining its computational scalability and ease of tuning. Instead of using fixed wavelet transforms, we treat the transform as an unknown latent quantity and learn an optimal transform by placing a Bayesian prior on the space of such transforms induced by random recursive partitioning on the image and compute the {\em maximum a posteriori} (MAP) estimate of the transform under our model. A compressed image can then be produced under the inferred transform that tailors to image-specific features. Moreover, such computation is as efficient as the classical wavelet-based methods---scaling linearly with the size of the image.

\begin{figure}[h!]
 \centering
 \includegraphics[width=0.9\linewidth]{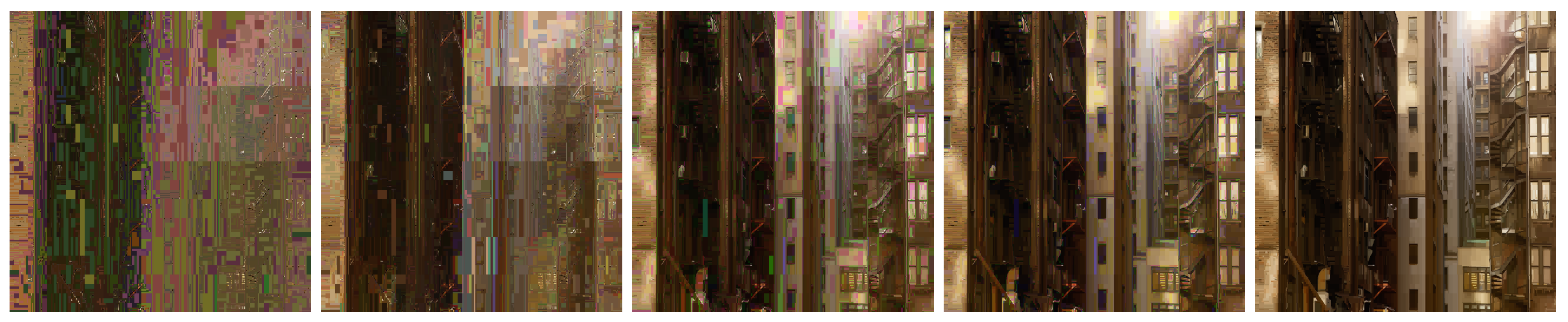}
 \caption{The reconstructed image is gradually improved from left to right when more data bits are transmitted (left to right: 300, 3000, 30000, 40000, 50000 bits). The original image is in Figure~\ref{rec_2c_1}.}\label{bitstream}
\end{figure}

Aside from achieving excellent compression efficiency (to be demonstrated in our numerical experiments), our method, called Compression via Adaptive Recursive Partitioning (CARP), enables simultaneous transformation learning and compression. Posterior distributions of the transformation and intensities of the images are jointly obtained, allowing information borrowing between these two steps in contrast to JPEG and JPEG2000. CARP is model-based and interpretable. Building on hierarchical Bayesian models, CARP outputs a tree structure that represents a learned optimal map to vectorize the image, where more comprehensive descriptions on the vectorization are additionally available via a probabilistic distribution. CARP enjoys three additional advantages. First, CARP is a general-purpose compression method that compresses $m$-dimensional images in a unified fashion for $m = 2, 3, 4$, etc., as opposed to specializing on one dimension (such as 2D or 3D) in most of the existing literature, making it readily applicable to a variety of image/video types. Second, it does not require a separate training stage on external data and involves minimal tuning. The Bayesian hierarchical modeling strategy uses hyperpriors on the parameters to allow automatic tuning on those parameters, leaving only one free parameter for the user to specify, which corresponds directly to the desired compression rate of the image. This makes CARP very easy to use, especially by common users, without requiring expert knowledge of the underlying method. Third, it allows progressive transmission to gradually improve the image quality as data bits are incrementally transmitted, which resembles the progressive principle in JPEG2000 \cite{jpeg2000overview}. In particular, the wavelet coefficients can be transmitted from coarse to fine scales in the decoding and reconstruction process to allow reconstruction of images at incremental resolutions. 
For example, Figure \ref{bitstream} demonstrates that the reconstructed building image is gradually improved when more data bits are transmitted.

\begin{figure}[h!]
\centering 
	\begin{tabular}{cc}
		 \includegraphics[width=0.5\linewidth, height=6cm]{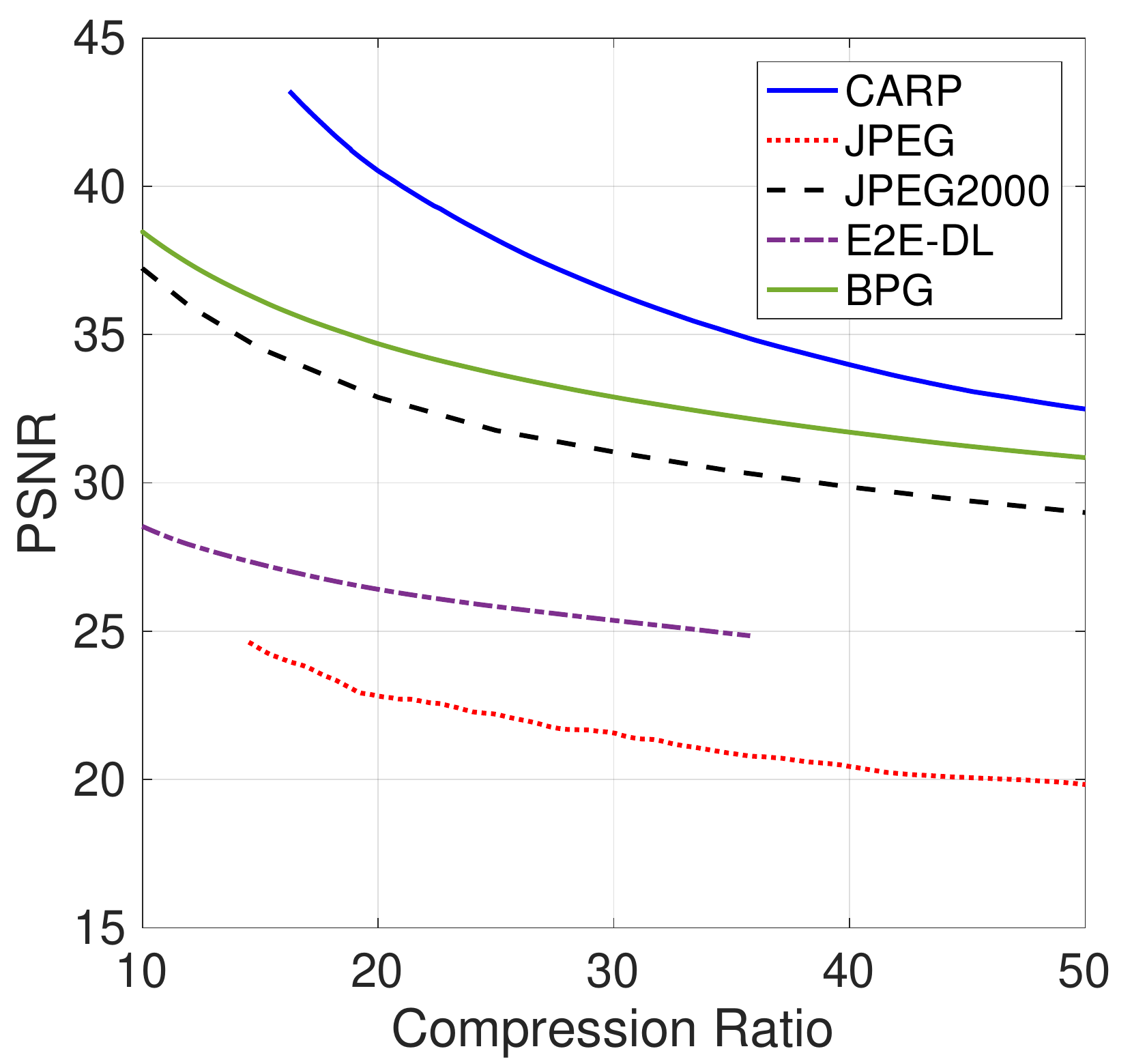} &
		  \includegraphics[width=0.5\linewidth, height=6cm]{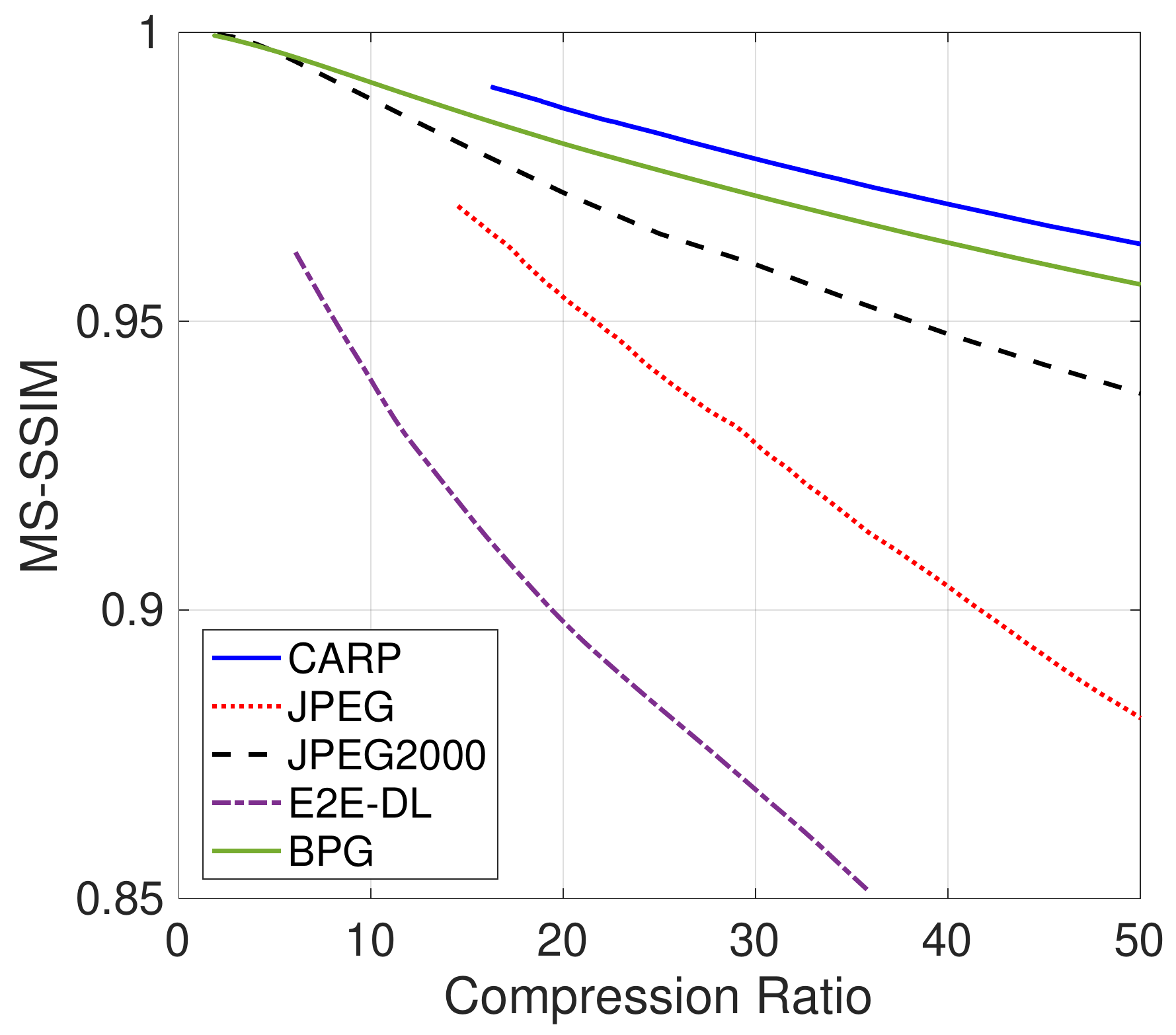} \\
		  {\small (a) Average PSNR on 2D still images} & {\small (b) Average MS-SSIM on 2D still images} \\
		  \includegraphics[width=0.5\linewidth, height=6cm]{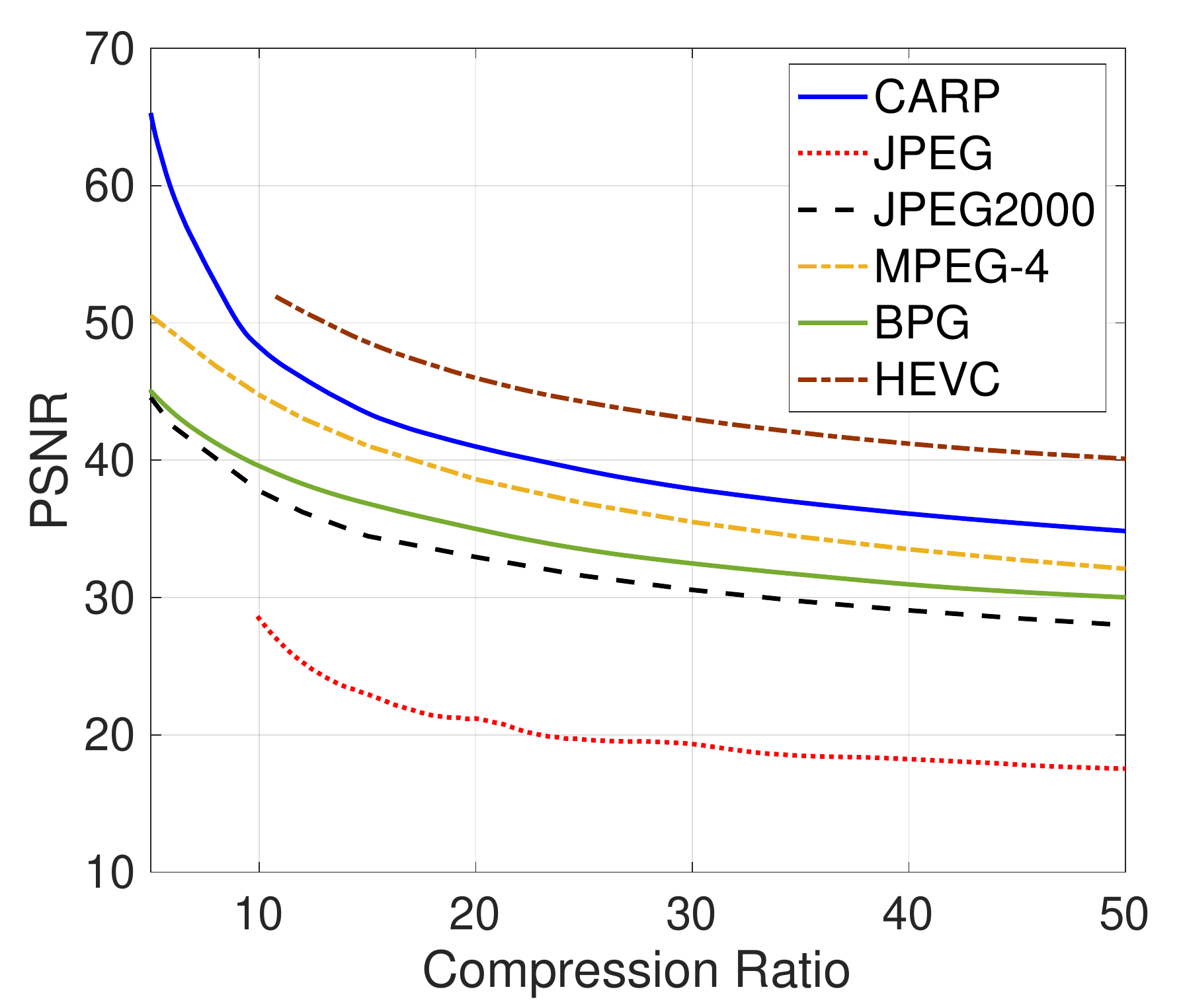} &
		  \includegraphics[width=0.5\linewidth, height=6cm]{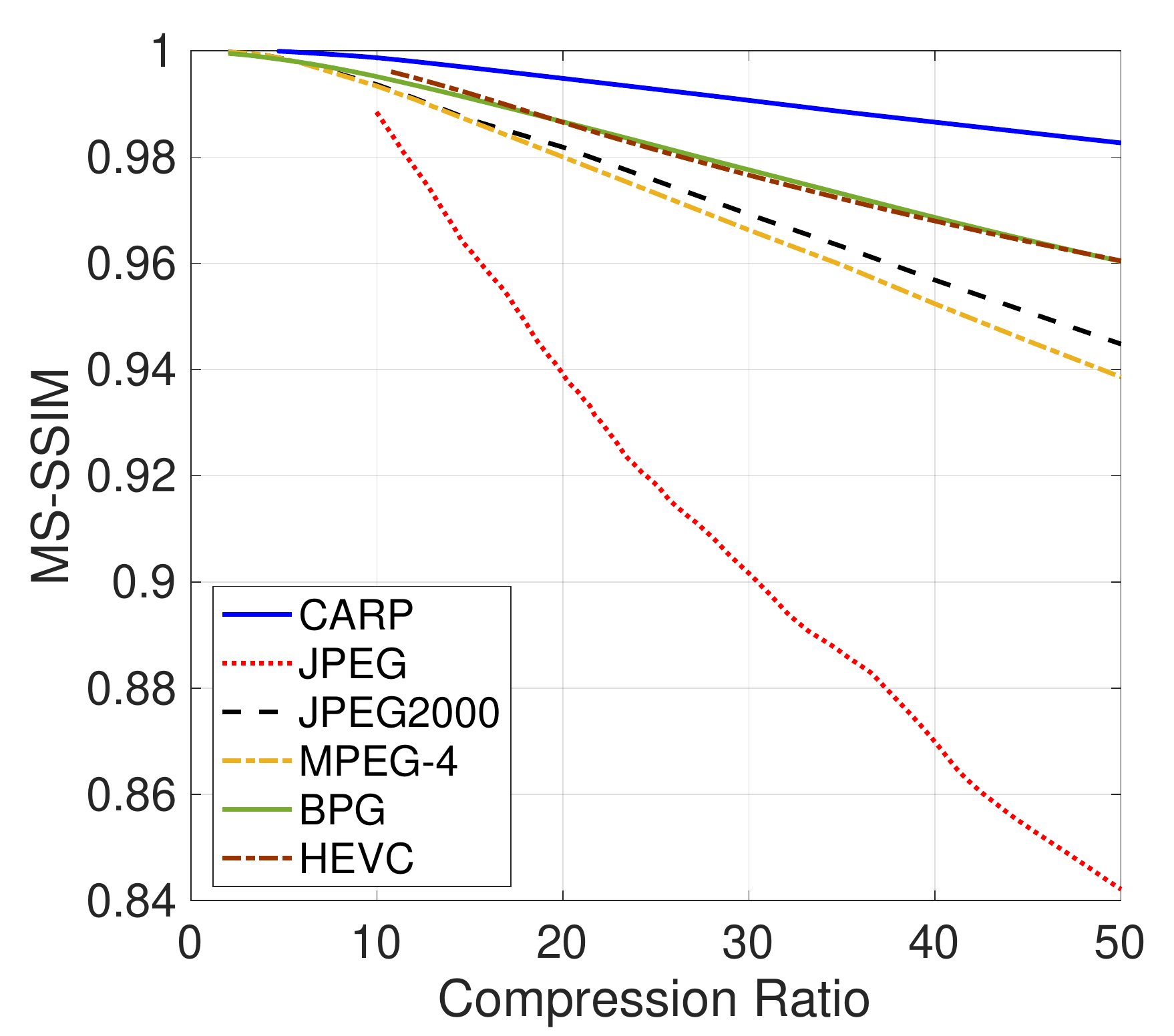} \\
		  {\small (c) Average PSNR on YouTube video images} & {\small (d) Average MS-SSIM on YouTube video images}\\
		  \includegraphics[width=0.5\linewidth, height=6cm]{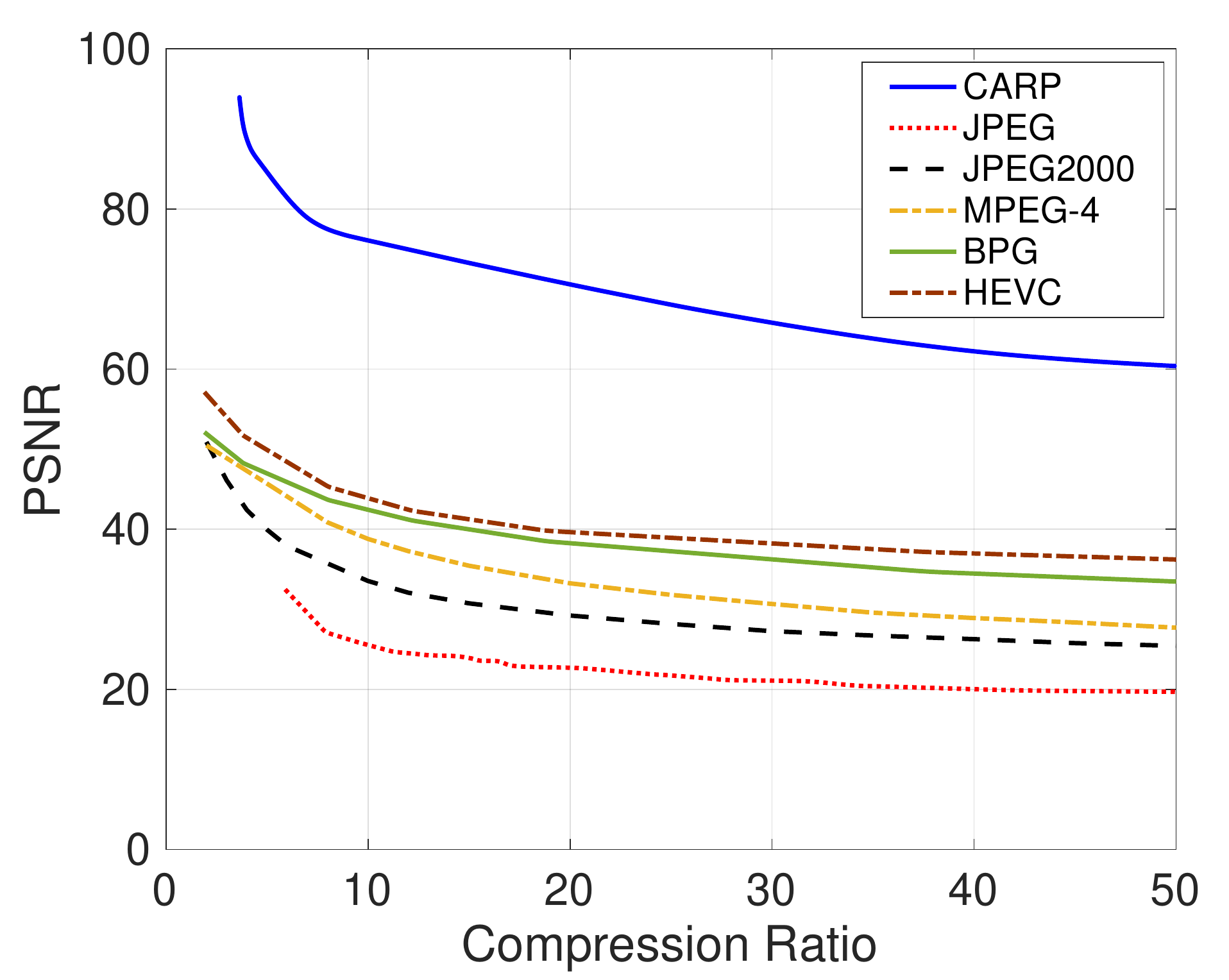} &
		  \includegraphics[width=0.5\linewidth, height=6cm]{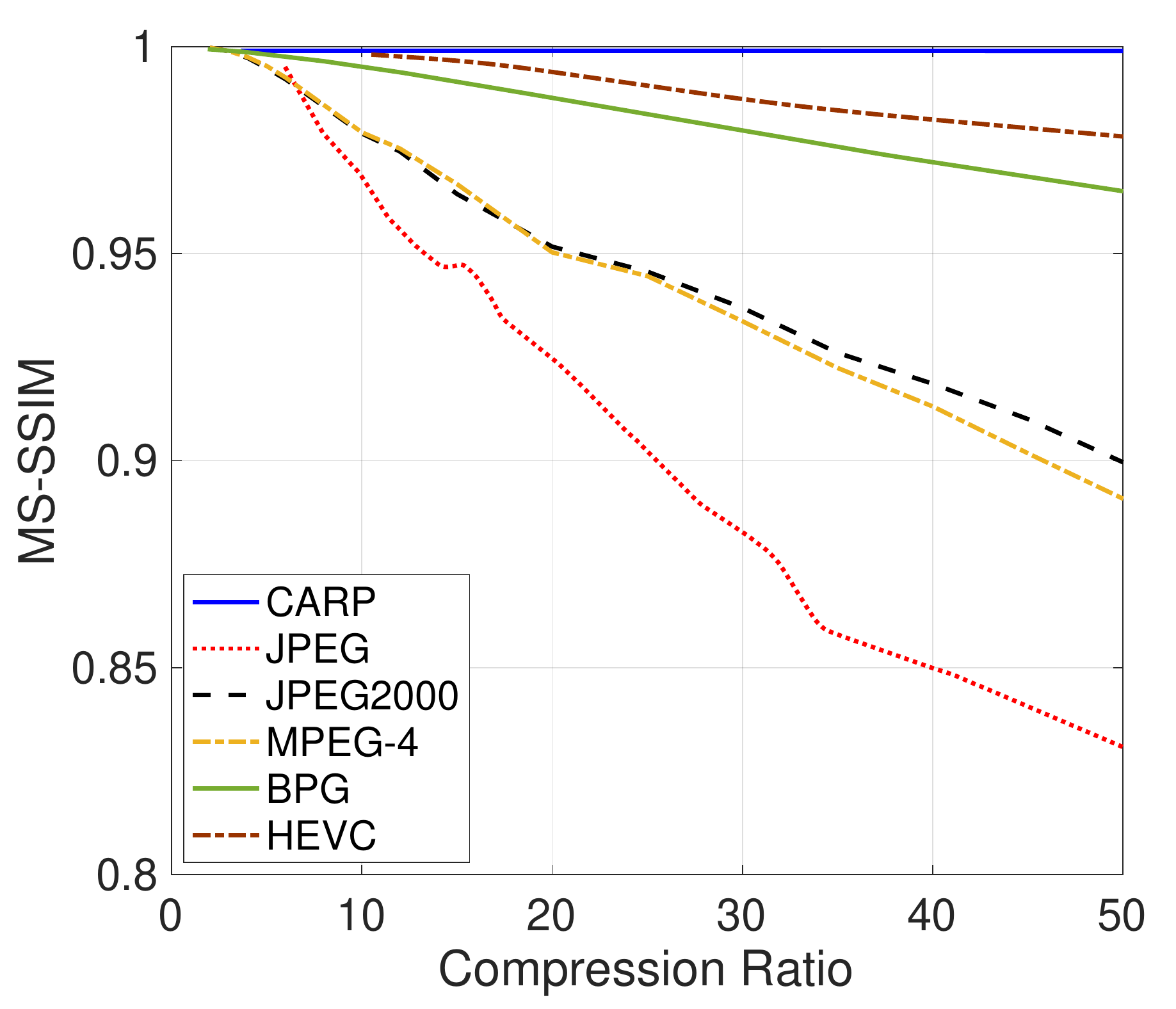} \\
		  {\small (e) Average PSNR on surveillance video images} & {\small (f) Average MS-SSIM on surveillance video images}
	\end{tabular}
\caption{Performance summary of CARP and competitors in three databases consisting of still images and videos at different resolutions. Each plot of the first column presents the average of peak-signal-to-noise-ratio (PSNR) while each plot of the second column presents the average of MS-SSIM at various compression ratios. Details are given in Section~\ref{sec:experiments}.} 
\label{fig:summary} 
\end{figure}

To validate the efficiency and robustness of our method, we use a variety of benchmark databases to compare CARP with several state-of-the-art compression methods for 2D and 3D videos. Figure~\ref{fig:summary} summarizes the average performance on three image/video types under the metric of peak signal-to-noise ratio (PSNR) and multi-scale structural similarity (MS-SSIM)~\cite{wang2003multiscale}, the latter being a widely used metric to assess perceived image quality and measure the structural similarity between two images; in all of them CARP dominates the state-of-the-art competitors---including JPEG, JPEG2000, End-to-End deep learning (E2E-DL), BPG, MPEG4, HEVC under both PSNR and MS-SSIM, with only one exception in Figure~\ref{fig:summary}(c) where CARP gives slightly smaller PSNR than HEVC while outperforming HEVC under MS-SSIM. We note that in Figure~\ref{fig:summary}(a) and \ref{fig:summary}(b), the averages of metrics are calculated over a subset of 70 images from the image dataset on which the methods being compared are able to achieve a wide range of compression ratios (from 15 to 35).  
Also, because we have used a pre-trained model for E2E-DL \cite{Balle17a}, the performance gap between E2E-DL and other methods could be narrowed had the CNNs been trained on images that are particularly suited for the 2D still image database. In Section~\ref{sec:experiments} we present more detailed numerical results that compare the image-specific performance of the methods, which show that CARP can dominate the competitors in nearly all of the individual images in some data sets we have examined. 

This paper is an extended version of our previous work~\cite{liu2020carp}. The organization of the paper is as follows. Section~\ref{sec:literature} surveys related work in image and video compression. In Section~\ref{sec:method}, we present the proposed CARP method,  including probabilistic image representation using a hierarchical Bayesian framework, posterior inference, and details in implementing the method. Section~\ref{sec:experiments} carries out extensive experiments on image and video compression. Section~\ref{sec:conclusion} concludes the paper. The code for average performance and implementing CARP is available on GitHub: \url{https://github.com/xylimeng/CARP}. 

\section{Related work} \label{sec:literature} 
For 2D images, perhaps the most well-known image compression algorithms are JPEG \cite{wallace1992jpeg} and its successor JPEG2000 \cite{skodras2001jpeg}. The JPEG standard uses a discrete cosine transform (DCT) on each 8 by 8 small block of pixels. A quantization table is applied, and Huffman encoding is used on DCT blocks for compression. Compared to the JPEG standard, 
JPEG2000 uses a multi-scale orthogonal wavelet decomposition with arithmetic coding. JPEG2000 standard defines a new image-coding scheme using state-of-the-art compression techniques based on wavelet technology. In particular, the discrete wavelet transform (DWT) decomposes images into their resolution and frequency
contents. The DWT can be performed with a non-reversible Daubechies (9,7) taps filter, which provides for higher, but lossy, compression. In addition to DWT, another feature for JPEG2000 is quantization. JPEG2000 quantizer follows an embedded dead-zone scalar approach. The quantizer step size used to scale the coefficients is independently selected for each wavelet sub-band. For higher dimensional data (larger than 2), JPEG2000 is still suitable once the data is reduced into a lower dimensional space. For example, JPEG2000 can be applied to 3D image data slice by slice, and then the compressed results are combined. However, the spatial correlation on the third axis is ignored in this method.

However, both JPEG and JPEG2000 are suboptimal for image compression \cite{li2018learning} due to the non-adaptive image transformation and a separate optimization on codecs. BPG \cite{bpg2017} is often advantageous over JPEG. Specifically, BPG can achieve a larger compression ratio than JPEG for the same reconstruction quality, and it supports up to 14 bits per color channel instead of up to 8 bits as in JPEG. In general, BPG is often advantageous over JPEG and maintains the properties of H.265 because one BPG image takes a single frame out of an H.265 video stream.

Besides JPEG, JPEG2000 and BPG, there is a growing literature in developing deep learning-based methods \cite{Liu2018DeepIC,Balle17a,NIPS2017_6714,li2018learning,baig2017learning,minnen2018joint} for image compression. Among these methods, end-to-end deep learning-based approaches are particularly appealing, which go directly from the input to the desired output with optimized codecs \cite{Balle17a,li2018learning}. For example, a pre-trained model over a database of training images was proposed in \cite{Balle17a} with all the required components for end-to-end implementation, including a nonlinear analysis transformation, a uniform quantizer, and a nonlinear synthesis transformation. 

Videos have a different structure than 2D images due to the extra temporal dimension.
Although a video can be compressed frame by frame by some existing 2D image compression methods (e.g., JPEG, JPEG2000 and BPG), the critical temporal redundancy is undesirably ignored. Thus, most video compression algorithms in Moving Picture Experts Group (MPEG)  \cite{watkinson2012mpeg} exploit both spatial and temporal redundancy. For example, MPEG-4 absorbs many features of different standards using both DCT and motion compensation \cite{chen2001design} techniques to achieve this goal. In addition, to reach a higher compression ratio, MPEG-4 only stores and encodes the inter-frame changes instead of the entire original frame. However, the redundancy detection strategy in MPEG-4 is localized to capturing the difference of adjacent frames, and thus might be not globally optimal and hurdle a more efficient compression. HEVC \cite{hevc2012}, standing for High Efficiency Video Compression, is an extension of MPEG-4 designed by an intra-frame coding strategy of applying intra-picture prediction and loop filters to optimally use parallel processing and improve the quality of the reconstruction frames.

Recursive partitioning induces a permutation of the pixels and has been used previously in other applications. In particular, \cite{anila2010usage} adopts peak transform to obtain spatial permutation. \cite{warp} uses random recursive partitioning to induce a prior on the permutations of image pixels, leading to an effective algorithm for image denoising using posterior Bayesian model averaging. 
In this work we use random recursive partitioning to induce a probability model on wavelet transforms, but instead aim at learning an optimal transform to represent the image thereby achieving efficient compression. 
The 1D vector returned by Bayesian multi-scale learning is passed to existing encoding methods to generate compressed representation, followed by the corresponding decoder. The literature \cite{chujoh2010video,borer2007image,normille1993encoding} has provided a rich menu of encoding/decoding methods designed for lower-dimensional images. 

\section{Method} \label{sec:method}
\subsection{CARP: The framework} 
CARP is a framework for image compression via adaptive recursive partitioning. It uses an optimal permutation of the image pixels inferred from a Bayesian probabilistic model to reduce the dimensionality of a $m$-dimensional image, thereby achieving a parsimonious representation that effectively preserves image information. More specifically, CARP utilizes a prior distribution on the space of permutations induced by random recursive partitioning along a bifurcating tree \cite{warp}. This random recursive partitioning incorporates the latent pruning variables to probabilistically terminate the partitioning within the partition blocks where the pixel intensities are similar enough, which means a further partition is unnecessary and permuting the entire image block is the fastest and most efficient way. The MAP estimate, i.e., the posterior mode of the posterior distribution on the recursive partitioning, produces a representative permutation (or vectorization) of the image pixels that can be readily fed into encoding methods to generate compressed representation, followed by the corresponding decoder to reconstruct the compressed image. 

\begin{figure}[h!]
	\centering
	\includegraphics[trim = 0 0 0 0, clip = TRUE, width=0.8\linewidth]{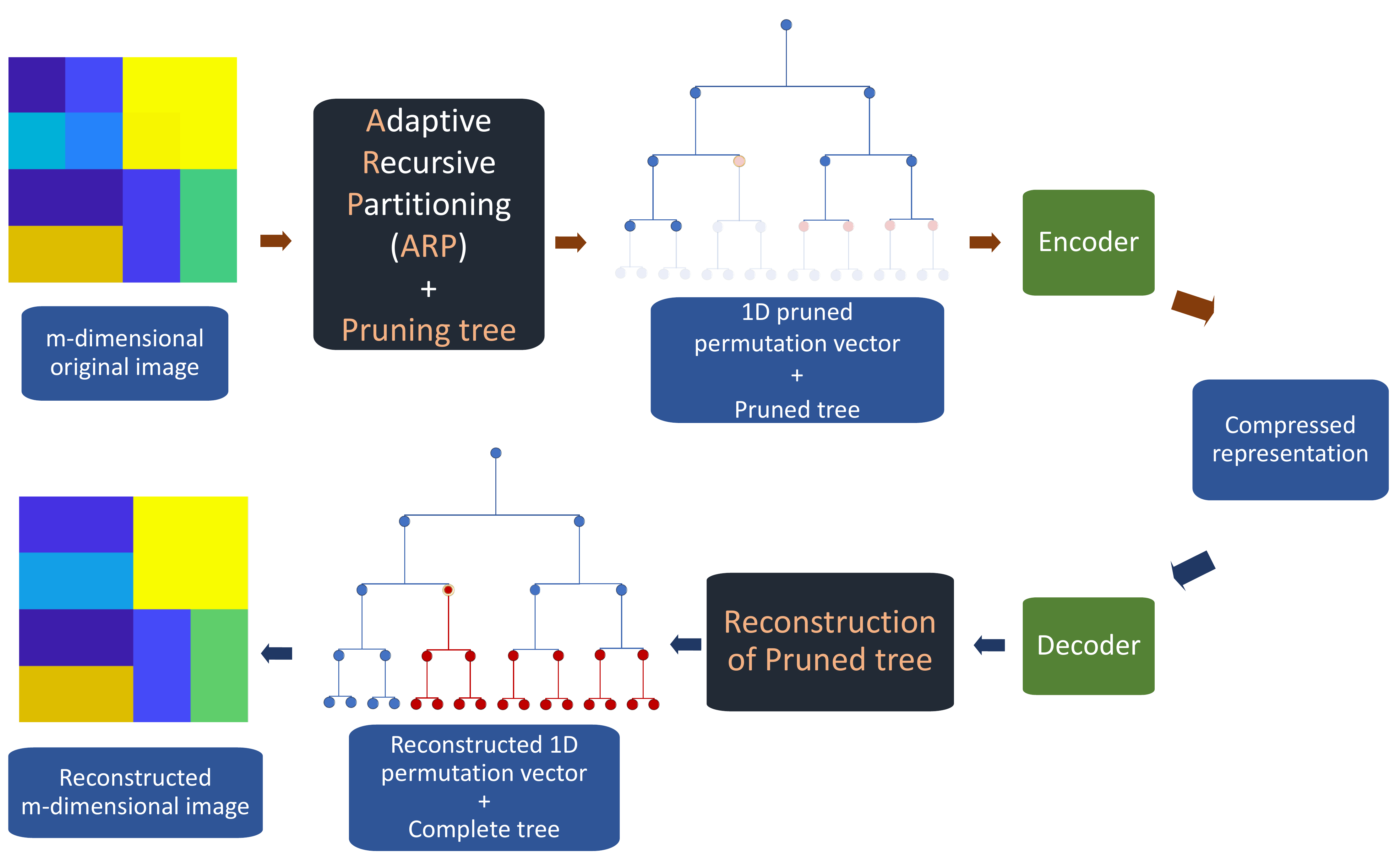}
	\caption{The workflow of CARP.}\label{workflow}
\end{figure}

\begin{figure}[h!]
	\centering
	\includegraphics[trim = 0 0 0 0, clip = TRUE, width=0.8\linewidth]{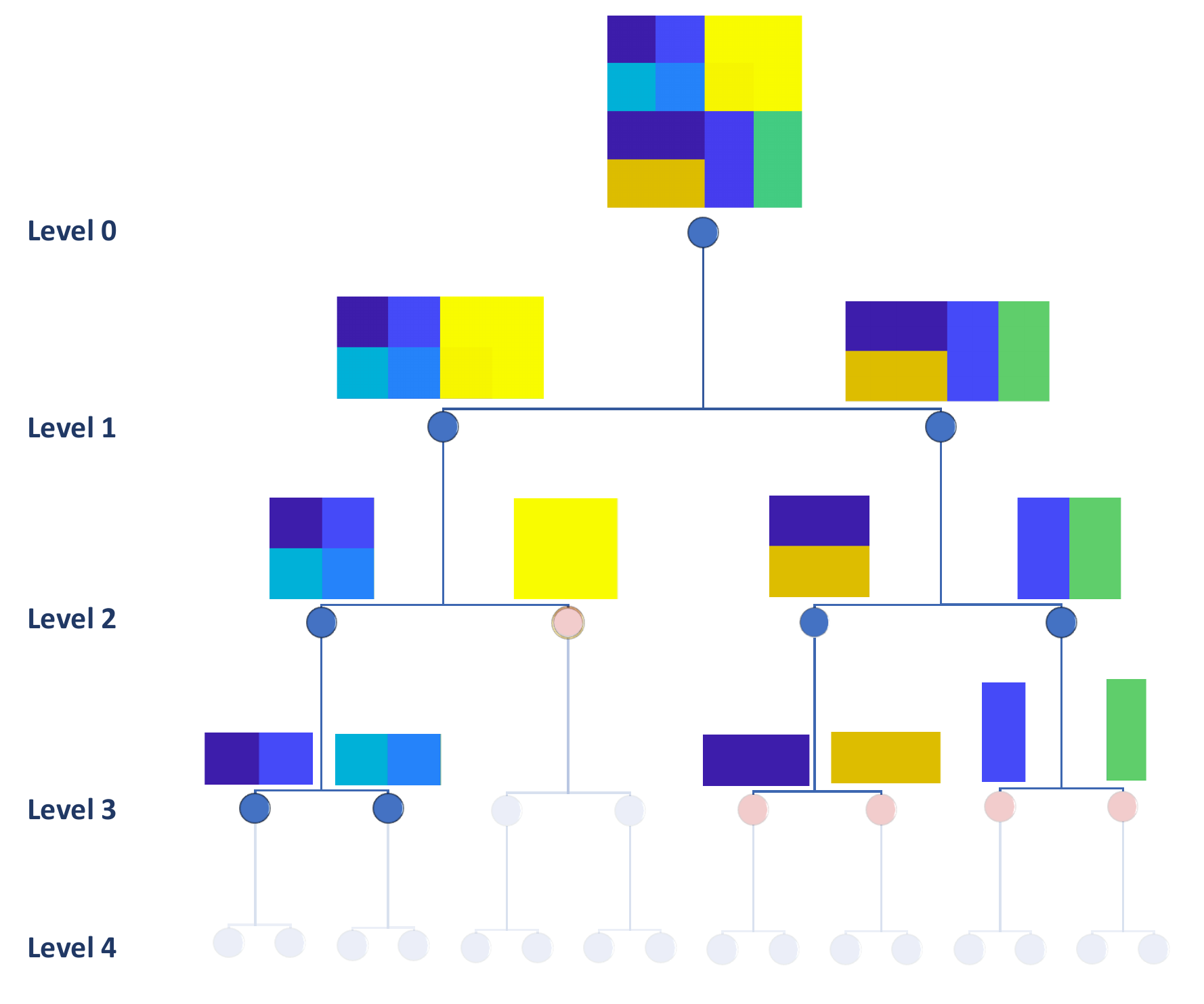}
	\caption{Toy example to demonstrate RDP and pruning: the partition tree is obtained by MAP.}\label{PrunWorkflow}
\end{figure}

Figure~\ref{workflow} presents the architecture of CARP, where the two black boxes pinpoint the key techniques used in CARP. In this work, we use the 1D discrete wavelet transform (DWT) and Huffman symbol encoding algorithm as the encoder, and use the inverse DWI and Huffman symbol decoding algorithm as the decoder. We next describe details of the pipeline of CARP, annotated by a toy example given by the 4 by 4 image in Figure~\ref{workflow} to ease demonstration. 
Figure~\ref{PrunWorkflow} shows the pruning tree workflow via an image example. For any m-dimensional image, it has a tree structure depends on its dimension and partition rule. If a node as a parent (in red) is pruned as shown in Figure~\ref{PrunWorkflow}, its children are also pruned and the intensity of each voxel in its corresponding block will be set by the average value of this block.

CARP takes an $m$-dimensional image $\by = \{y(\bx): \bx \in \om\}$ observed at an $m$-dimensional rectangular ``pixel'' space $\om$ of the form  
\[
\om=[0,n_1 - 1]\times [0,n_2 - 1]\times \cdots \times [0,n_m - 1], 
\]
where the notation $[a,b]$ is the set $\{a,a+1,\ldots,b\}$ for two integers $a$ and $b$ with $a\leq b$. 
This means CARP can be readily applied to images of various dimensions including but not limited to 2D still images and 3D videos. Without loss of generality, we assume $n_i=2^{J_i}$ in the $i$th dimension for $i=1,2,\ldots,m$; an image of general size can be upsized to such dimensions through padding. The total number of pixels is $n = 2^J$, where $J = \sum_{i = 1}^m J_i$. In our toy example in Figure~\ref{PrunWorkflow}, we have $m = 2$, $J_1 = J_2 = 2$, and $J = 4$. 

The effectiveness of the compression highly depends on the representation power of the transform in use, especially its adaptivity to local and spatial features in an image. In CARP, this is achieved by a Bayesian probabilistic modeling strategy, where adaptivity to image features is incorporated into a wavelet transform based multi-scale image processing framework. In particular, we use random recursive partitioning on $\om$ to induce models on wavelet transforms that incorporate such adaptivity. In the following section, we describe some basic concepts related to recursive dyadic partitioning, which will form the building blocks for the model used in CARP.

\subsection{Recursive dyadic partitioning with pruning}
While multi-scale wavelet transforms enjoy excellent scalability, a deterministic transform may fail to efficiently adapt to the rich spatial and local features present in a multi-dimensional image. We enrich the representation power and effectiveness of wavelets by a convolution between a classic 1D wavelet transform and a class of permutation of the index space $\om$ of the pixels. 

The space consisting of all permutations of pixels in $\om$ is massive. Considering all $n!$ permutations of pixels in $\om$ is not only computationally prohibitive but wasteful as well because the vast majority of permutations ignores the spatial features in the image. In CARP, we only consider the class of permutations induced by a recursive dyadic partitioning (RDP) on $\om$, which includes a rich class of permutations for effective representation of the image while allowing scalable learning of the optimal permutation among this class---with computational complexity $O(n)$. An RDP on $\om$ denoted by $\T$ as it is essentially a bifurcating tree, consists of a sequence of nested partitions on $\om$, i.e., $\T=\cup_{j=0}^{J} \T^{j}$ with the partition $\T^j$ being the set of all blocks at level $j$ for $j = 0, \ldots, J$. Specifically, we start with $A_{0, 0} = \om$ and $\T^0 = \{A_{0, 0}\}$. For each $j = 0, \ldots, J - 1,$ $\T^{j+1}$ is obtained by dividing each set in $\T^{j}$ into two halves  along a divisible dimension, i.e., $A_{j,k} = A_{j+1,2k}\cup A_{j+1,2k+1}$ for $k=0,1,\ldots,2^{j}-1$. The last level $\T^J$ contains all the single elements in $\om$, which are referred to as \textit{atomic} nodes.

From now on we shall refer to the partition blocks as ``nodes'' in the partition tree $\T$. Two children nodes are formed by dividing a parent node into two halves in one of its dimensions, and $\T^J$ consists of the leaf nodes, each of which contains a single element in $\om$. Note that each RDP induces a unique permutation of $\om$, with the order of the pixels given by the binary coding sequence tracking the left/right children that each pixel belongs to along the corresponding branch in the tree. Figure~\ref{PrunWorkflow} shows one RDP from Level 0 ($\om$) down to Level 4, where all pixels at Level 4 are ordered according to the induced permutation of $\om$.

Parsimonious representation is crucial for image compression. We improve upon RDPs in this regard by incorporating early stopping to prune the partition tree induced by an RDP. Indeed, a complete RDP $\T$ might not be necessary to represent an image when there are homogeneous nodes with almost constant intensities therein. For example in Figure~\ref{PrunWorkflow}, the entire sub-branch started from the yellow node at Level 2 can be compressed before $\T$ reaches the last level. In CARP, we prune $\T$ by an early stopping to help effectively compress similar blocks in an image. 
Note that all descendants of a pruned node are pruned by design.  

Each RDP turns the pixel space $\om$ into a vector of the same length as the number of pixels, decoupling the image intensity to a 1D vector and spatial structure described by a partition tree with pruning. We next describe generative models for RDPs and the image given RDPs. 

\subsection{Generative modeling of RDPs} \label{sec:prior.RDP}
We use a hierarchical Bayesian model to adaptively learn an optimal RDP through maximizing the posterior distribution. Our emphasis is to obtain a probabilistic description of RDPs with pruning. To this end, we adopt a generative distribution called ``random RDP'' (RRDP) proposed in \cite{wongandma:2010,ma:2013,warp} as the prior on the RDP. While RRDP has been successfully applied to various tasks such as testing of conditional association and image reconstruction, there is no work to consider using RRDP for compression. For any node $A$ in $\T$, $\blam$ denote a mapping from $A$ to a hyperparameter $\blam(A)$ that specifies the probability to divide $A$ in each of its divisible dimensions. RRDP specifies the probability of partitioning $A$ in its $i$th dimension for $i = 1, \ldots, m$ using a vector-valued hyperparameter $\blam(A) = (\lambda_1(A), \ldots, \lambda_m(A))$. By default, we set the value of $\blam$ to be such that all divisible dimensions of each $A$ have equal probability to be divided. 

To prune a tree $\T$, let $\lambda_0(A)$ be the probability of node $A$ being pruned. This probability is the success probability in a Bernoulli distribution if one introduces an indicator accompanying each node in $\T$ to indicate whether or not it is pruned. However, it turns out that we can achieve the equivalent pruning by introducing an absorbing state in a hierarchical modeling manner through Markov trees; see Section~\ref{sec:model.pixel} for the detailed formulation. 

\subsection{Data generating model given RDPs: Markov-tree wavelet regression} \label{sec:model.pixel}

To achieve a parsimonious image representation, the next component of our model aims at pruning the tree in nodes where the pixel intensities are similar enough. 
To this end, we use wavelet shrinkage. In particular, given an RDP, we adopt a wavelet regression model as the data generative mechanism for the image. There is a rich literature on how to effectively carry out thresholding and shrinkage on the wavelet coefficients \cite{Chipman1997,Clyde+George:00,Brown2001,Johnstone+Silverman:05,Morris2006}, and we shall use a Bayesian wavelet regression model with a Markov tree prior on the wavelet coefficients to achieve adaptive shrinkage \cite{crouse:1998markovtree, warp}. An important benefit of adopting a Bayesian model for the image given the RDP is that we can now combine it with our Bayesian model on the RDPs to form a coherent hierarchical model, allowing inference to be carried out in a principled manner (through maximizing the posterior distribution) without {\em ad hoc} strategies to ``stitching'' together separate algorithmic pieces. 

Specifically, conditional on an RDP tree $\T$ and following an application of Haar wavelet transform to the vectorized image under $\T$, the Bayesian wavelet regression model is as follows
\begin{align} \label{eq:dwt}
    w_{j,k} & = z_{j,k}+ u_{j,k} \\
      	 z_{j,k} \,|\,S_{j,k} & \ind  \left\{ 
    \begin{array}{lr}
    \delta_0(\cdot) & \text{if $S_{j, k} = 0$ or 2}\\
    \text{Normal}(0, \tau_j^2 \sigma^2) &  \text{if } S_{j, k} = 1
    \end{array} \right. 
\end{align} for $j=0,1,\ldots,J-1$ and $k=0,1,\ldots,2^{j}-1$. Here $w_{j,k}$, $z_{j,k}$, $u_{j,k}$ are the $k$th wavelet coefficient, signal, and ``noise'' at the $j$th scale in the wavelet domain, respectively. The ternary latent state variable $S_{j,k}$ indicates whether $z_{j,k}$ is from $\delta_0(\cdot)$ (a point mass at 0) if $S_{j,k} = 0$ or 2, or a normal distribution with mean 0 and variance $\tau_j^2 \sigma^2$ if $S_{j, k} = 1$. To achieve adaptive pruning, we model $S_{j,k}$ jointly by a Markov tree model \cite{crouse:1998markovtree} such that if $S_{j-1,\lfloor k/2\rfloor} = 2$ then $S_{j, k} = 2$ with probability 1. Thus $S_{j,k}=2$ is an ``absorbing state'' representing the pruning of a branch of $\T$. If $S_{j-1,\lfloor k/2\rfloor} \neq 2$ then $S_{j, k} = (0, 1, 2)$ with probabilities 
$$
(\rho(A_{j, k}) \{1 - \lambda_0(A_{j,k})\}, \{1 - \rho(A_{j, k})\}\{1 - \lambda_0(A_{j,k})\}, \lambda_0(A_{j,k})), 
$$
where $\rho(A)$ is the so-called spike probability in Bayesian spike-and-slab models. We assume $u_{j, k} \sim N(0, \sigma^2)$ independently across $j$ and $k$. 

It is worth noting that in the context of compressing noiseless images, the ``noise'' term $u_{j,k}$ quantifies the extent of local variations in pixel intensities to which one can ignore to produce a compressed image, and therefore its standard deviation $\sigma$ becomes a parameter for setting how aggressively (in terms of the compression ratio) one wants to compress the image through pruning the tree. This tuning parameter $\sigma$ will be in a monotone relation to the final compression rate of the image, and thus can be set by the user to achieve the desired rate of compression.

\subsection{Posterior inference and image representation via MAP}

For image compression, we need to find a single representative RDP that most effectively represent features in the image. One strategy is to draw a random sample of $\T$ from its posterior distribution, and use the induced permutation for compression. However, a more appealing approach is to resort to a particular non-random representative sample. To this end, we maximize the posterior probability of $\T$ based on its marginal posterior distribution. In other words, we aim to find the {\em maximum a posteriori} (MAP) estimate for $\T$, which we denote as $\hat{\T}$.

We first need to find the marginal posterior distribution for $\T$. A remarkable observation is that under the above model, the posterior of $\T$ is conjugate---it is still an RRDP distribution with pruning, but with updated posterior selection probabilities $\tilde{\blam}$ and updated pruning probabilities $\tilde{\lambda}_0$, where we use tilde to indicate the posterior updated values for the parameters $\blam$ and $\lambda_0$. Such conjugacy can be obtained in a general Markov tree setting as long as the joint distribution of $(w_{j,k}, z_{j, k})$ given $S_{j,k}$ is independent for all $j$ and $k$, which has been studied in Theorem 2 of~\cite{warp}. We next derive such conjugacy specifically for our model in~\eqref{eq:dwt} and provide recipes for inference on $\T$.

Let $\A$ be the set collecting the nodes generated by all possible RDPs. Let $\Psi(A)$ be the marginal likelihood of node $A$ for $A \in \A$. Hereafter by the marginal likelihood ``of a node'' we refer to the marginal likelihood from all data with locations in $A$ if the parent node of $A$ is not pruned (i.e., integrating out those $\T$'s that contain the node $A$ with respect to their prior distribution). We start with introducing a few quantities associated with each node $A$ that will be used for describing the posterior. Let $D(A)$ be the set collecting all divisible dimensions of $A$, $y(A) = \sum_{\bx\in A}y(\bx)$ the sum of observations with locations in $A$, and SST$(A)$ the corrected sum of squares of all data with locations in $A$, i.e., SST$(A)$ = $\sum_{x\in A}(y(\bx)- y(A)/|A|)^2$ with $|A|$ being the number of locations in $A$. For each $d \in D(A)$ such that $A$ is divided in its $d$th dimension, let $(A_l^{(d)}, A_r^{(d)})$ be the pair of left and right children nodes, and $w_d(A)  = \{y(A^{(d)}_{l}) - y(A^{(d)}_{r})\}/\sqrt{|A|}$ be the the Haar wavelet coefficient on $A$.

The marginal likelihood $\Psi(A)$ can be obtained in a recursive manner at a complexity $O(n)$. To see this, first decompose  $\Psi(A)$ according to the $|D(A)| + 1 $ possible actions on $A$: 
\begin{equation}
    \label{eq:psi.1}
\Psi(A) = \lambda_0 \Psi_0(A) + (1 - \lambda_0)\sum_{d \in D(A)} \lambda_d \Psi_d(A),
\end{equation}
where $\Psi_d(A)$ is the marginal likelihood of $A$ if $A$ is divided in its $d$th dimension given $A$ is not pruned, and $\Psi_0(A)$ is the marginal likelihood of $A$ if $A$ is pruned. When $d = 0$, all wavelet coefficients in $A$ and its descendants have mean 0, and thus 
\begin{equation}\label{eq:psi.2}
\Psi_0(A)=({2\pi\sigma^2})^{-\frac{|A|-1}{2}}\exp\left\{-\frac{\text{SST}(A)}{2\sigma^2}\right\}. 
\end{equation}
When $d \in D(A)$, $\Psi_d(A)$ is independently decomposed into the marginal likelihood of $w_d(A)$ that is spelled out in~\eqref{eq:dwt} and the marginal likelihoods of its two children nodes, that is, 
\begin{equation} \label{eq:psi.3}
\Psi_d(A) = \{\rho(A) N(w_d(A); 0, (1 + \tau_j^2) \sigma^2) + (1-\rho(A)) N(w_d(A); 0, \sigma^2)\} \Psi(A_{l}^{(d)})\Psi(A_{r}^{(d)}), 
\end{equation}
where $N(a; b, c^2)$ is the density function of Normal($b, c^2$) evaluated at $a$. 
A bottom-up algorithm is then readily applicable to obtain all $\Psi(A)$'s by combining~\eqref{eq:psi.1},~\eqref{eq:psi.2}, and~\eqref{eq:psi.3} and letting $\Psi(A) = 1$ for all atomic nodes $A$. 

Once $\Psi(A)$ is calculated, a direct application of Bayes' theorem gives 
$\tilde{\lambda}_0(A)=\lambda_0(A)\Psi_{0}(A)/\Psi(A)$ and $\tilde{\lambda}_d(A) = \lambda_d(A) \Psi_d(A)/\{\sum_{d \in D(A)} \lambda_d(A) \Psi_d(A)\}$, which respectively are the posterior probability of pruning $A$ and the posterior probability for $A$ to be divided in dimension $d$ if $A$ is not pruned. These two mappings completely describe the marginal posterior distribution of $\T$.

We then compute the MAP tree $\hat{\T}$ by standard bottom-up dynamic programming as follows, which again incurs complexity $O(n)$.
Define a recursive mapping $\kappa:\A\rightarrow [0,1]$:
 \begin{eqnarray}
 \kappa(A) =\left\{\begin{array}{cc}
 1, &\text{if A is atomic},\nonumber\\
 	\max\{\tlam_0(A), \tlam_{-0}^{\max}(A)\}, & \text{otherwise,}
 \end{array} \right.
 \end{eqnarray}
 where $$\tilde{\lambda}_{-0}^{\max}(A)=(1-{\tilde{\lambda}}_0(A))\max_{d\geq 1}\{{\tilde{\lambda}}_d(A) \kappa(A^{(d)}_l)\kappa(A^{(d)}_{r})\}.$$
Once the mapping $\kappa$ has been computed on all $A\in\A$, $\hat{\T}$ can be generated inductively as follows. Note that $\hat{\T}$ consists of partition index sets $\{\hat{A}_{j,k}:0\leq k < 2^{j},0\leq j\leq J\}$ and a pruning indicator vector associated with each $\hat{A}_{j,k}$. 
For $j=0$, $\hat{A}_{0,0}=\om$. Suppose $\hat{\T}$ have been generated for all levels no more than $j$. For each $\hat{A}_{j,k}$ that is not pruned, define $\hat{d}_{j,k}={\rm argmax}_{d\geq 1} \tlam_d(A)\kappa(A^{(d)}_{l})\kappa(A^{(d)}_{r})$ with $A=\hat{A}_{j,k}$. 
Then, the block $A_{j,k}$ is pruned if $\tlam_0(A) > (1-{\tlam}_0(A)){\tlam}_d(A) \kappa(A^{(d)}_l)\kappa(A^{(d)}_{r})$ with $d = \hat{d}_{j,k}$ and $A = \hat{A}_{j,k}$; otherwise, the partition sets in level $j+1$ are
$\hat{A}_{j+1,2k} = A^{(\hat{d}_{j,k})}_{l}$ and $\hat{A}_{j+1,2k+1} = A^{(\hat{d}_{j,k})}_{r}$. Recall that all descendants of a pruned node are pruned by design; hence, for each $\hat{A}_{j,k}$ that is pruned, we randomly select a direction $d_0 \in D(\hat{A}_{j,k})$ and set $\hat{A}_{j+1,2k} = A^{(d_0)}_{l}$ and $\hat{A}_{j+1,2k+1} = A^{(d_0)}_{r}$. Note the reconstructed image is invariant to this random choice as all pixels in the pruned block will be reconstructed as a common constant. 

\subsection{Encoder/decoder and compressed structures}
Given the permutation of the original image induced by $\hat{\T}$, under which the order of each pixel is given by binary coding of the branch under $\T$ to which each pixel belongs, we have a vectorization of the original image. Within a pruned node, the ordering of the pixels is arbitrary. At this point, one has the flexibility of choosing the favorite encoder and decoder of this vectorized image. 

In addition, the encoding part in Figure~\ref{workflow}  includes a symbol encoder, which is used to reduce the coding redundancy. 
For the symbol encoder, we adopt the Huffman encoding method to reduce coding redundancy. Specifically, the Huffman table is derived from the estimated probability or frequency of occurrence for each possible value of the source symbol. Furthermore, the reduced symbols are stored as the compressed representation, while this Huffman table is also used in the decoder part to perform the inverse operation of the symbol encoder.

In our following numerical experiments, we use the 1D Haar DWT and a symbol encoder as the encoder part while a symbol decoder and the inverse DWT as the decoder part, respectively, due to their computational scalability. 

\subsection{Empirical Bayes for setting hyperparameters} 

We specify the two mappings $\rho(\cdot)$ and $\eta(\cdot)$ as well as parameters $\tau_{j}$ by reparameterizing them using five hyperparameters $(\alpha, \beta, C, \tau_0, \eta_0)$: $\rho(A) = \min(1, C 2^{-\beta j})$ for any node $A$ at the $j$th level, $\tau_{j} = 2^{-\alpha j} \tau_0$, and $\eta(A) = \eta_0$ for any node $A \in \A$. We use an empirical Bayes strategy to set the hyperparameters by maximizing the marginal likelihood $\Psi(\om)$ over a grid. 
We observe that specifying the hyperparameters at fixed values eliminates the need for computing the maximum likelihood estimates without sacrificing compression efficiency much. As such, our software allows both options. Under either option, a user just needs to specify a single parameter $\sigma$ to obtain images at desired compression ratios when applying CARP.

\section{Experiments} \label{sec:experiments} 

\begin{figure}[h!]
 \centering 
 \includegraphics[width=0.9\linewidth,height=4cm]{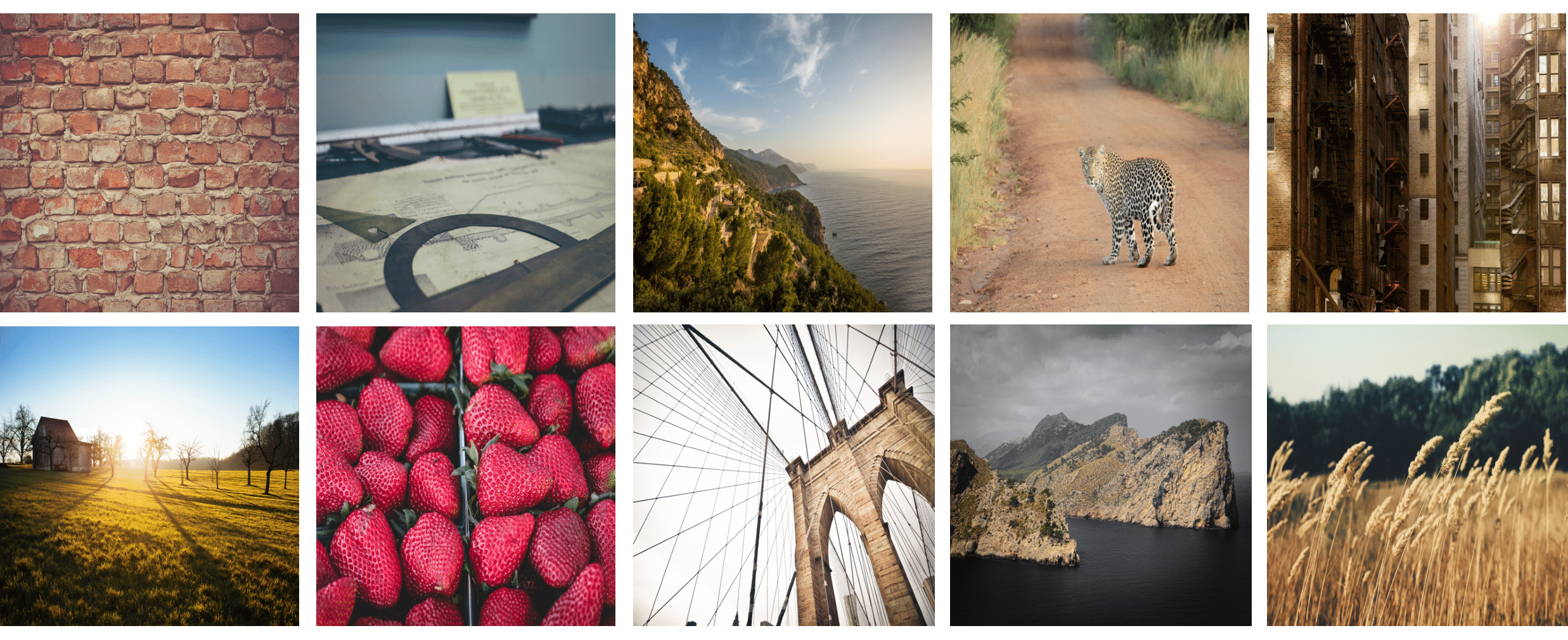}\\
 \includegraphics[width=0.92\linewidth,height=4cm]{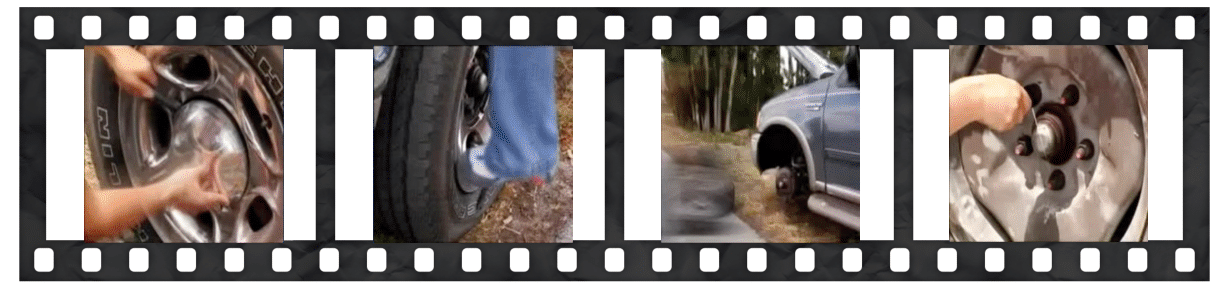}\\
 \includegraphics[width=0.91\linewidth,height=4cm]{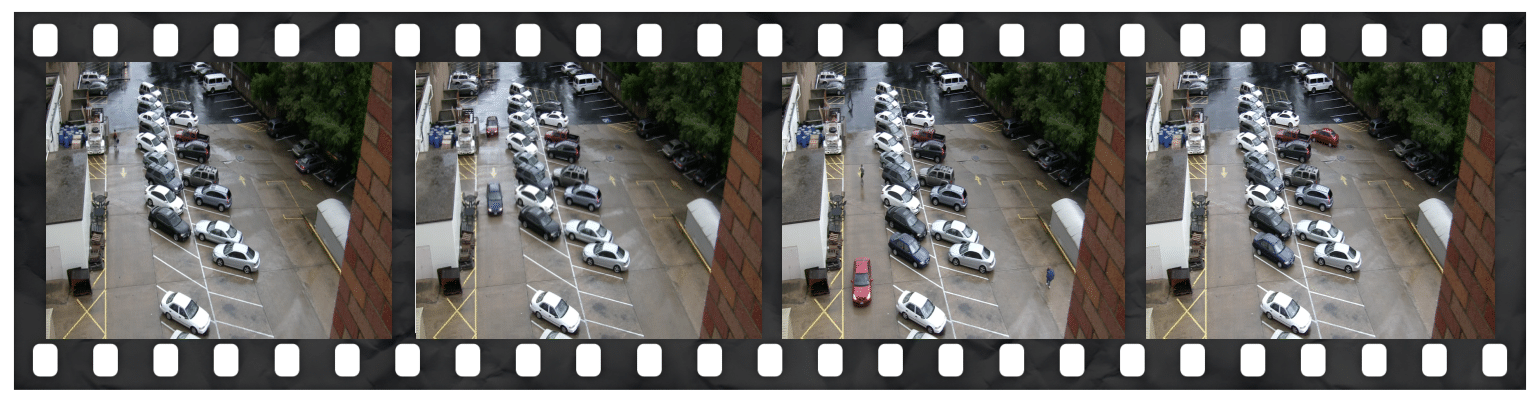}
 \caption{Selected example 2D still images (No.1-No.10) from CLIC challenge (top); Selected example frames of videos in the YouTube dataset (middle); Selected example frames of videos in the Surveillance video dataset (bottom).}\label{bpg}
\end{figure}

In this section, we compare CARP with several state-of-the-art compression methods using a variety of benchmark databases, including still images and videos of low and high resolutions. In particular, as shown in Figure \ref{bpg}, we use a 2D still image dataset from the 2020 CLIC workshop and challenge \url{http://challenge.compression.cc/tasks/},
a YouTube video dataset from~\cite{Alayrac16unsupervised}, and a surveillance video dataset from \cite{vezzani2010video}. CARP and its software implementation are readily applicable to all these types of images, while the competitors may tailor to images of a particular dimension. We thus compare CARP with a different batch of methods depending on the image type. 
In this section, we opt for fixed hyperparameters for simplicity. In particular, we use $\alpha = 0.5, \beta = 1, C = 0.05, \tau_0 = 1/\sigma$, and $\eta_0 = 0.4$.

\subsection{2D still images}
We compare CARP with popular compression methods designed for 2D images, including JPEG, JPEG2000, BPG, and a deep learning method. For BPG, we adopted the executable file in
\cite{bpg2017}; for the deep learning method, we used a pre-trained end-to-end optimized image compression method (`E2E-DL') in \cite{Balle17a}. CARP is applicable for colorful images by taking the mean of the results in all channels.
Here we randomly select 100 images from the 2020 CLIC workshop and challenge \url{http://challenge.compression.cc/tasks/} and resize the size of those images to $512\times 512$ to test each method. We provide these 100 images in the GitHub repository for CARP and display the first 10 images in the top rows of Figure \ref{bpg}.

\begin{figure}[ht!]
\centering 
	\begin{tabular}{cc}
		 \includegraphics[width=0.45\linewidth, height=3cm,trim={0.8cm 0.9cm 0 0},clip]{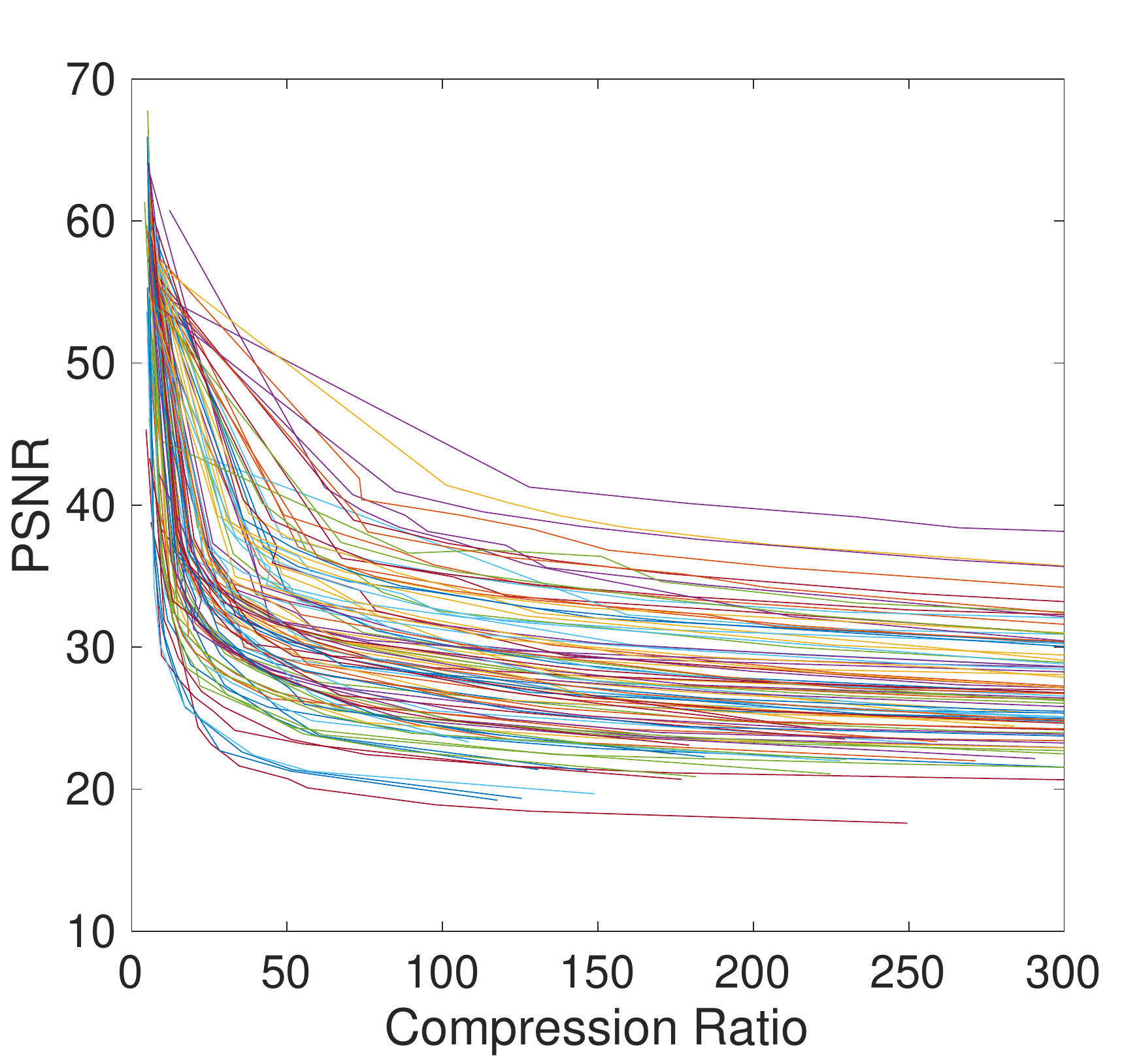} & 
		  \includegraphics[width=0.45\linewidth, height=3cm,trim={0.9cm 0.8cm 0 0},clip]{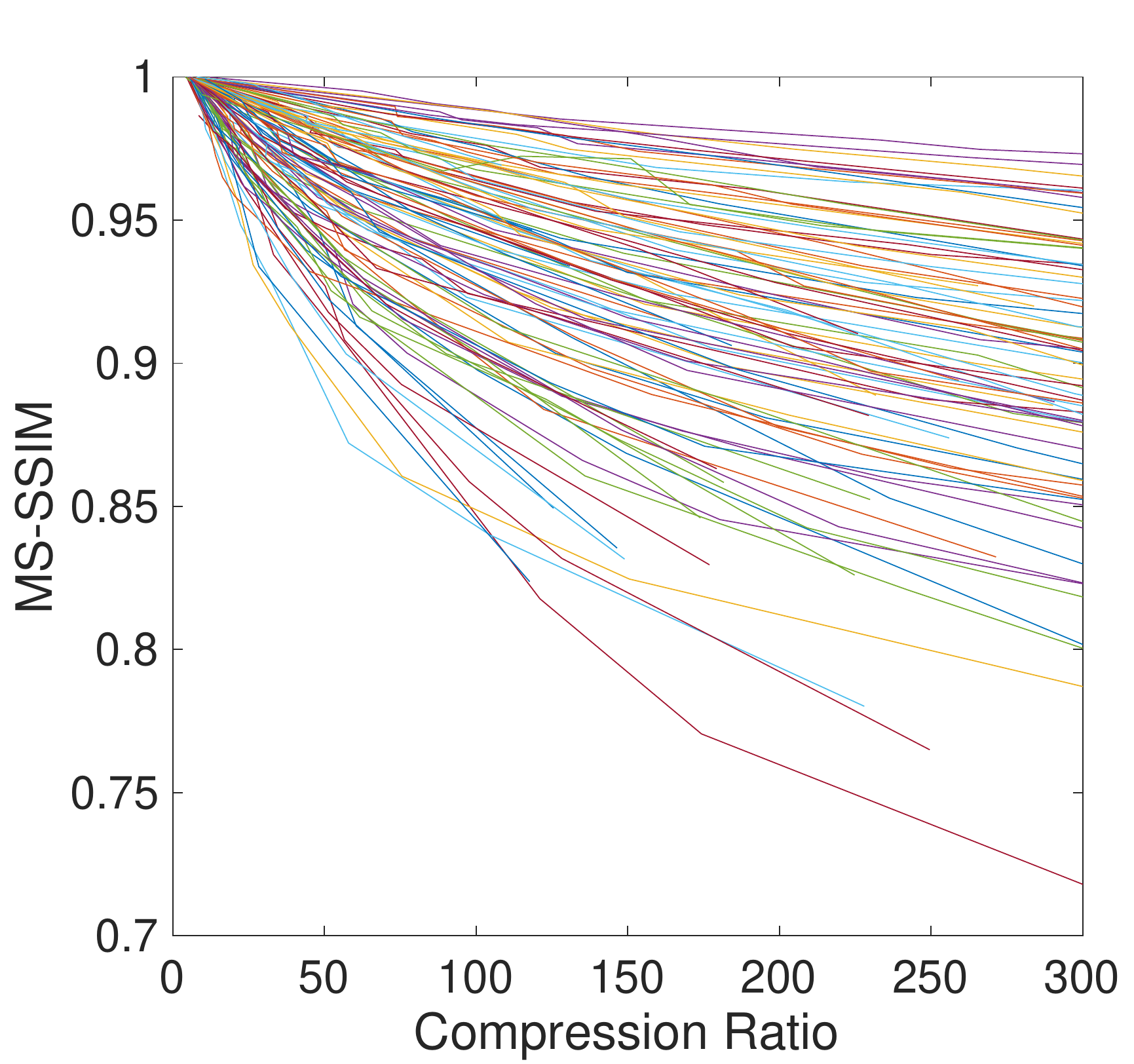} \\
		  {\small (a)  PSNR of CARP for 100 images} & {\small (b) MS-SSIM of CARP for 100 images} \\
		  \includegraphics[width=0.45\linewidth, height=3cm,trim={0.9cm 0.9cm 0 0},clip]{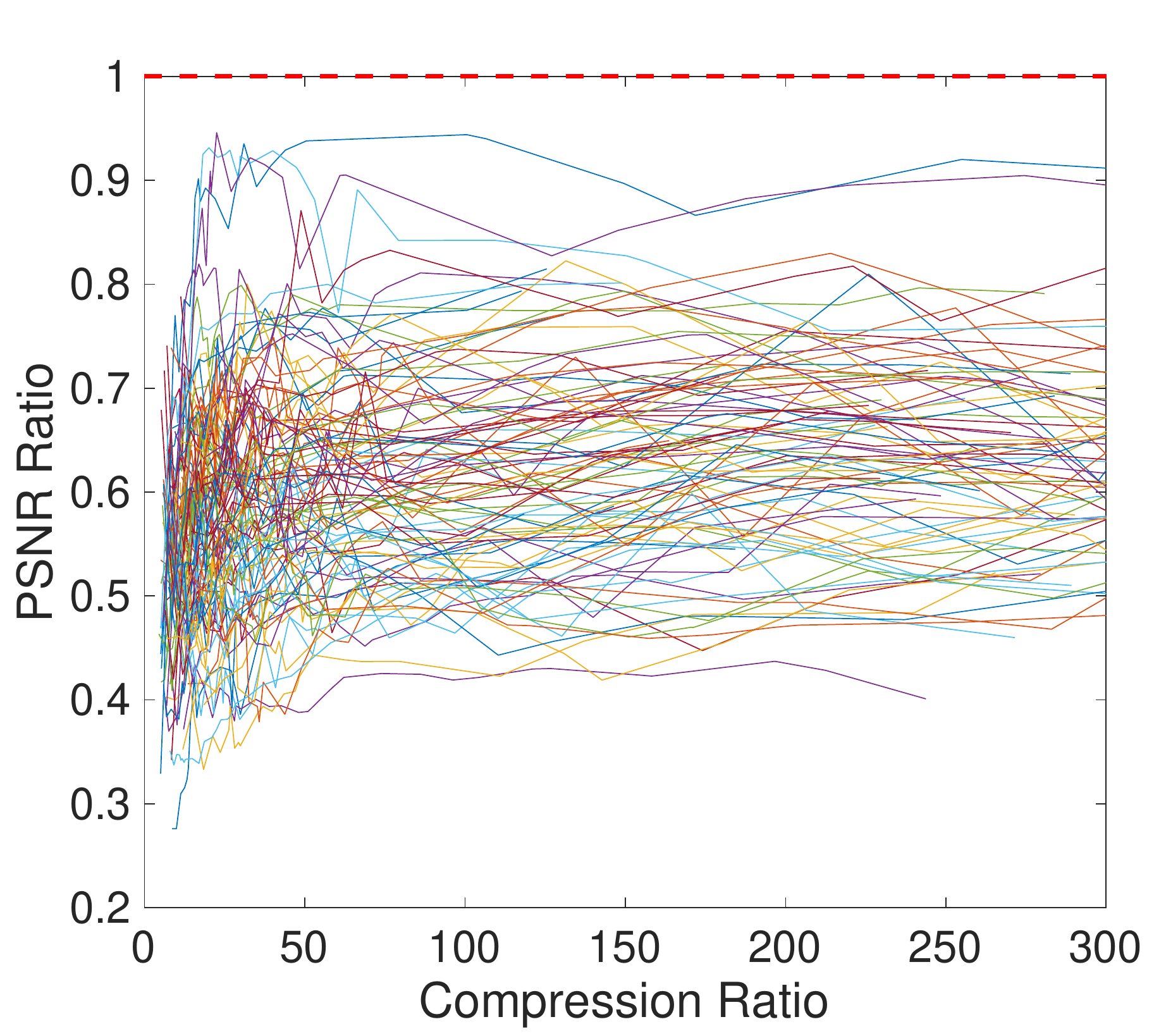} & 
		  \includegraphics[width=0.45\linewidth, height=3cm,trim={0.9cm 0.9cm 0 0},clip]{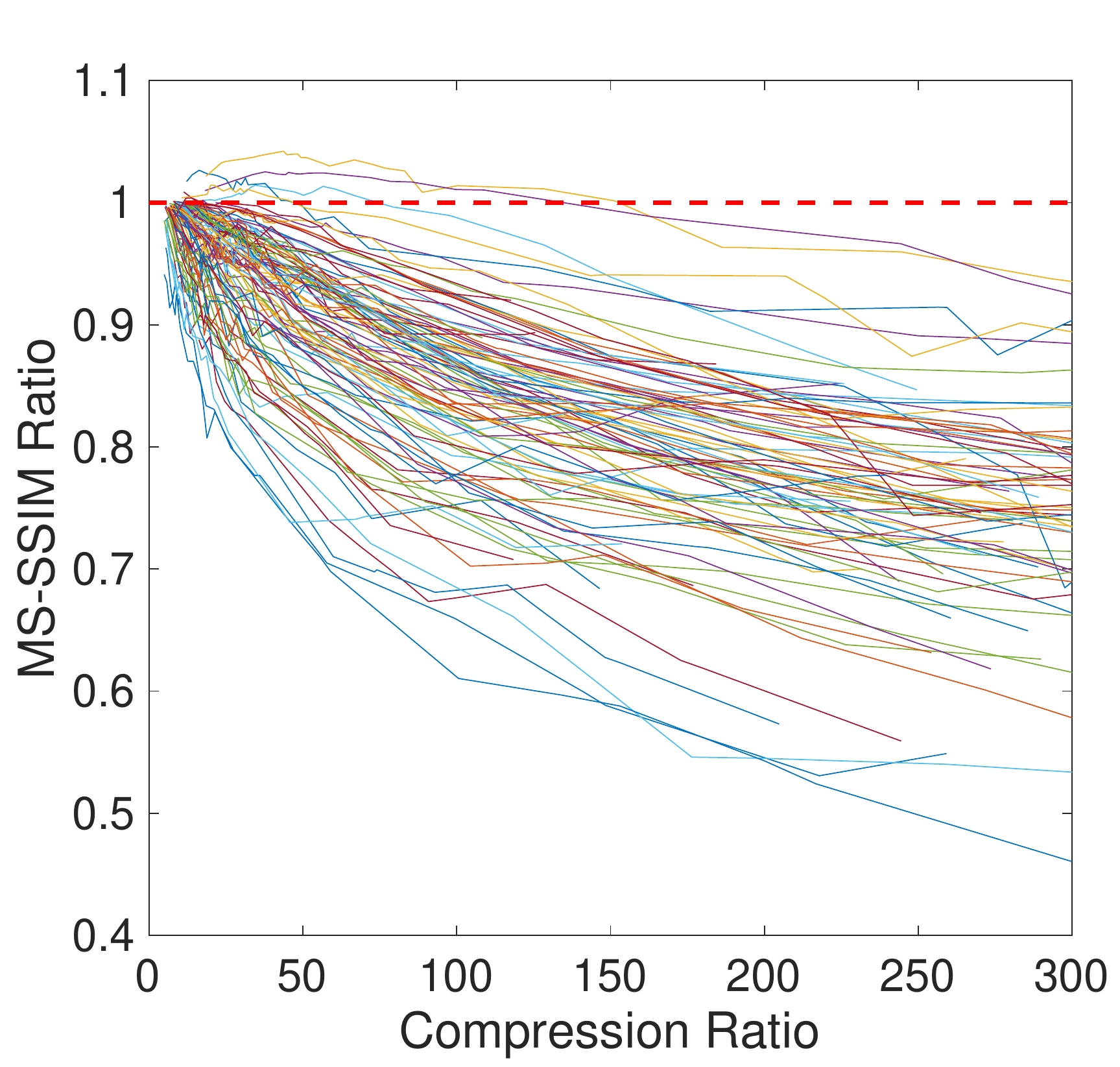} \\
		  {\small (c) PSNR ratio: JPEG/CARP} & {\small (d) MS-SSIM ratio: JPEG/CARP} \\
		  \includegraphics[width=0.45\linewidth, height=3cm,trim={0.9cm 0.8cm 0 0},clip]{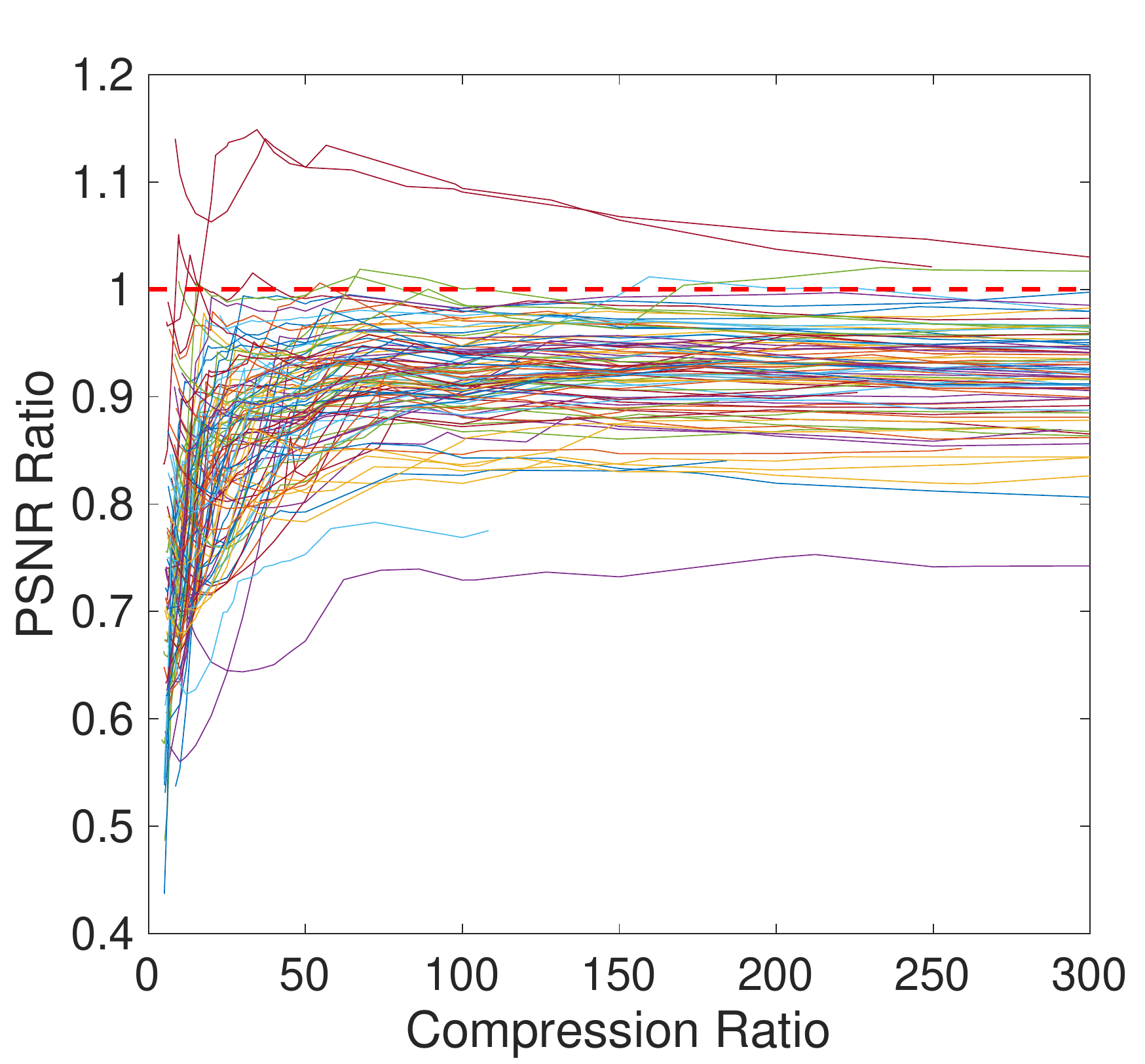} & 
		  \includegraphics[width=0.45\linewidth, height=3cm,trim={0.9cm 0.8cm 0 0},clip]{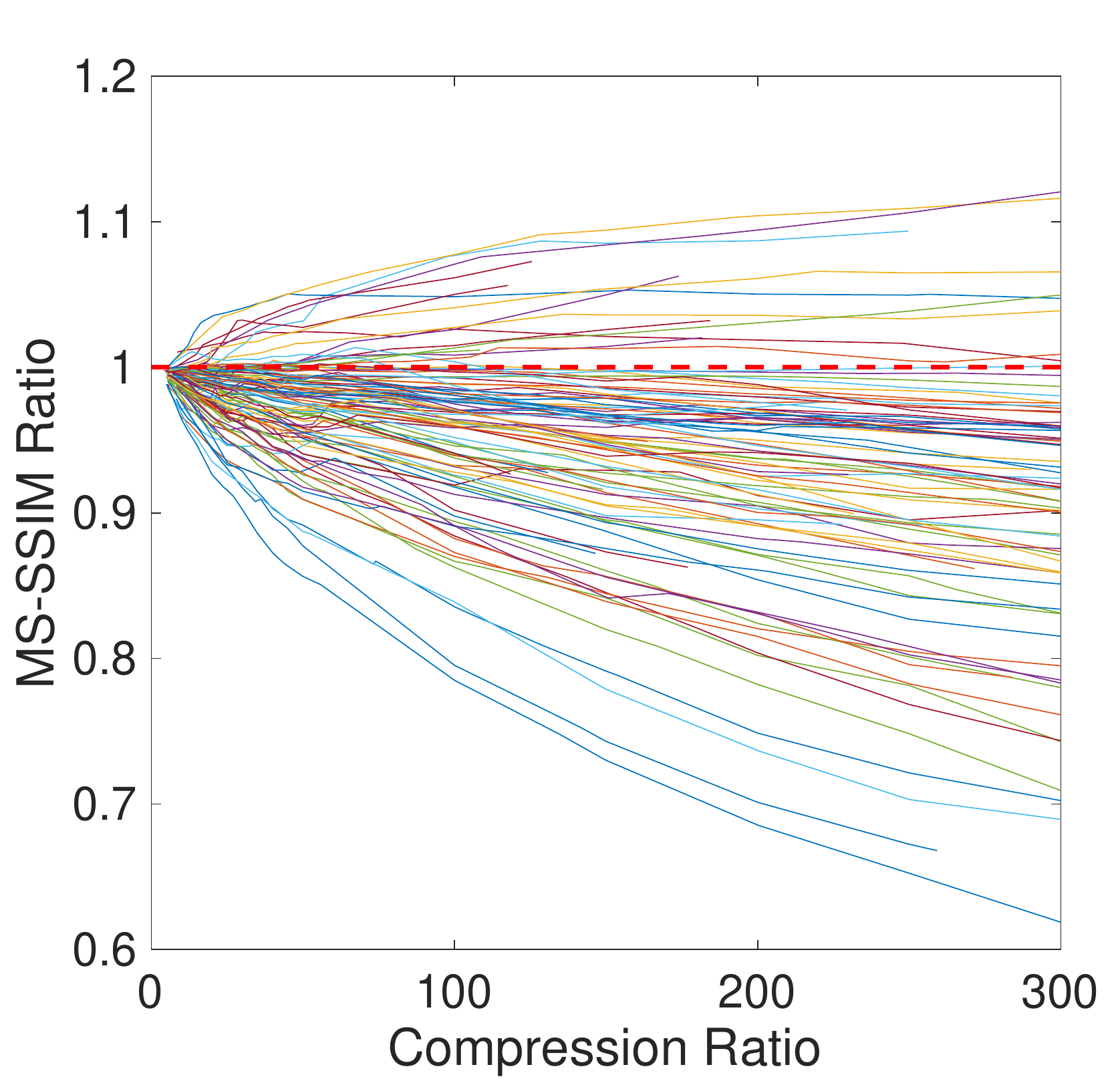} \\
		  {\small (e) PSNR ratio: JPEG2000/CARP} & {\small (f) MS-SSIM ratio: JPEG2000/CARP}\\
		  \includegraphics[width=0.45\linewidth, height=3cm,trim={0.9cm 0.9cm 0 0},clip]{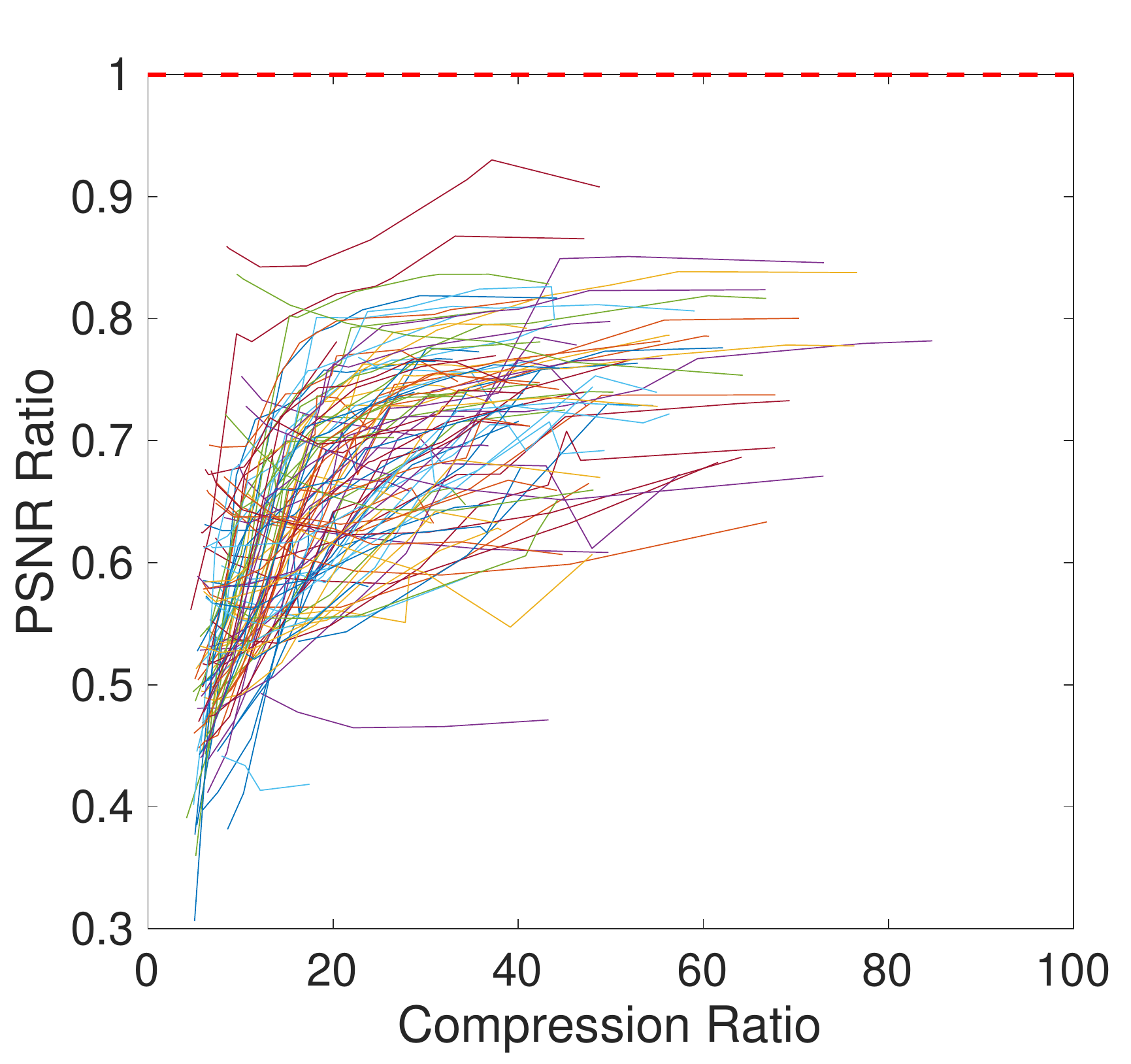} & 
		  \includegraphics[width=0.45\linewidth, height=3cm,trim={0.9cm 0.9cm 0 0},clip]{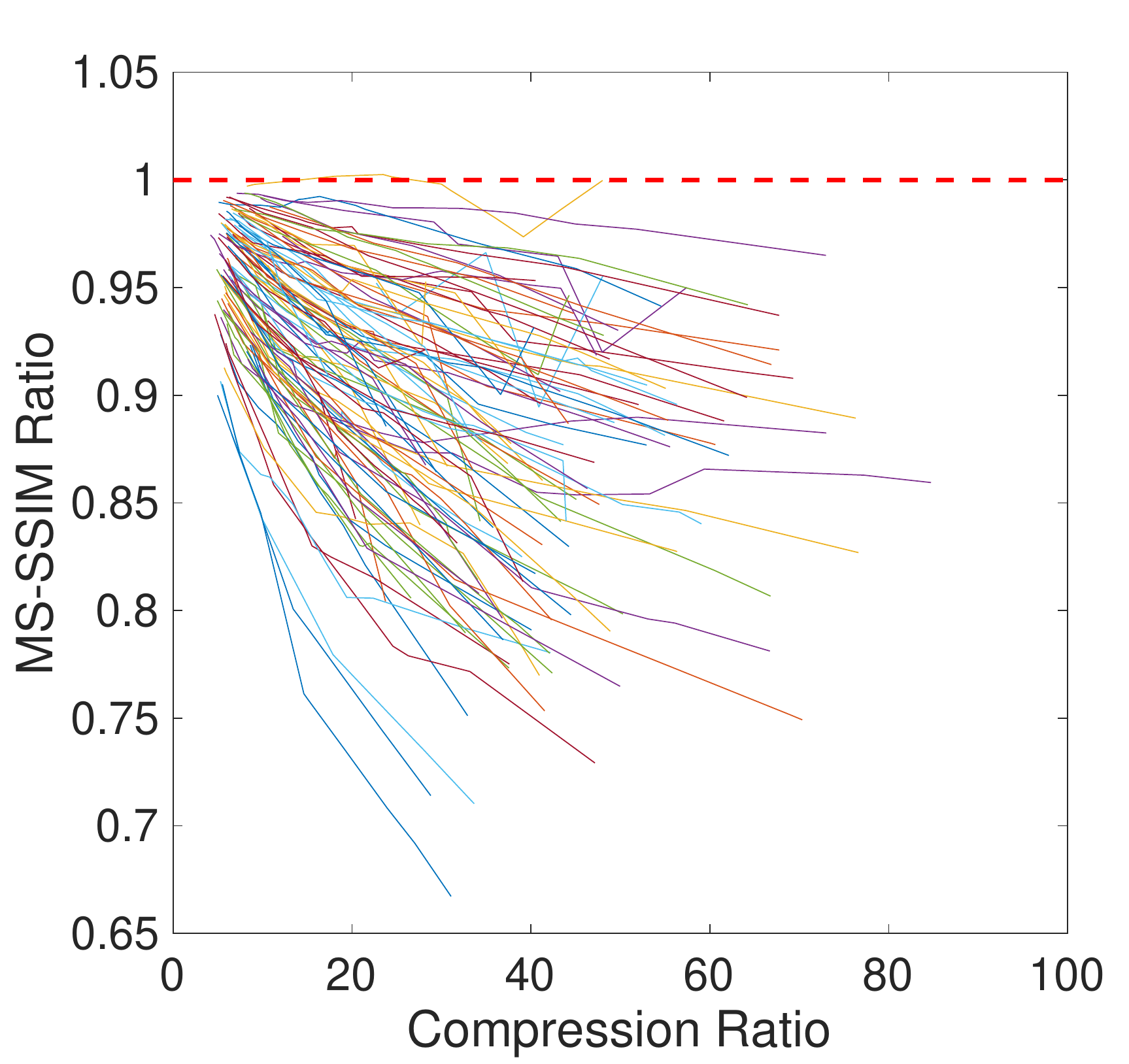} \\
		  {\small (g) PSNR ratio: E2E-DL/CARP} & {\small (h) MS-SSIM ratio: E2E-DL/CARP} \\
		  \includegraphics[width=0.45\linewidth, height=3cm,trim={1cm 0.9cm 0 0},clip]{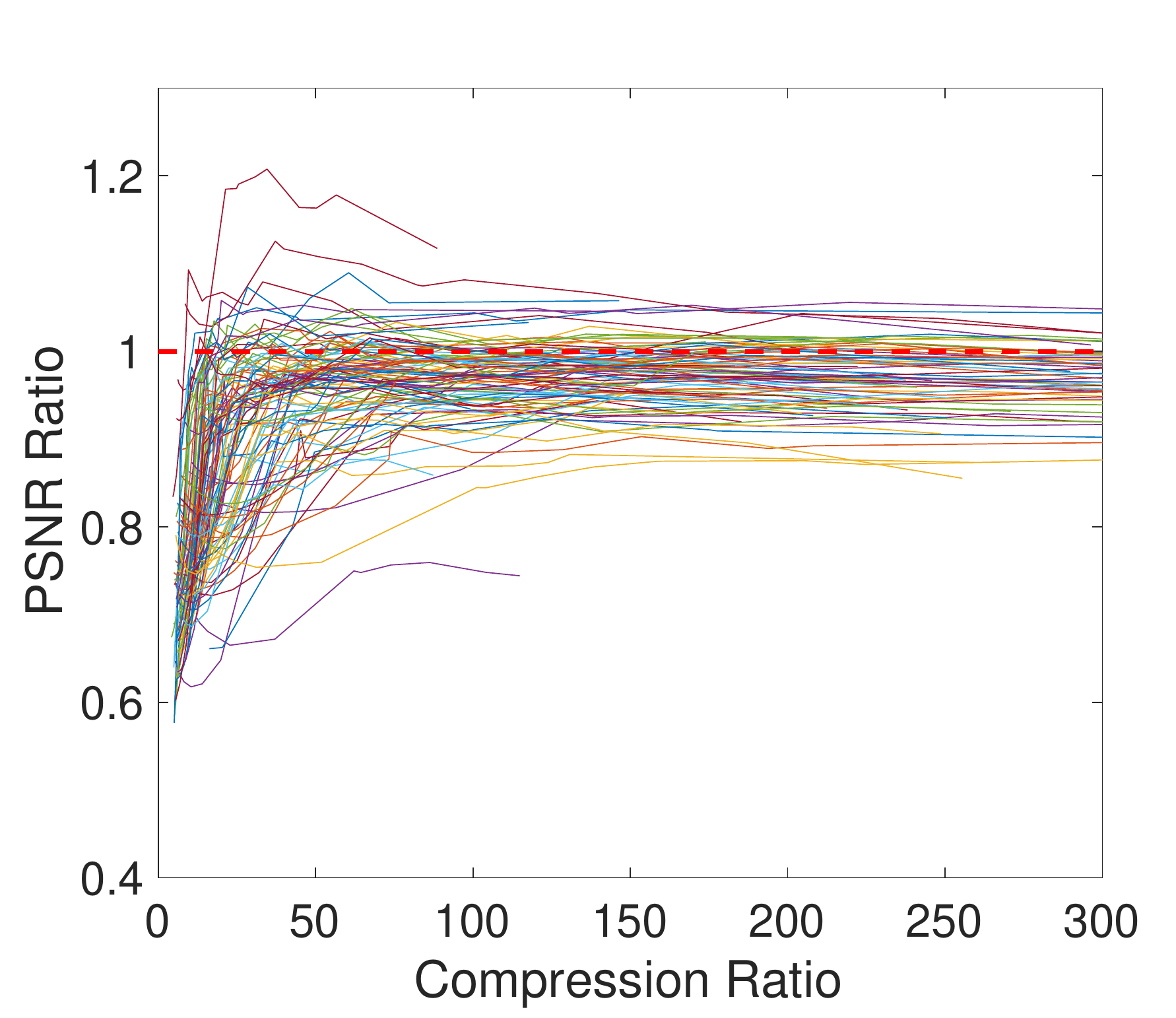} & 
		  \includegraphics[width=0.45\linewidth, height=3cm,trim={1.4cm 0.9cm 0 0},clip]{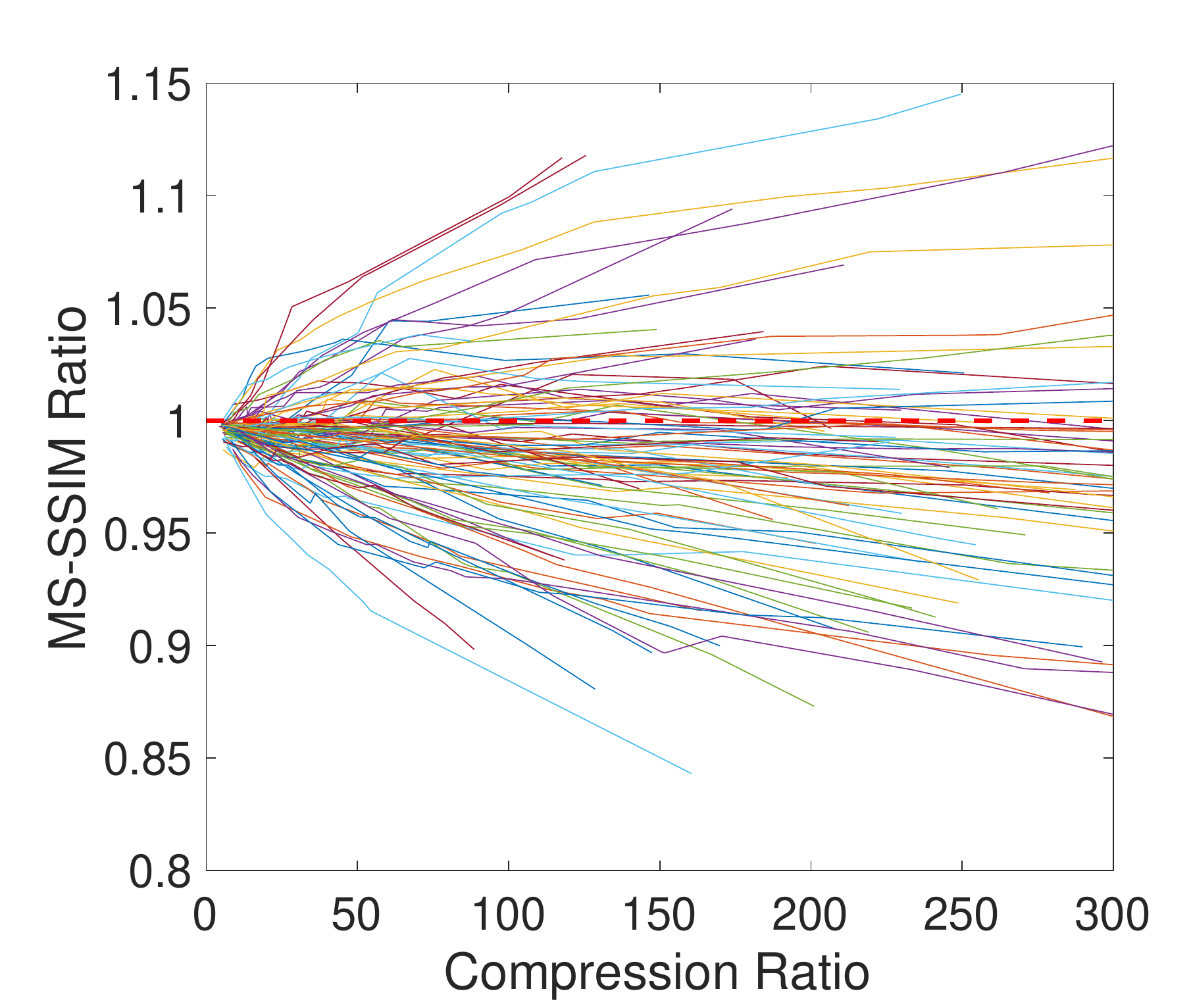} \\
		  {\small (i) PSNR ratio: BPG/CARP} & {\small (j) MS-SSIM ratio: BPG/CARP}
	\end{tabular}
\caption{2D still images: PSNRs of CARP for 100 individual images in (a) and MS-SSIM of CARP for 100 individual images in (b); PSNR ratio curves and MS-SSIM ratio curves for JPEG, JPEG2000, E2E-DL, and BPG relative to CARP in (c)- (j), respectively.}\label{mean100}
\end{figure}

\begin{figure}[ht!]
\centering 
	\begin{tabular}{ccc}
		 \includegraphics[width=0.33\linewidth, height=3cm]{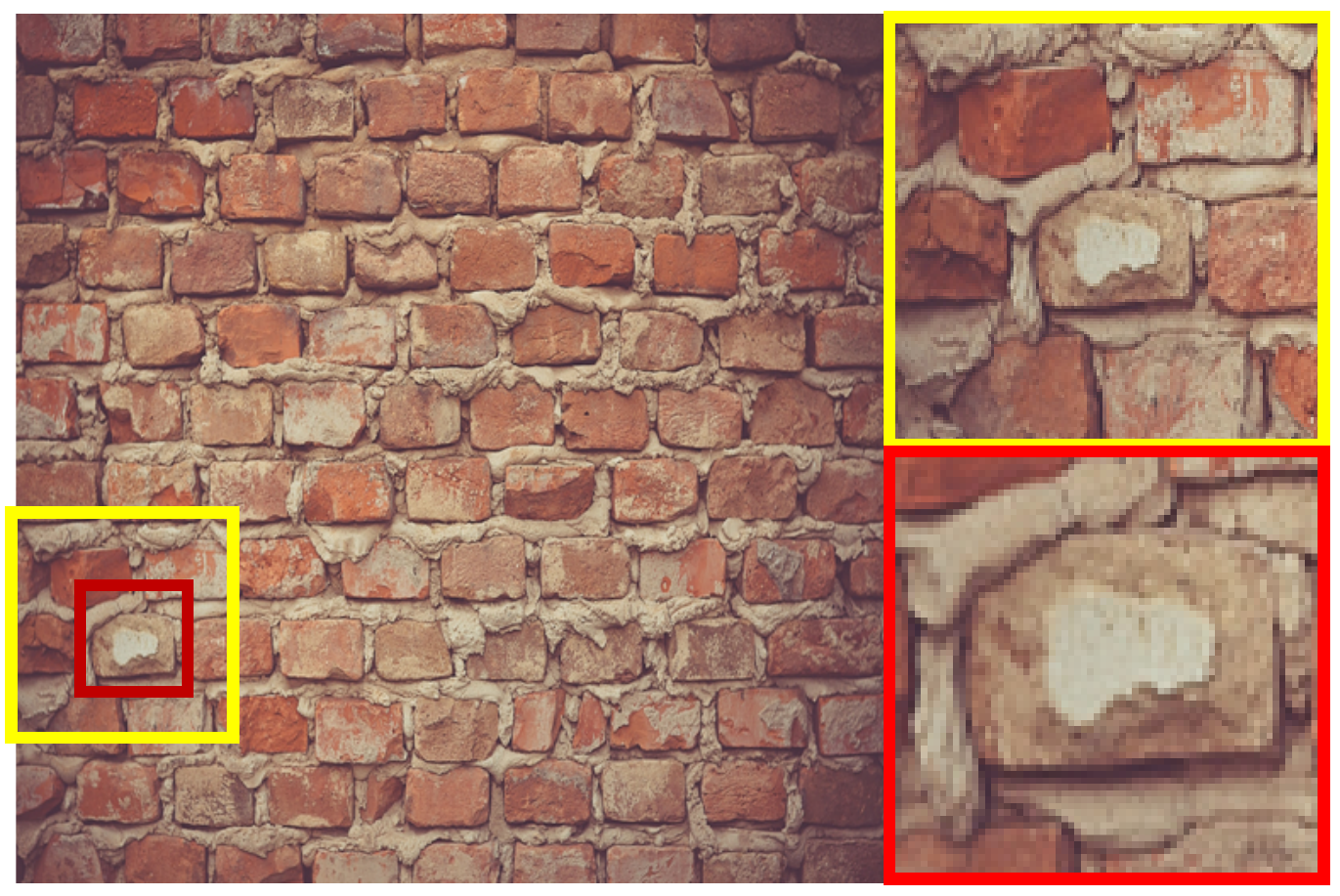}&\includegraphics[width=0.33\linewidth, height=3cm]{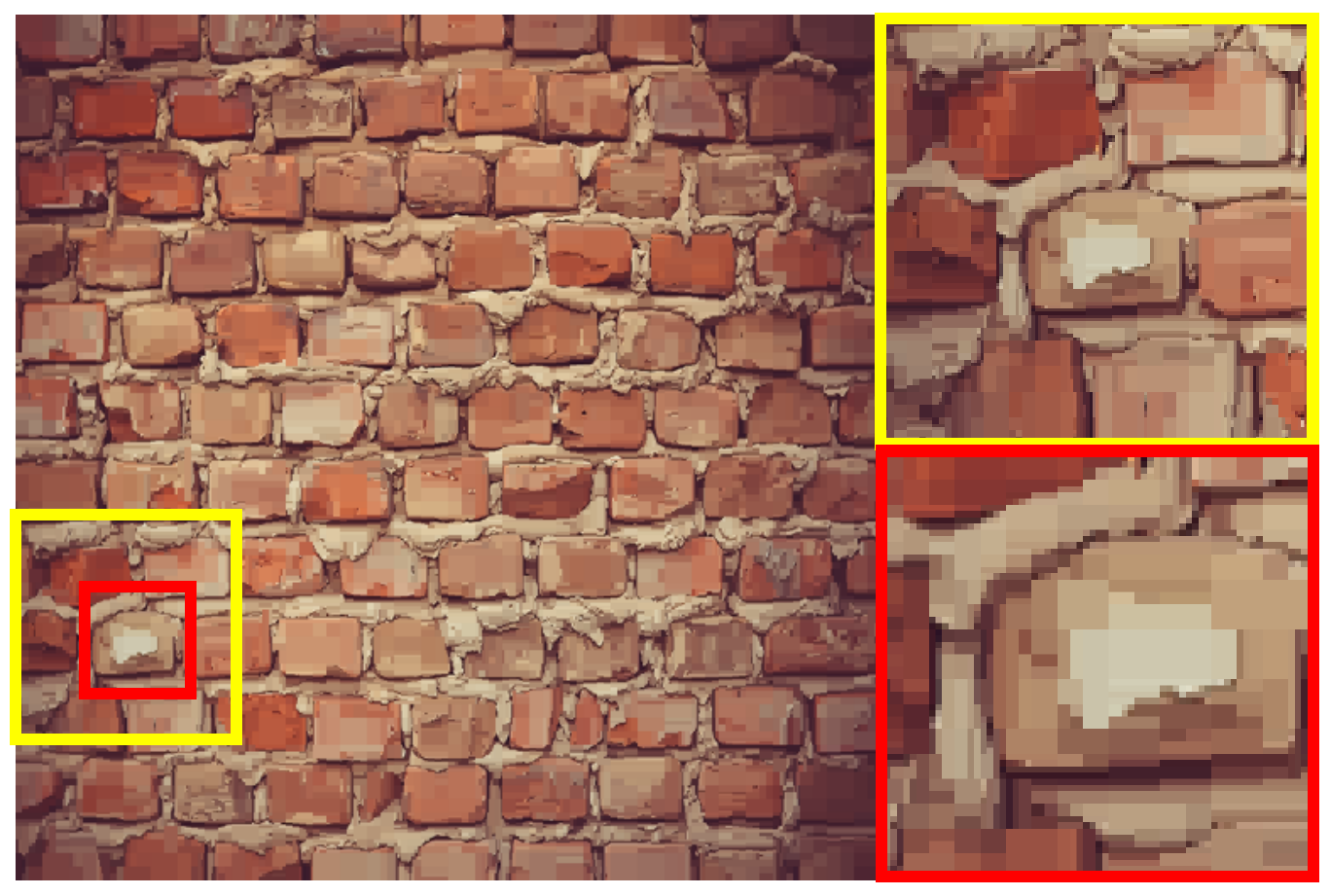}&\includegraphics[width=0.33\linewidth, height=3cm]{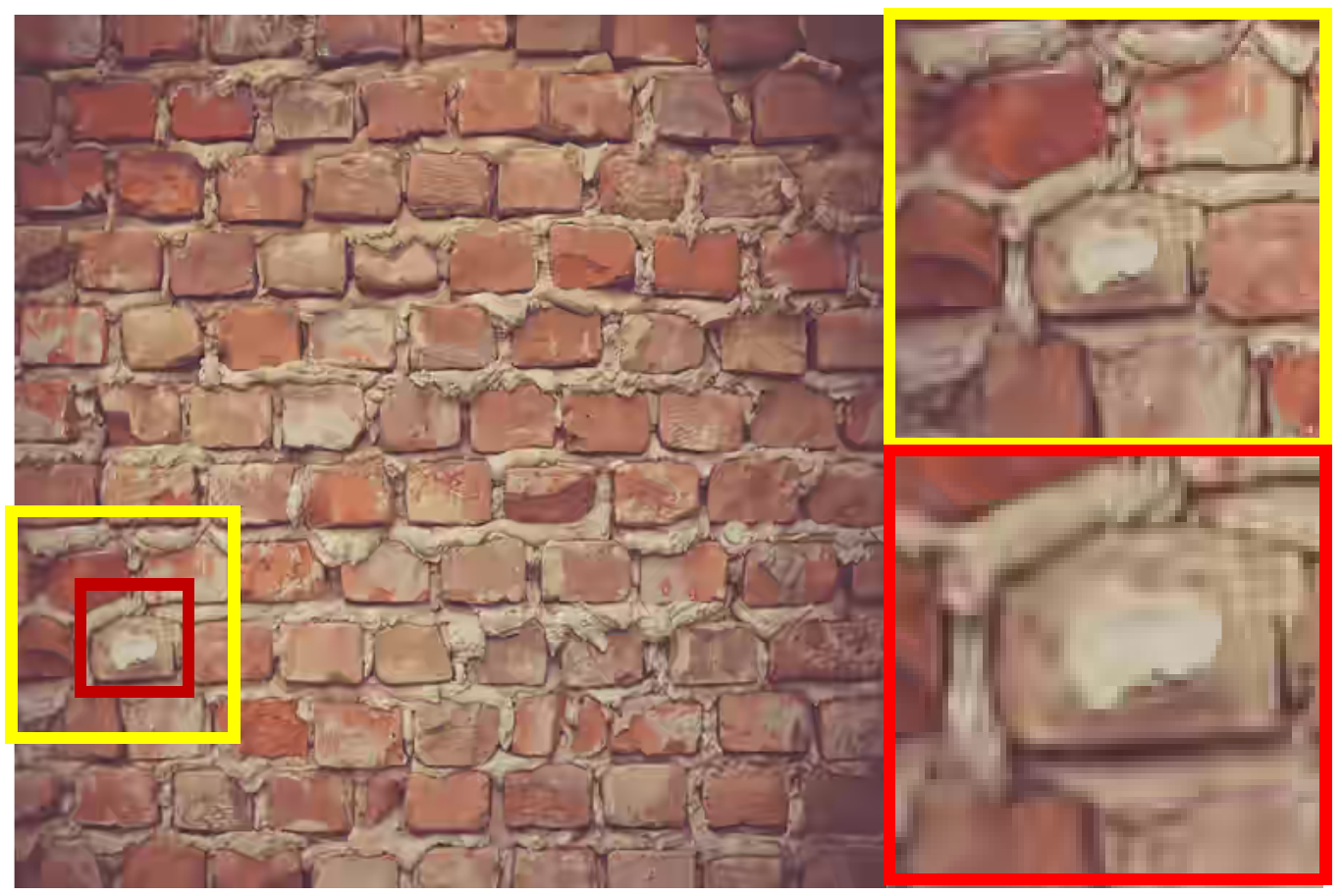}  \\
		 {\small (a) Original} & {\small (b) CARP}& {\small (c) BPG}\\
		 \includegraphics[width=0.33\linewidth, height=3cm]{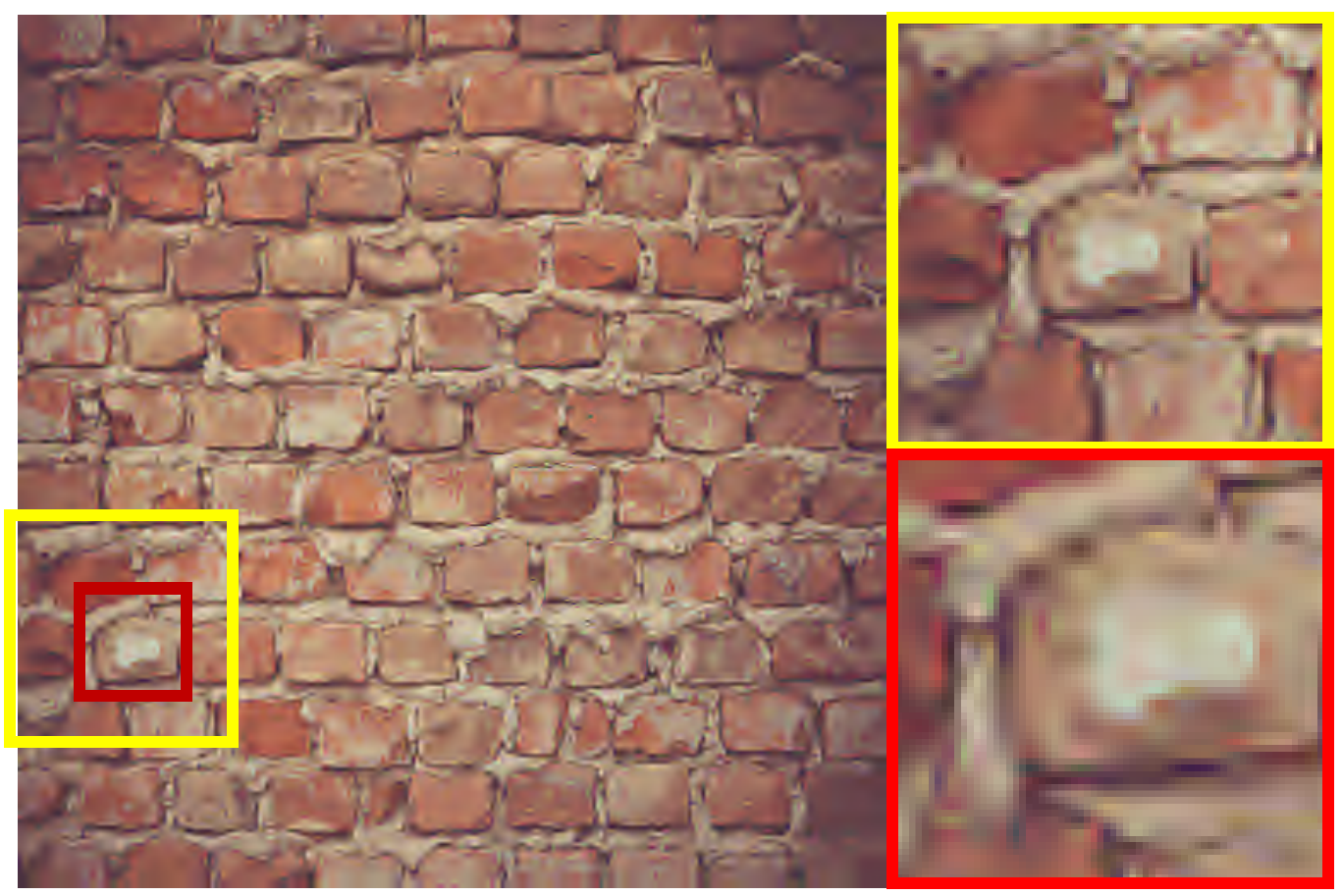}&\includegraphics[width=0.33\linewidth, height=3cm]{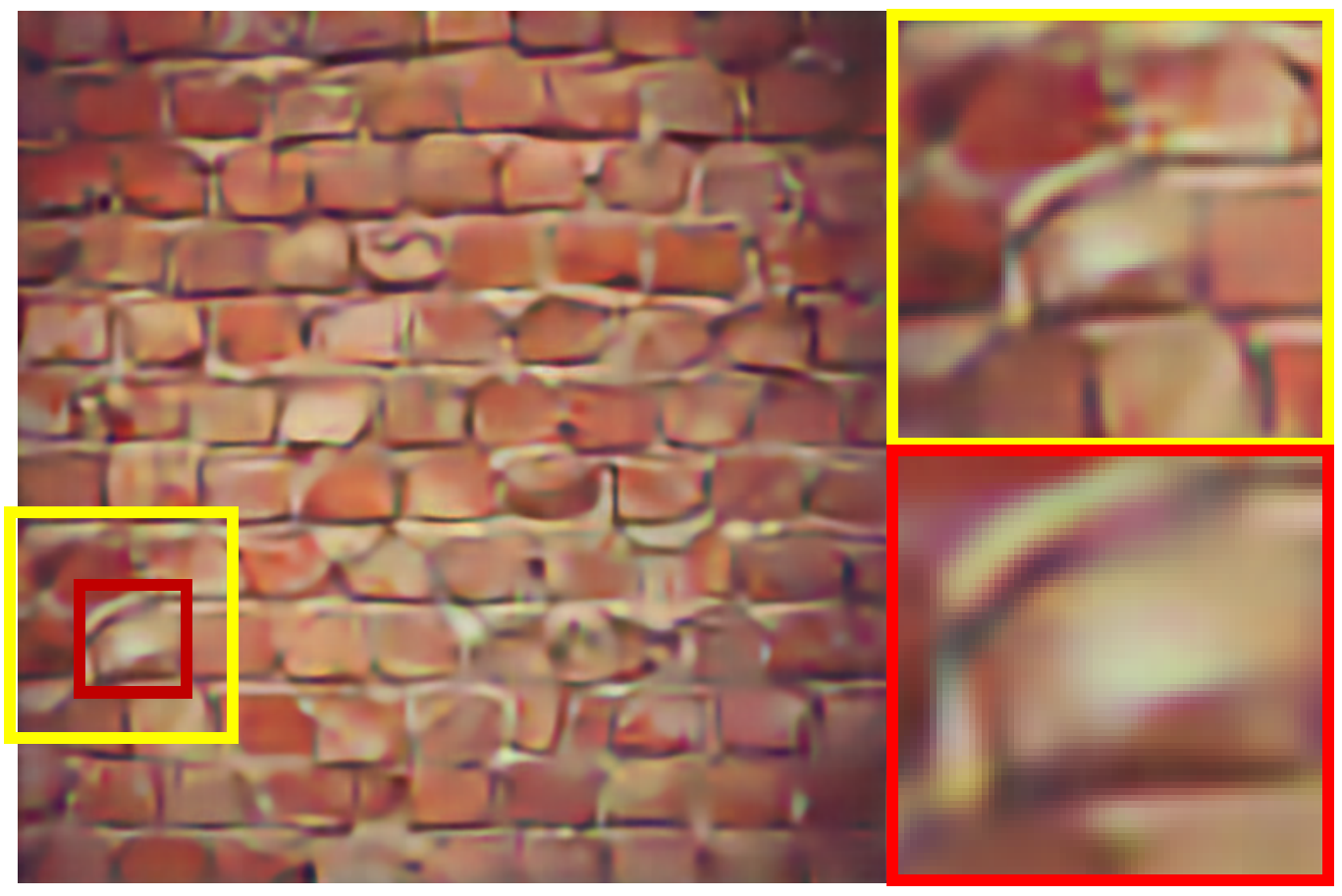}&\includegraphics[width=0.33\linewidth, height=3cm]{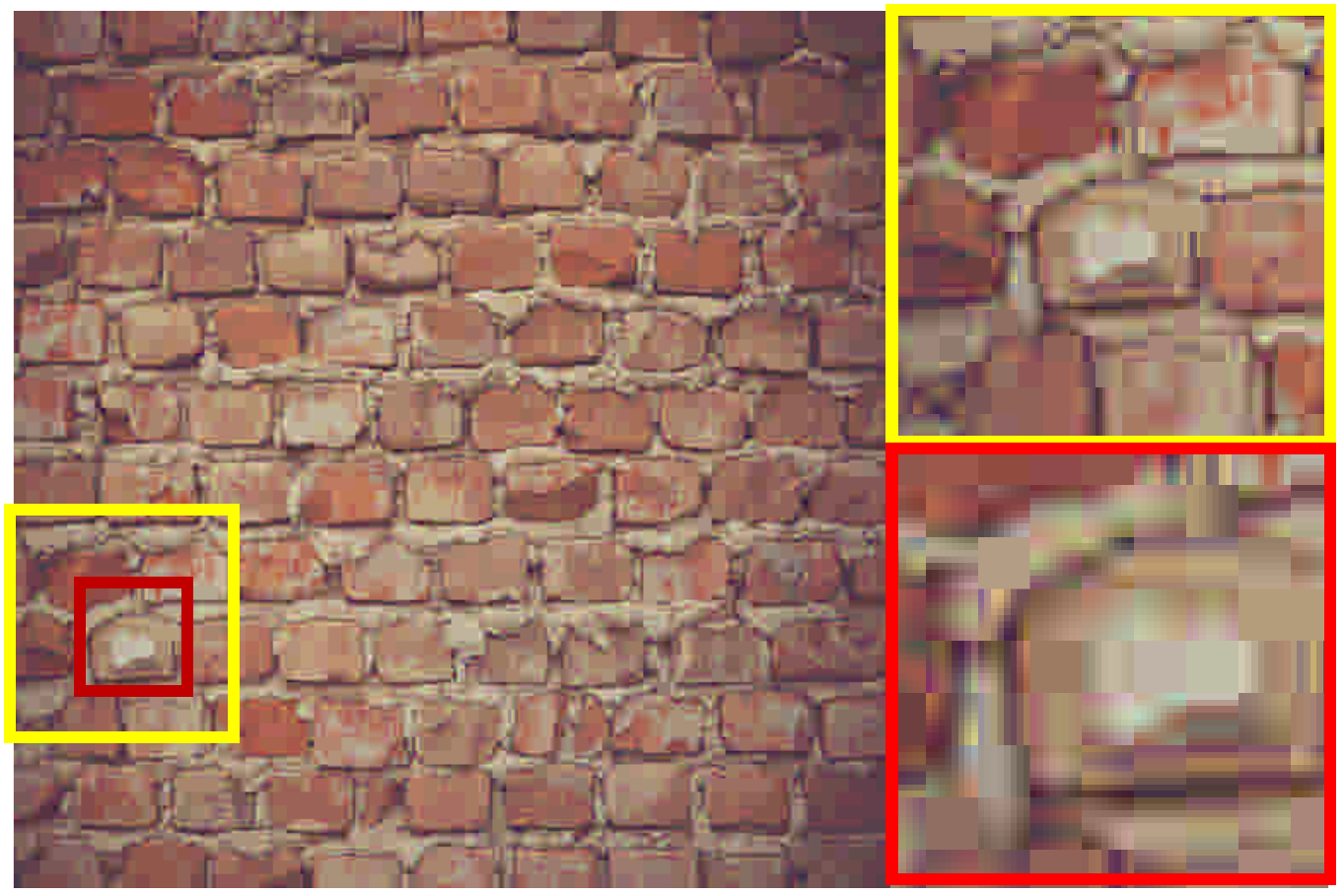}\\
		 {\small (e) JPEG2000} & {\small (f) E2E-DL}& {\small (g) JPEG}\\
		 \includegraphics[width=0.33\linewidth, height=3cm]{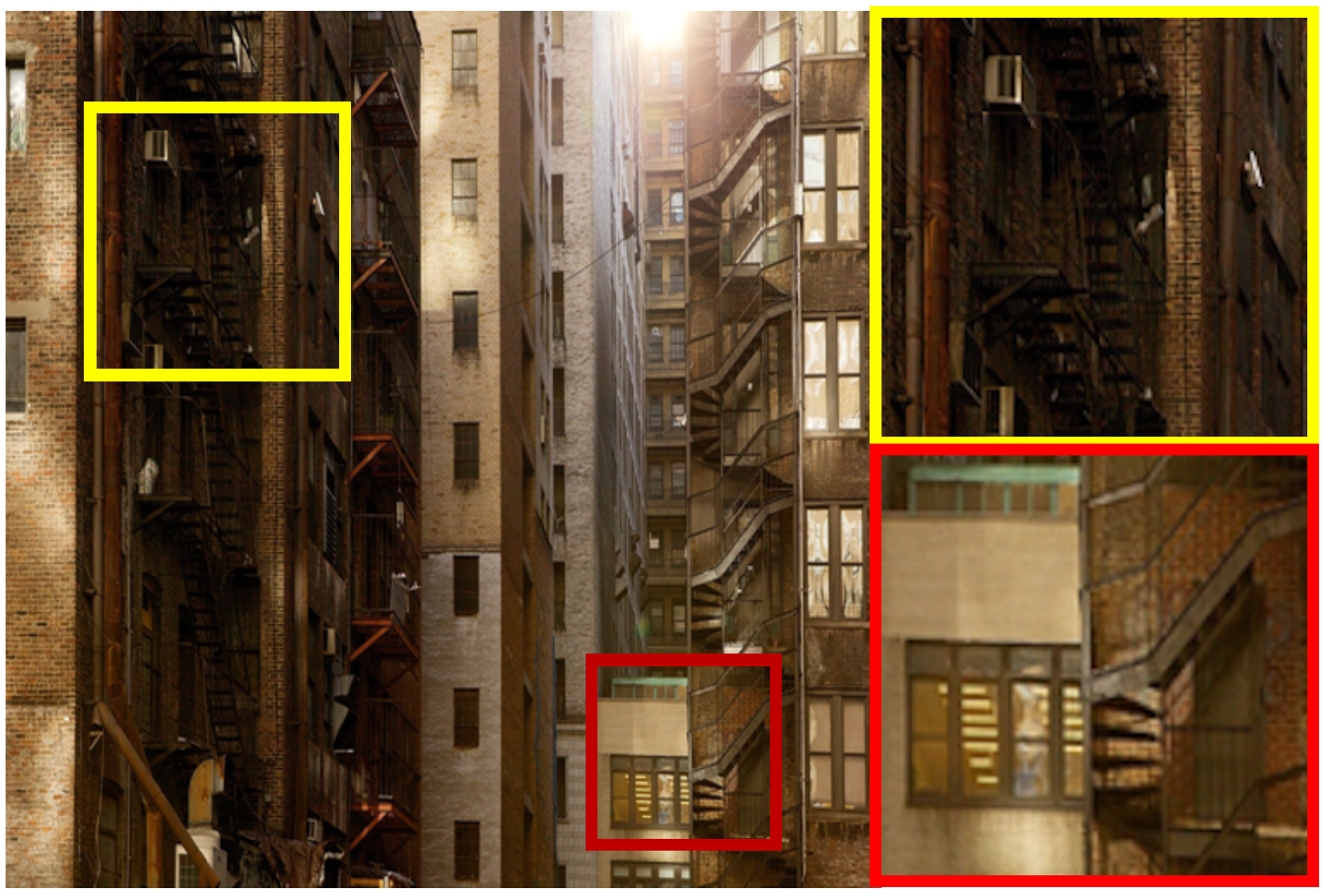}&\includegraphics[width=0.33\linewidth, height=3cm]{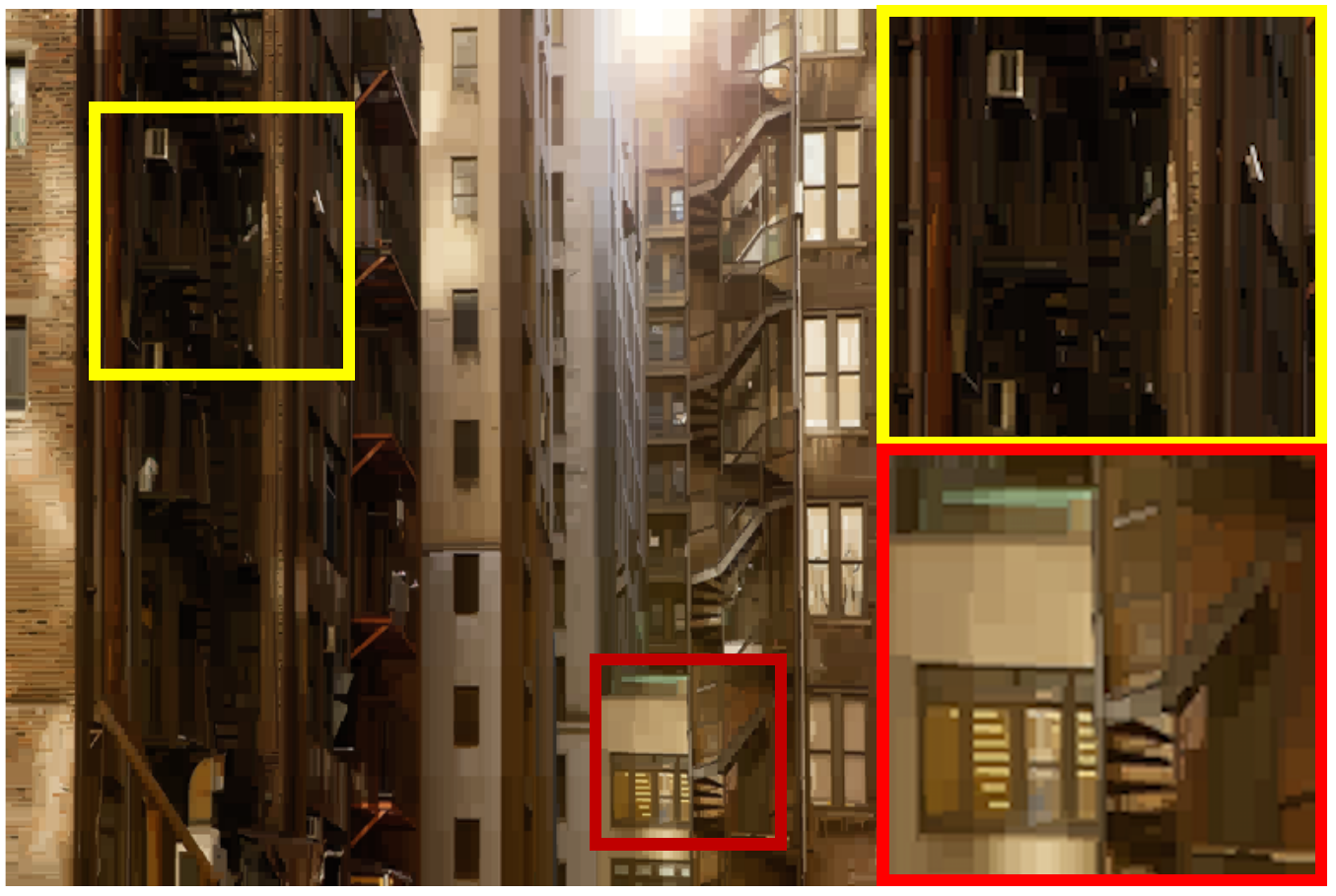}&\includegraphics[width=0.33\linewidth, height=3cm]{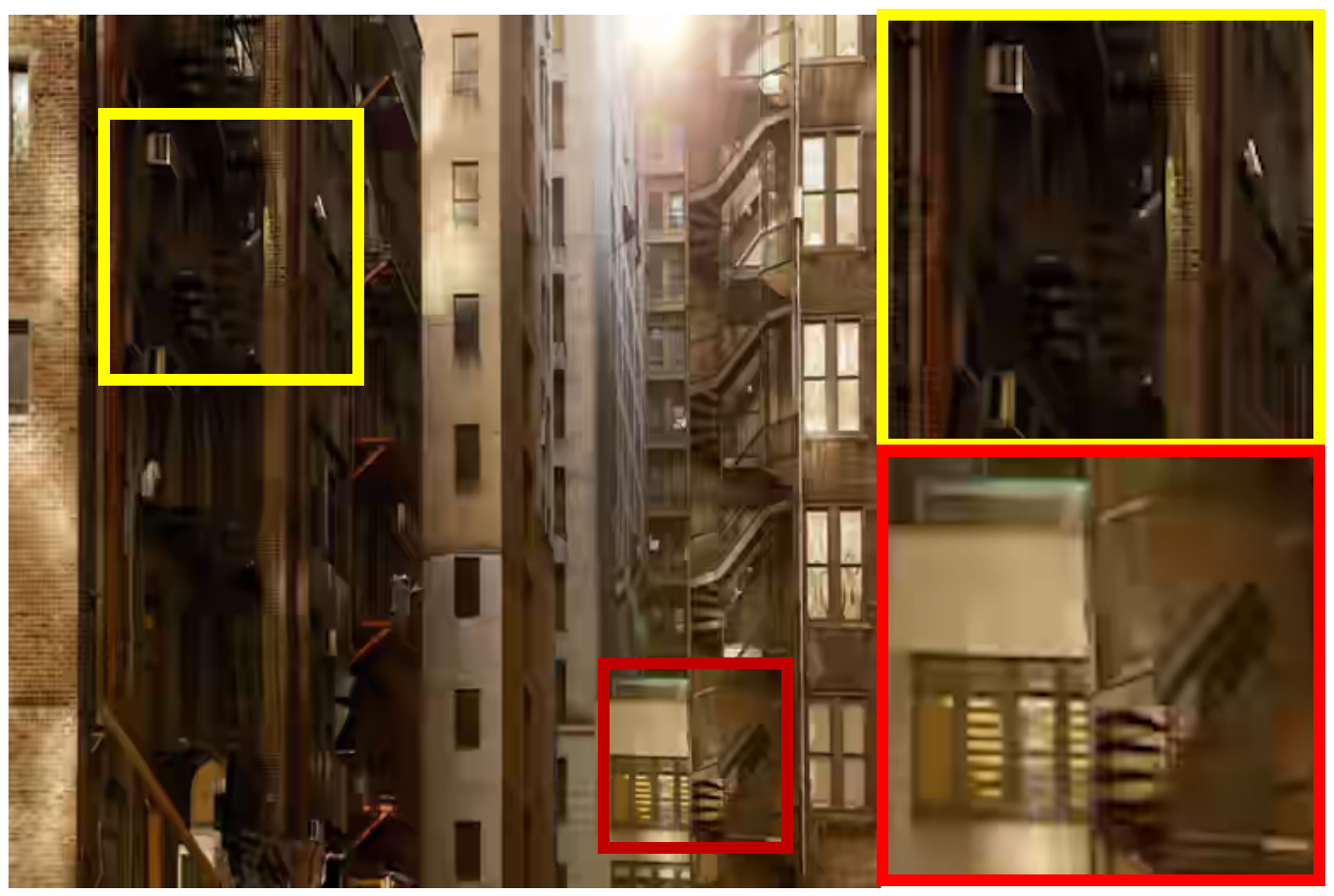}  \\
		 {\small (a) Original} & {\small (b) CARP}& {\small (c) BPG}\\
		 \includegraphics[width=0.33\linewidth, height=3cm]{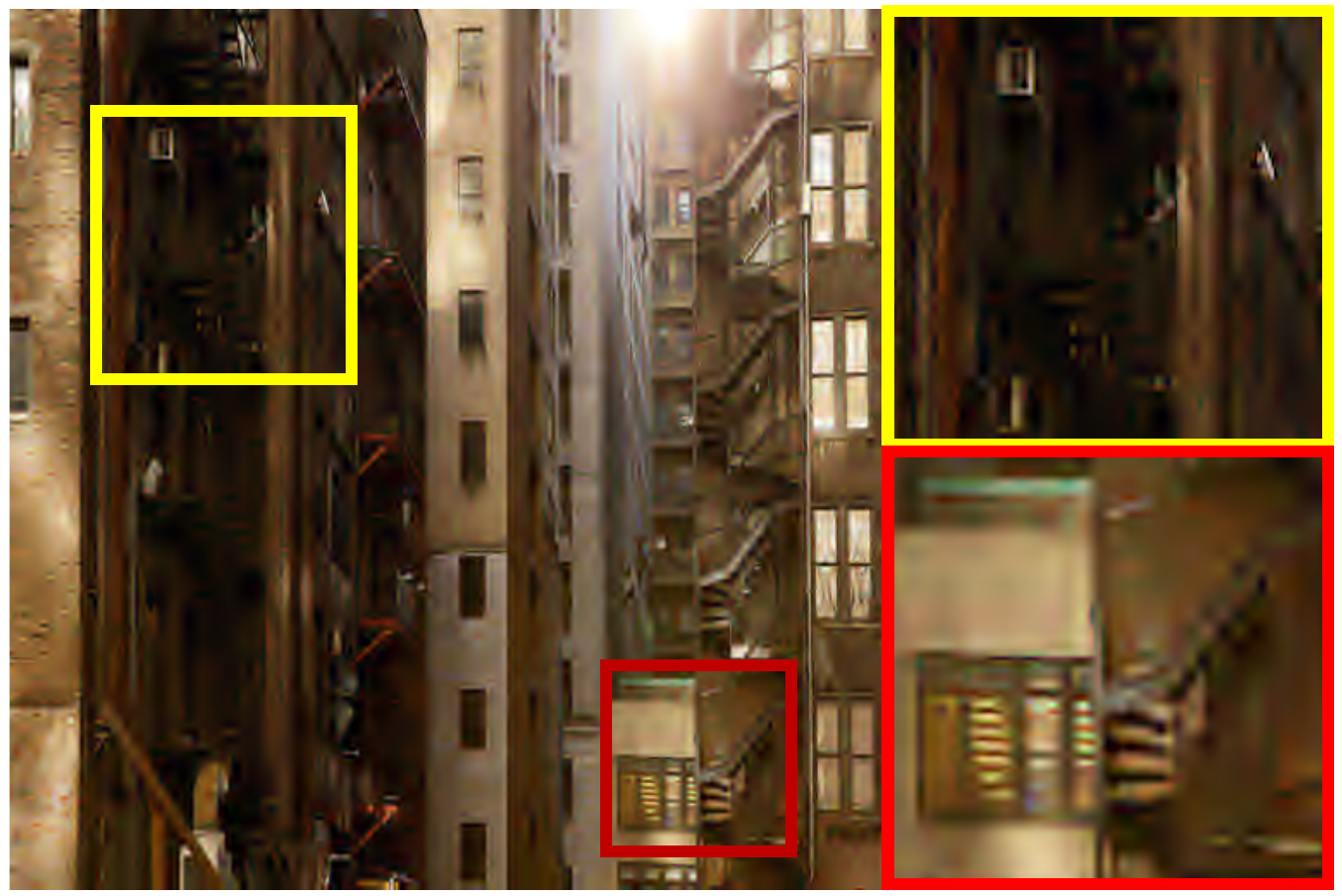}&\includegraphics[width=0.33\linewidth, height=3cm]{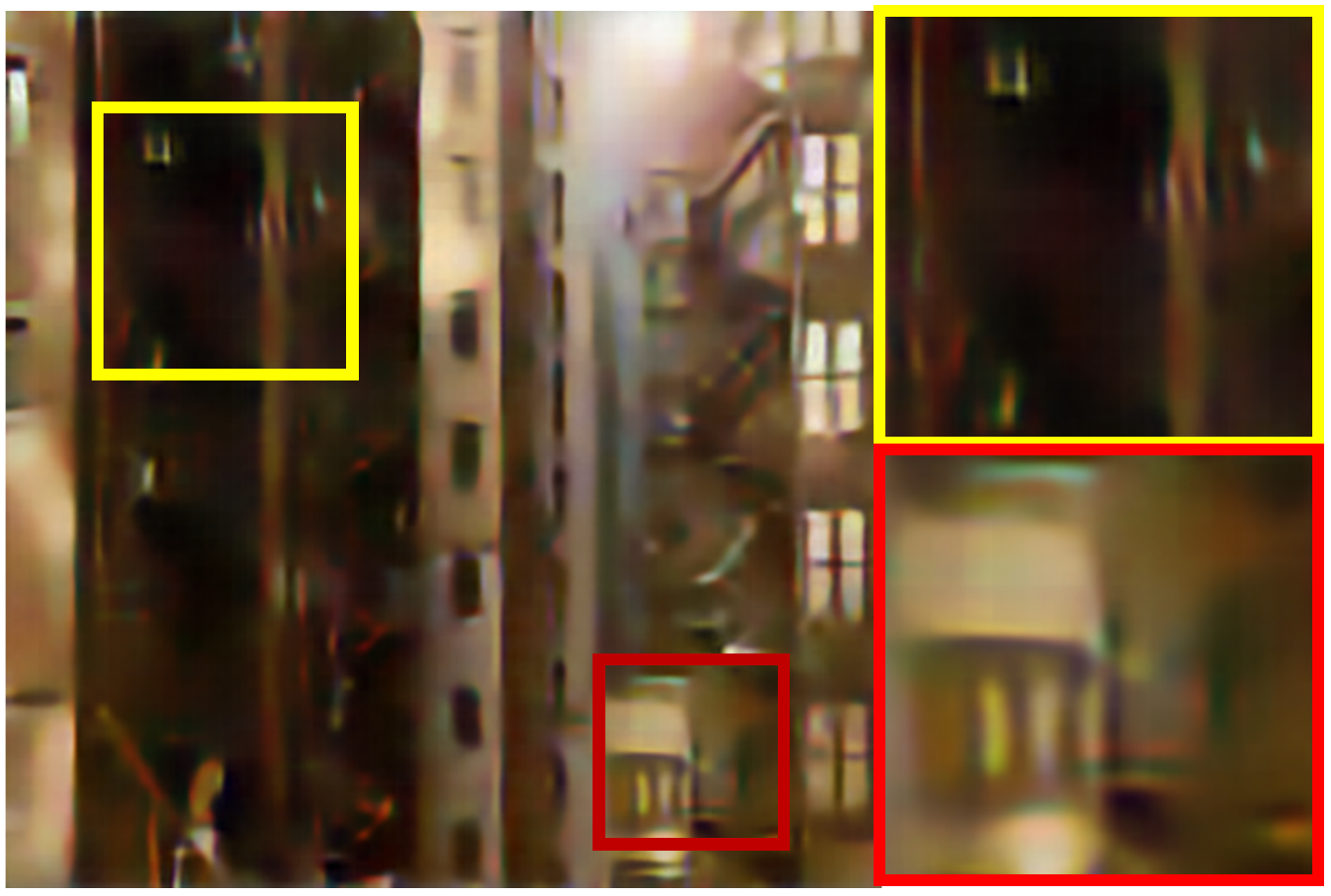}&\includegraphics[width=0.33\linewidth, height=3cm]{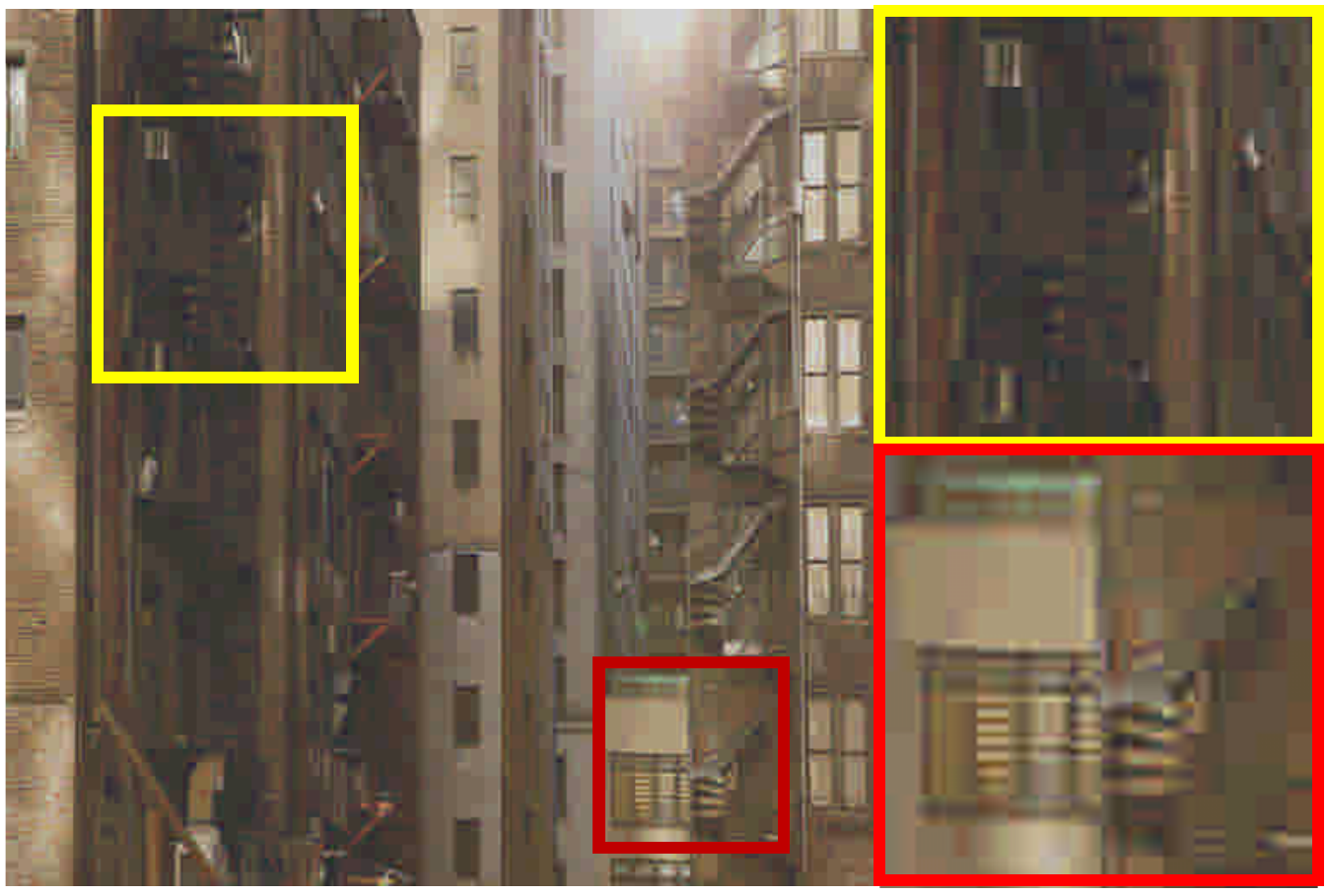}\\
		 {\small (e) JPEG2000} & {\small (f) E2E-DL}& {\small (g) JPEG}\\
		 \includegraphics[width=0.33\linewidth, height=3cm]{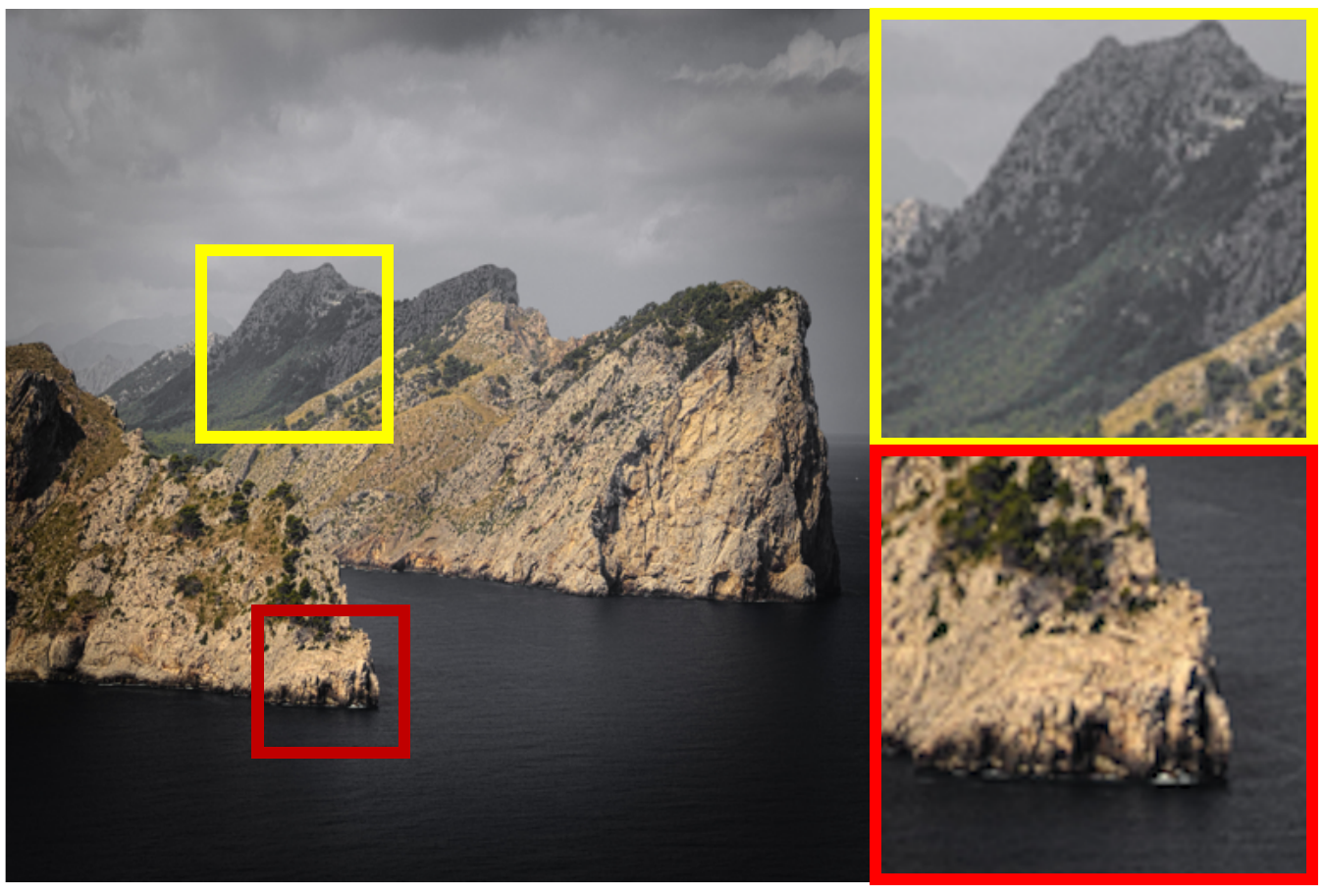}&\includegraphics[width=0.33\linewidth, height=3cm]{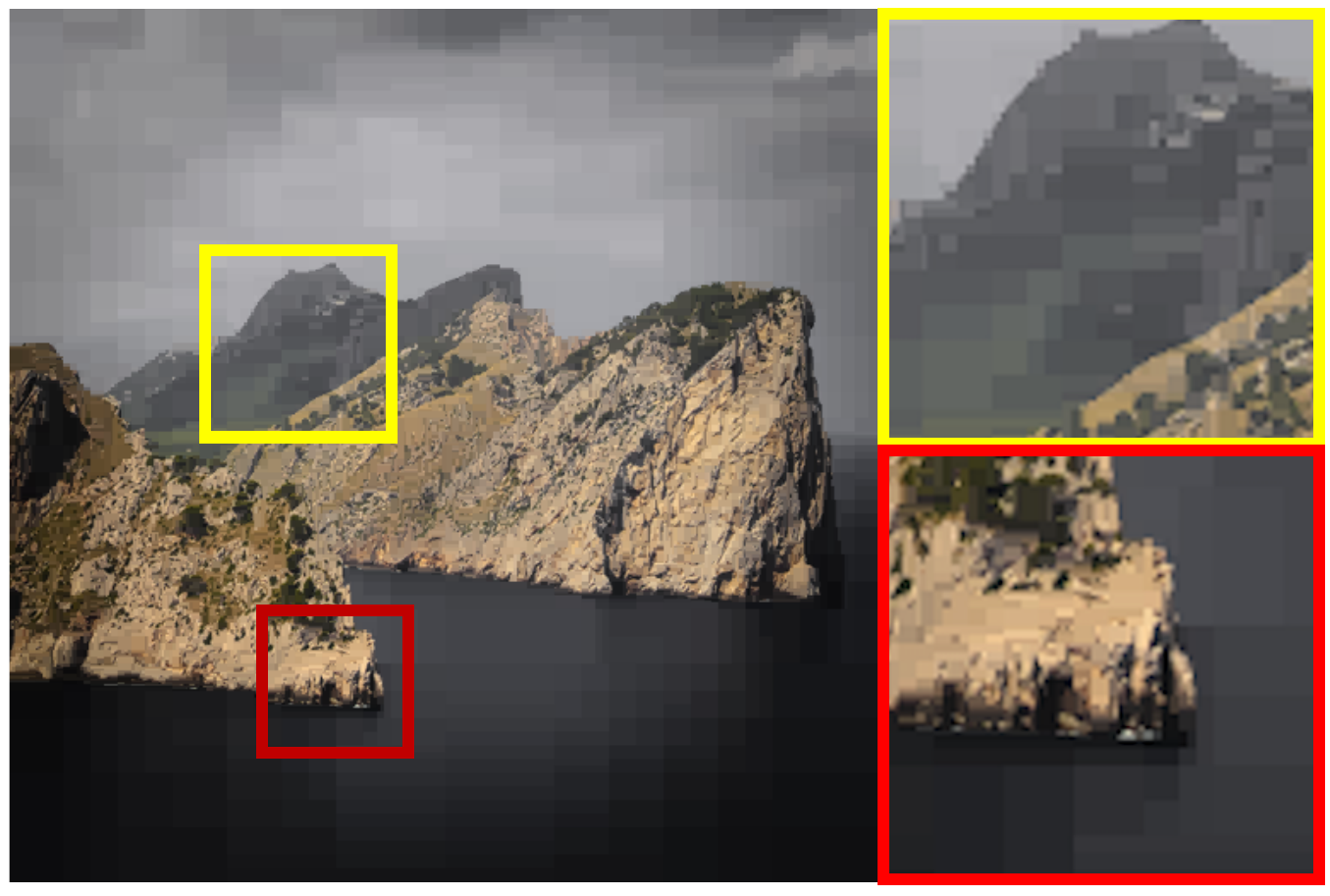}&\includegraphics[width=0.33\linewidth, height=3cm]{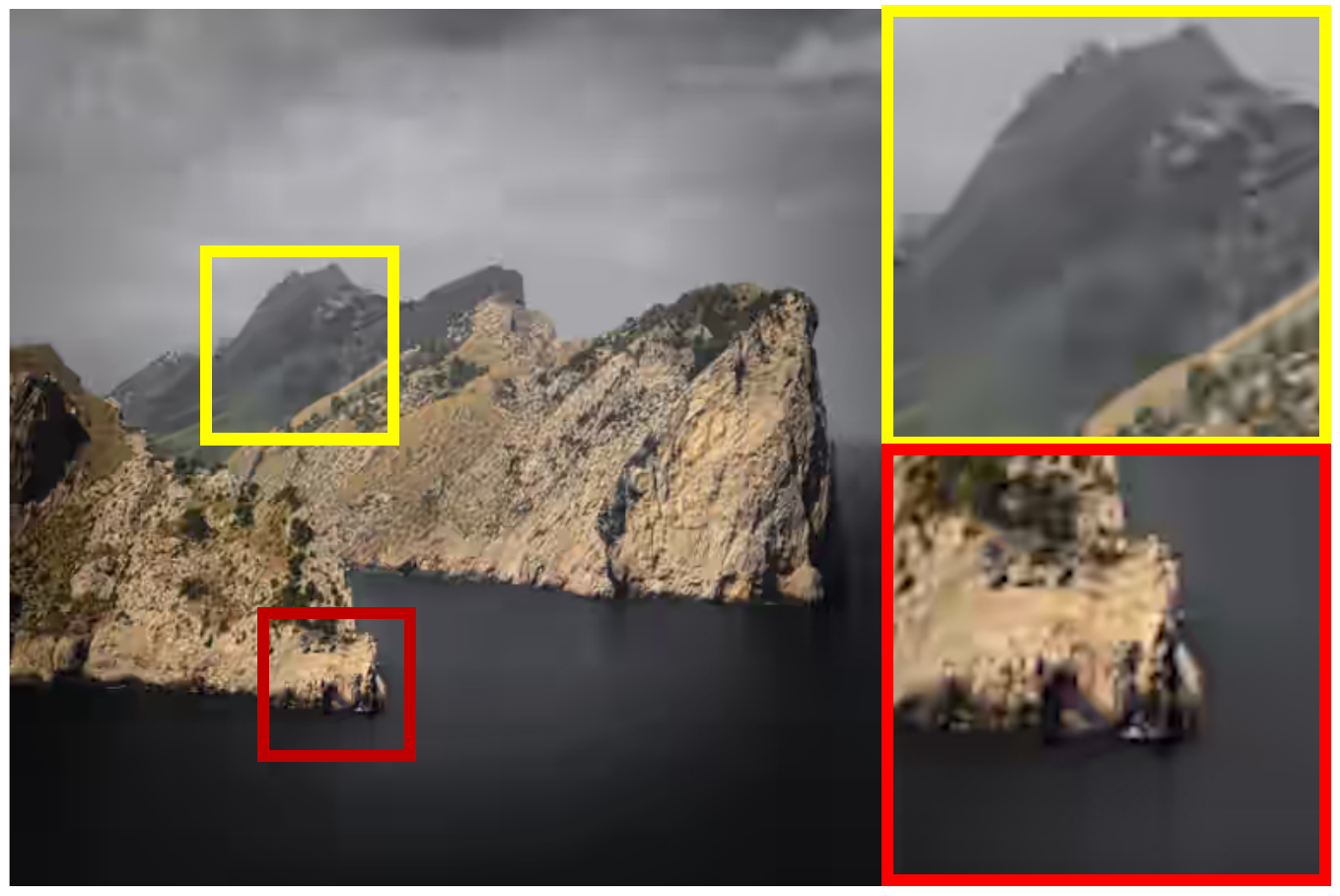}  \\
		 {\small (a) Original} & {\small (b) CARP}& {\small (c) BPG}\\
		 \includegraphics[width=0.33\linewidth, height=3cm]{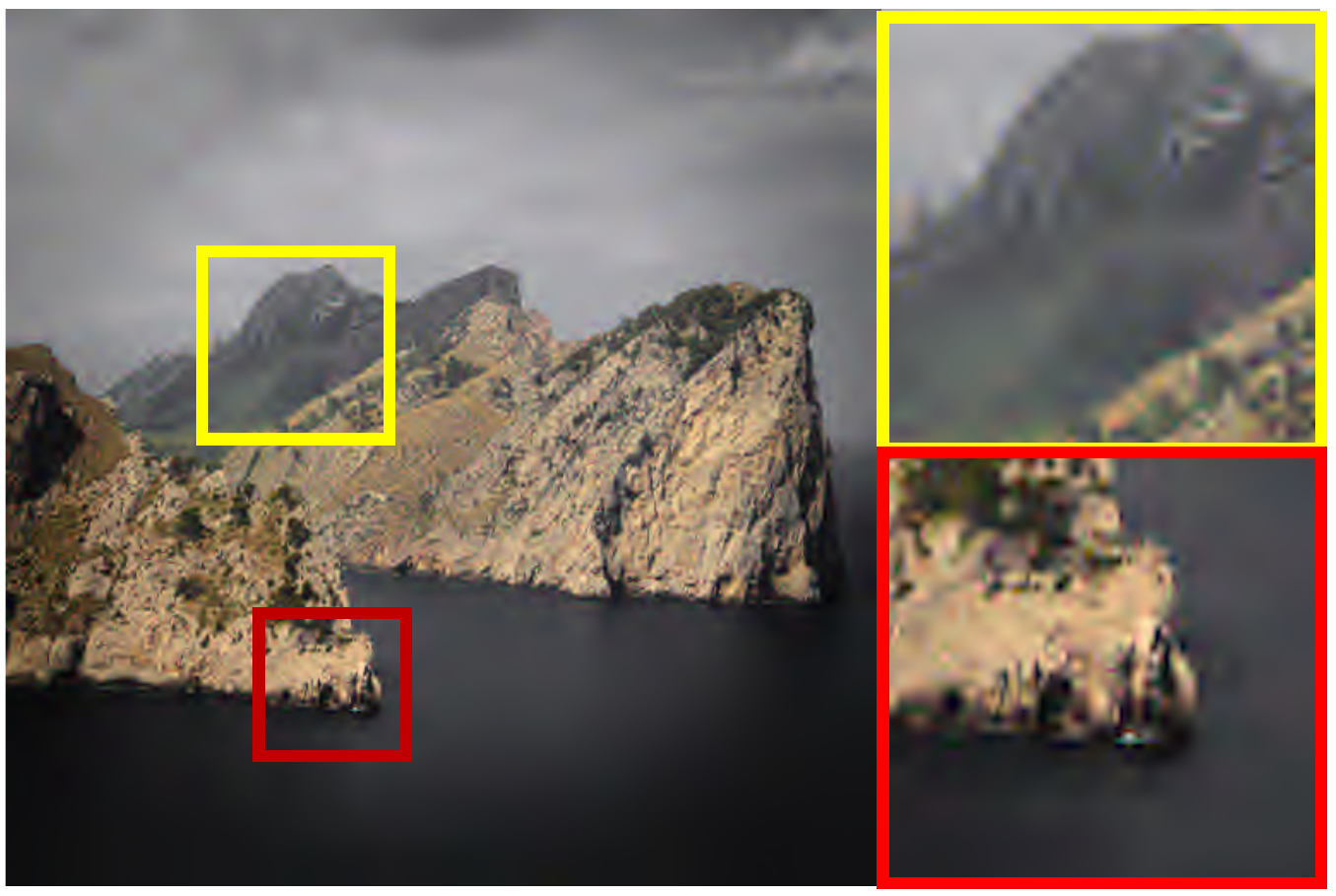}&\includegraphics[width=0.33\linewidth, height=3cm]{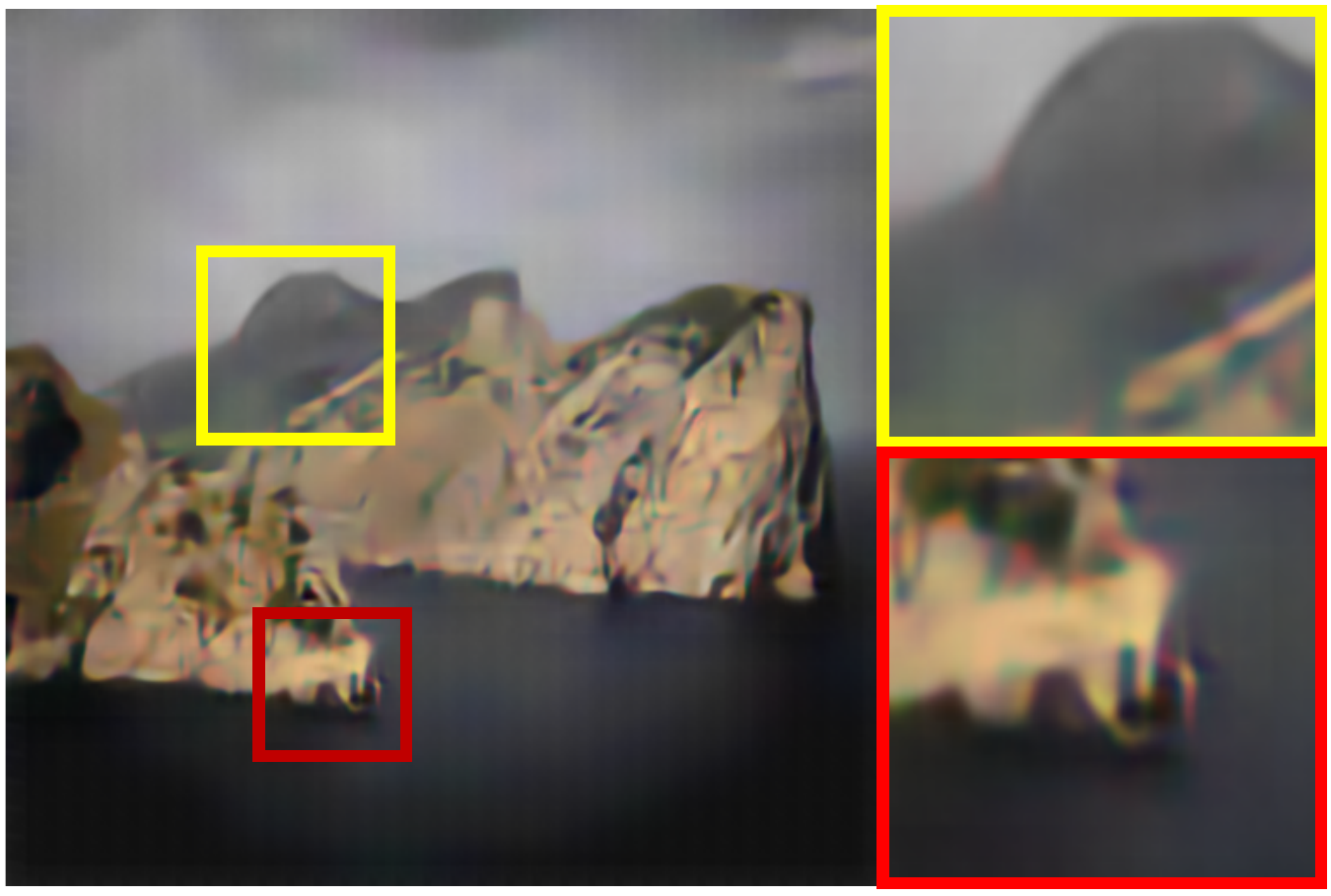}&\includegraphics[width=0.33\linewidth, height=3cm]{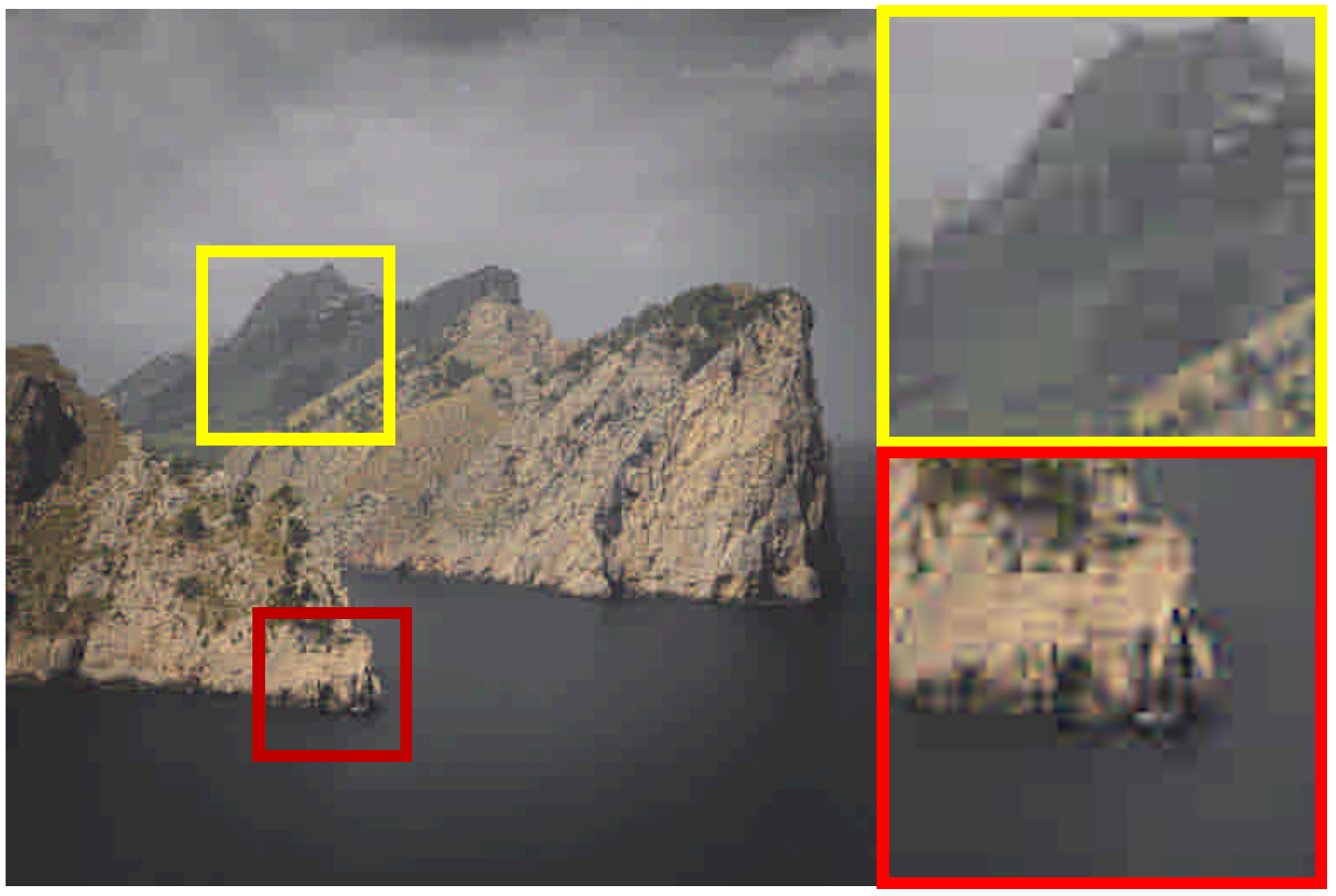}\\
		 {\small (e) JPEG2000} & {\small (f) E2E-DL}& {\small (g) JPEG}
	\end{tabular}
\caption{Comparison of reconstructed images among five different methods at specific compression ratio: Img1 at compression ration 45; Img5 at compression ratio 66; Img9 at compression ratio 45.} 
\label{rec_2c_1} 
\end{figure}

To assess each method, we use the peak signal-to-noise ratio (PSNR) and the multi-scale structural similarity index (MS-SSIM) of reconstructed images at various compression ratios, which is further supplemented by visual comparison. Specifically, at various compression ratios, each 2D image is compressed and reconstructed, then the PSNR and MS-SSIM are calculated using the reconstructed image. Figure \ref{fig:summary}(a) shows CARP gives the best average PSNR while Figure \ref{fig:summary}(b) shows CARP gives the best average MS-SSIM at all compression ratios. Figure \ref{mean100}(a) and Figure \ref{mean100}(b) plot the PSNR and MS-SSIM curves for CARP for all 100 images; we present the PSNR (i.e., Figure \ref{mean100} (c), (e), (g), (i)) and MS-SSIM ratio curve  (i.e., Figure \ref{mean100} (d), (f), (h), (i)) between each alternative method and CARP---with values under 1 indicating CARP outperforms the competitor for all 100 individual images. The performances of all MS-SSIM ratio curves are consistent with those in PSNR ratio curves. CARP almost uniformly outperforms all of the four competitors for nearly all individual images and at all compression ratios up to 300 at which we are able to apply the competitor, except on a handful of images for JPEG and JPEG2000 at very low compression ratios and a couple of images for BPG. For this database, E2E-DL underperforms CARP substantially, but we acknowledge that part of the substantial performance gap could be narrowed had the CNNs been trained on images that are particularly suited for the specific database. Like JPEG and JPEG2000, CARP does not require external pre-training and is applicable for compressing any single image. Moreover, the user does not need to specify any tuning parameter other than $\sigma$, which is equivalent to specifying the compression ratio. 

The locally adaptive nature of CARP enhances its ability to preserve local details in the images. As an illustration, we visualize reconstructions with a particular focus on detailed features in an image using three selected images in Figure \ref{rec_2c_1}.
The region of interest is marked in the original image, and we present zoom-in views of the interesting region with a yellow and red block in the reconstructed images from five different methods at one specific compression ratio. Overall, CARP and BPG clearly outperform the other three methods (JPEG, JPEG2000, and E2E-DL). In all three cases, it shows CARP preserves substantially more details in the reconstruction compared to JPEG and JPEG2000. Specifically, for the redbrick wall in the top of Figure \ref{rec_2c_1}, the reconstructed image by CARP in sub-figure (b) appears sharper and warmer than BPG in sub-figure (c); for the building in the middle of Figure \ref{rec_2c_1}, the stairs in the yellow and red block by CARP is much clearer than BPG, and a further zoom-in into the stairs shows that CARP preserves the most details among all methods; for landscape in the bottom of Figure \ref{rec_2c_1}, there are more recognizable details in the mountain in sub-figure (b) but more blurry in sub-figure (c).

The details in reconstructed images show that CARP preserves the most details among all methods.
On the one hand, the use of RDP in CARP cuts the image horizontally or vertically, which leads to its superiority when the target image has repeated patterns along the horizontal or vertical direction, e.g., the buildings in Img5 (the middle two rows of Figure \ref{rec_2c_1}). For an image with frequent and large jumps in intensity between adjacent pixels, CARP tends to underperform BPG. 
On the other hand, the pruning option in CARP increases the efficiency of image compression, which helps detect the blocks in an image with similar intensities and convert the tiles into pixel vectors in the fastest way, e.g., patches of the large dark bottom region in Img9 (the last two rows of Figure \ref{rec_2c_1}). 

\subsection{YouTube video dataset}

We use the YouTube dataset in~\cite{Alayrac16unsupervised}, which consists of instructional videos for five different tasks including making a coffee, changing a car tire, performing
cardiopulmonary resuscitation (CPR), jumping a
car and re-potting a plant. The dataset has 150 videos with an average length of about 4,000 frames (or 2 minutes).

\begin{figure}[H]
\centering 
	\begin{tabular}{cc}
		 \includegraphics[width=7cm, height=3cm,trim={1.0cm 0.8cm 0 0},clip]{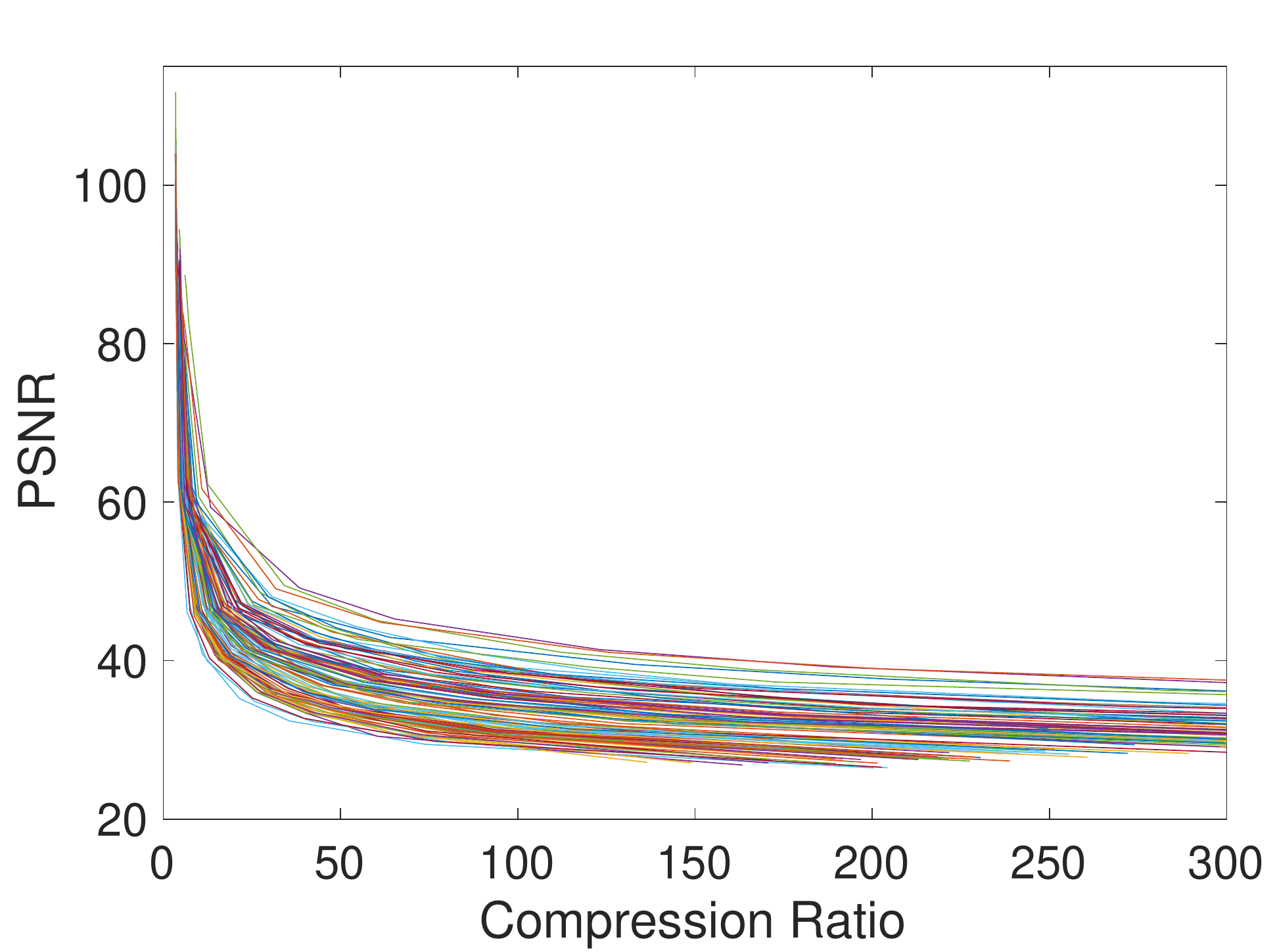} & 
		  \includegraphics[width=7cm, height=3cm,trim={1.0cm 0.8cm 0 0},clip]{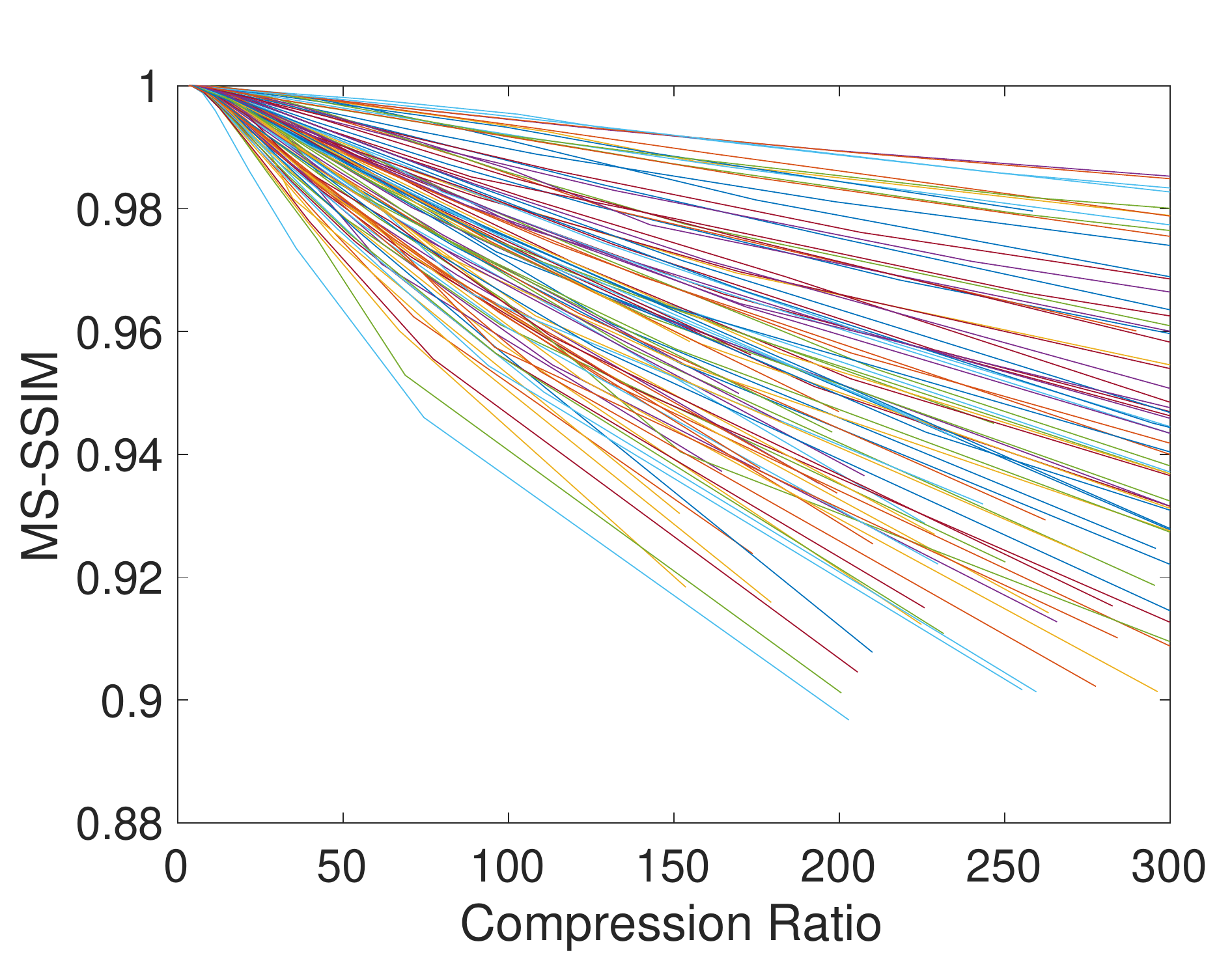}\\
		  {\small (a) PSNR of CARP for 100 videos} & {\small (b) 
		  MS-SSIM of CARP for 100 videos} \\
		  \includegraphics[width=7cm, height=3cm,trim={1.0cm 0.8cm 0 0},clip]{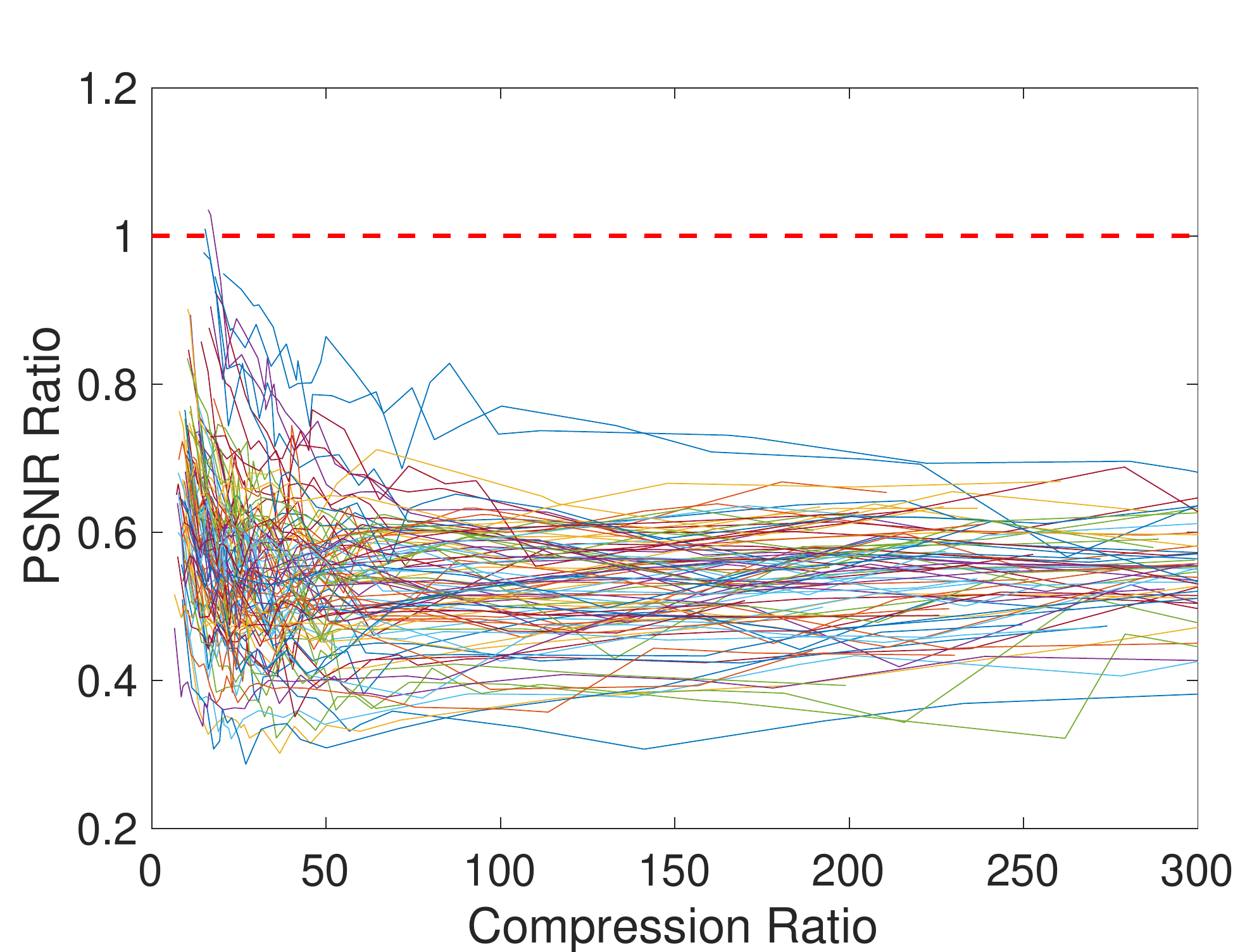} &
		  \includegraphics[width=7cm, height=3cm,trim={1.1cm 0.9cm 0 0},clip]{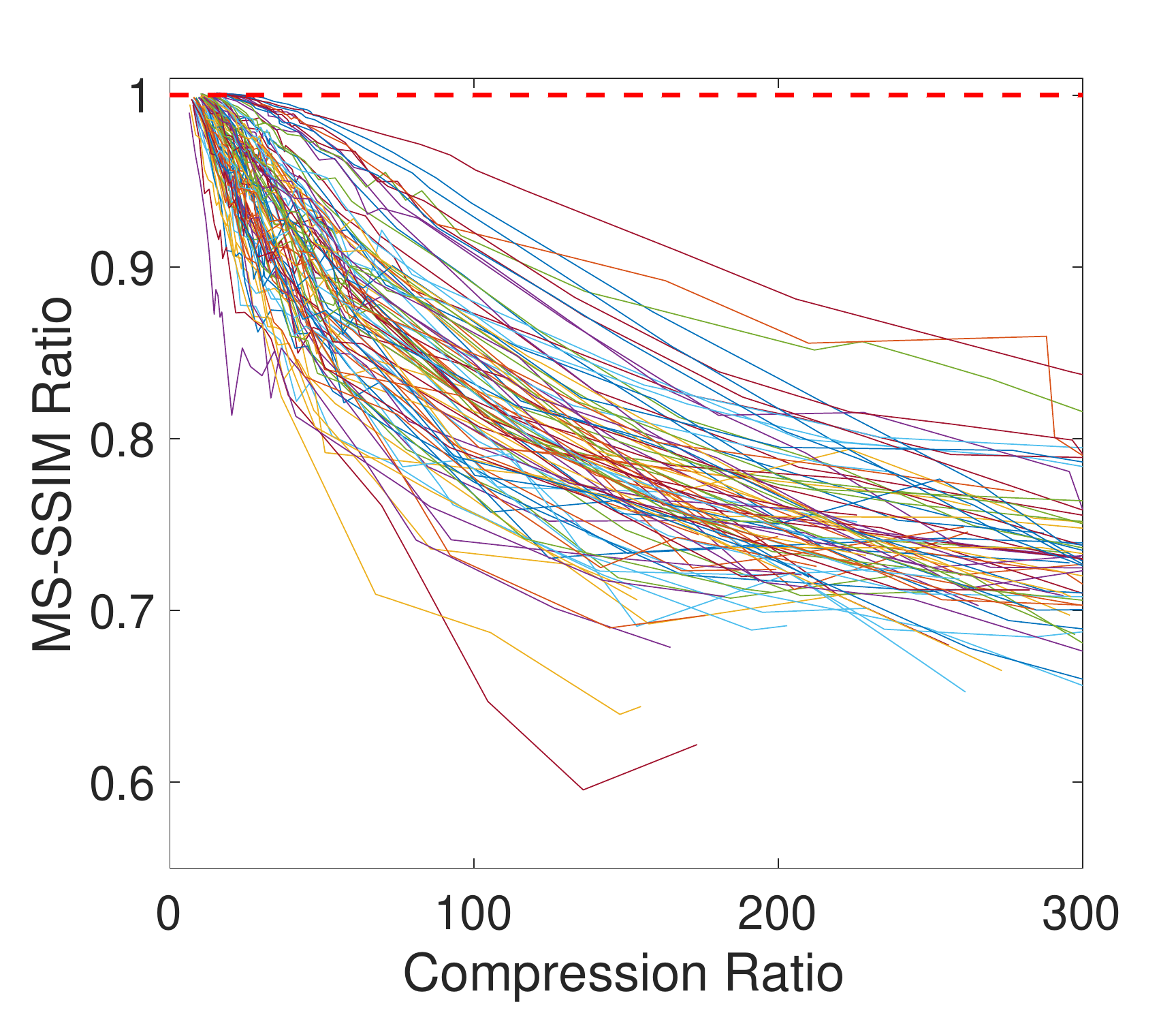}\\
		   {\small (c) PSNR ratio: JPEG/CARP} & {\small (d) MS-SSIM ratio: JPEG/CARP}\\
		  \includegraphics[width=7cm, height=3cm,trim={1.0cm 0.8cm 0 0},clip]{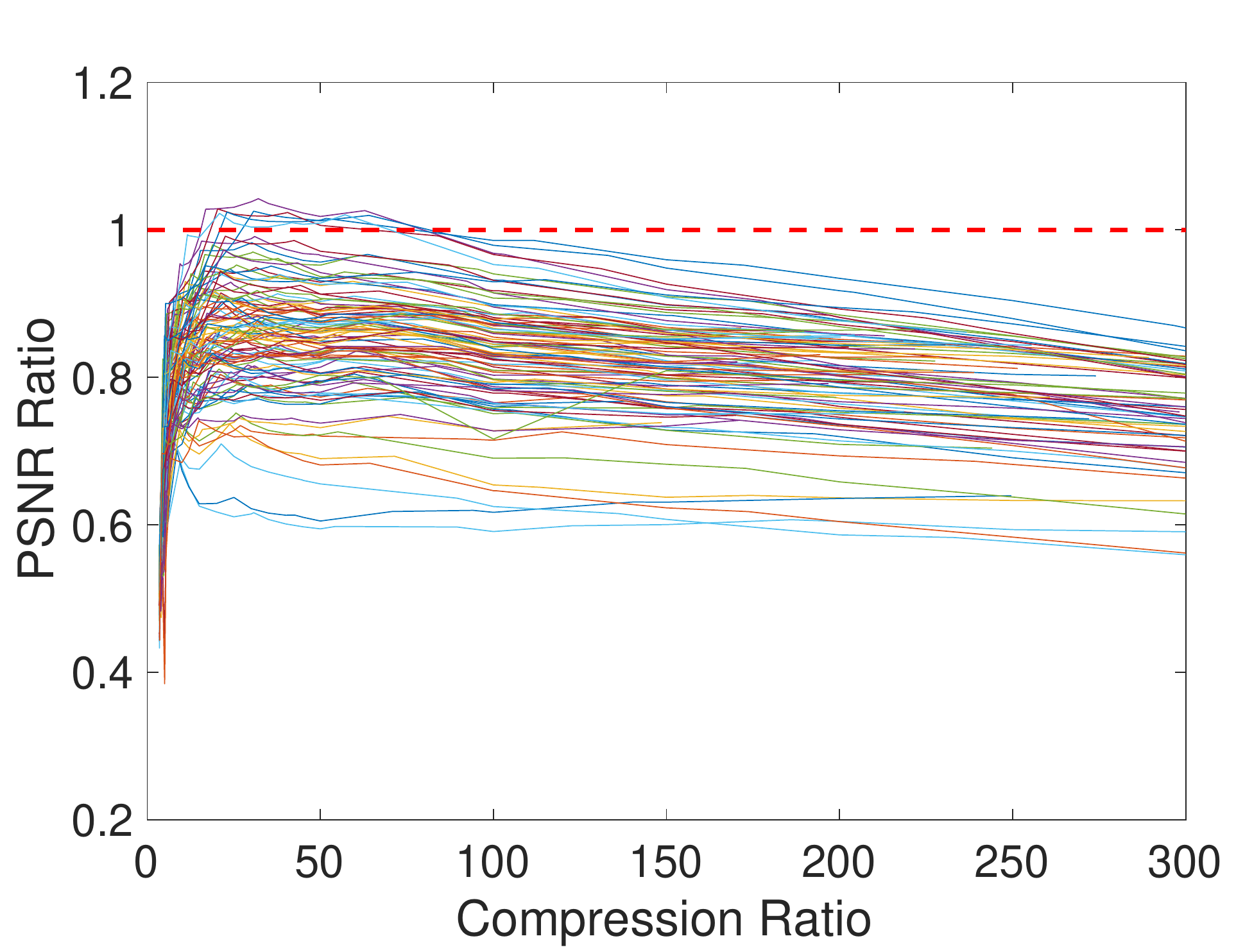}&
		  \includegraphics[width=7cm, height=3cm,trim={1.1cm 0.9cm 0 0},clip]{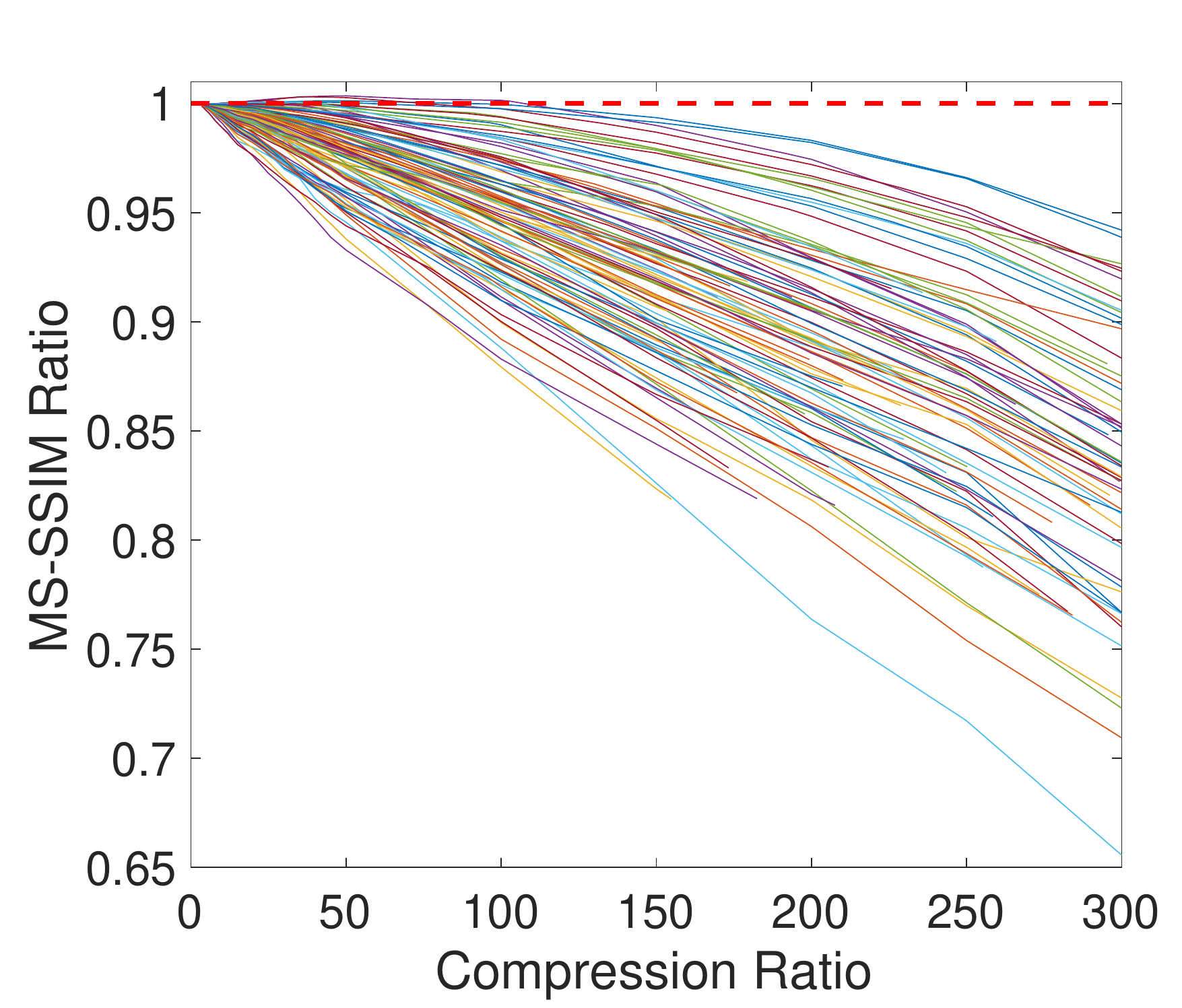} \\
		  {\small (e) PSNR ratio: JPEG2000/CARP} & {\small (f) MS-SSIM ratio: JPEG2000/CARP} \\
		  \includegraphics[width=7cm, height=3cm,trim={1.0cm 0.8cm 0 0},clip]{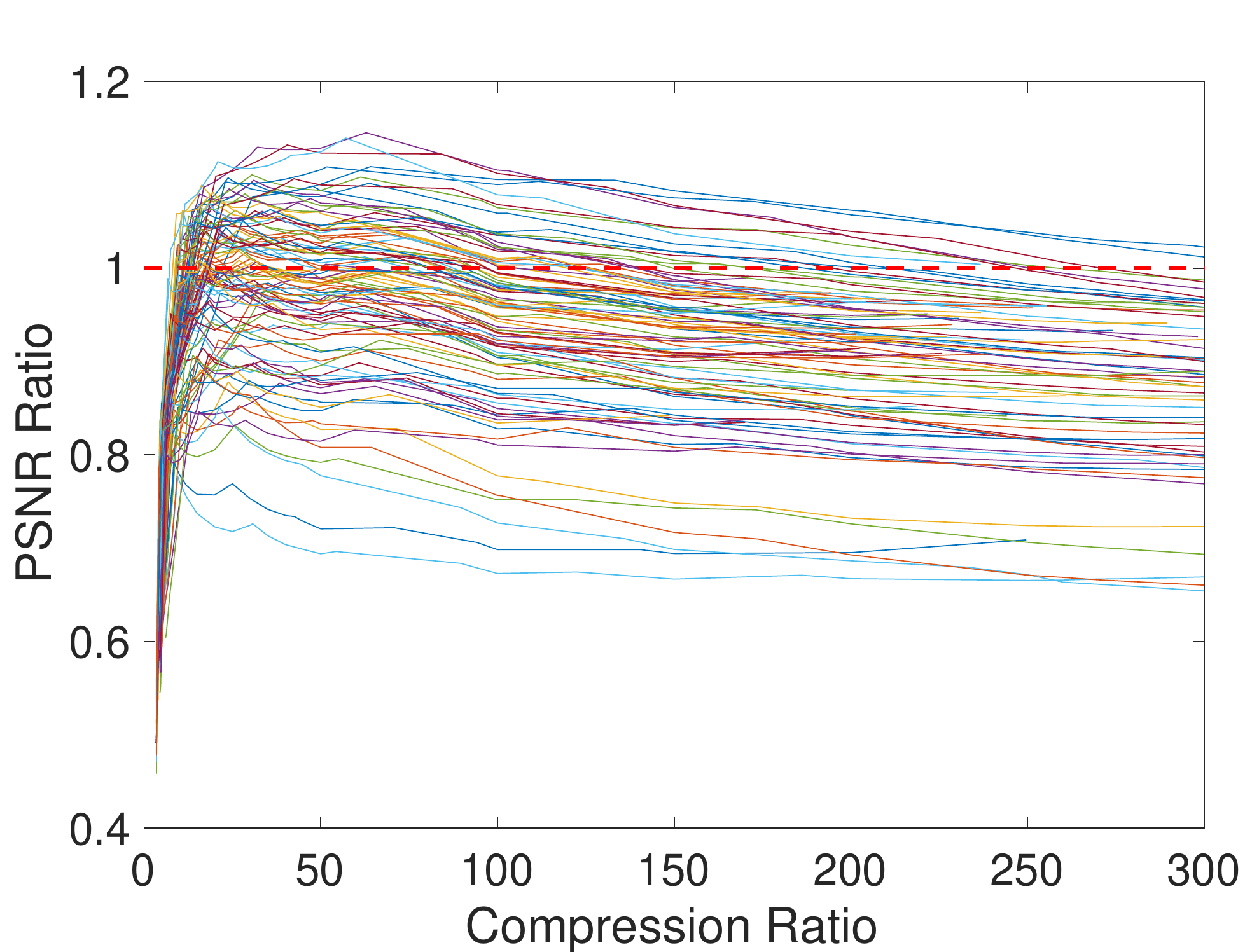} &
		  \includegraphics[width=7cm, height=3cm,trim={1.0cm 0.9cm 0 0},clip]{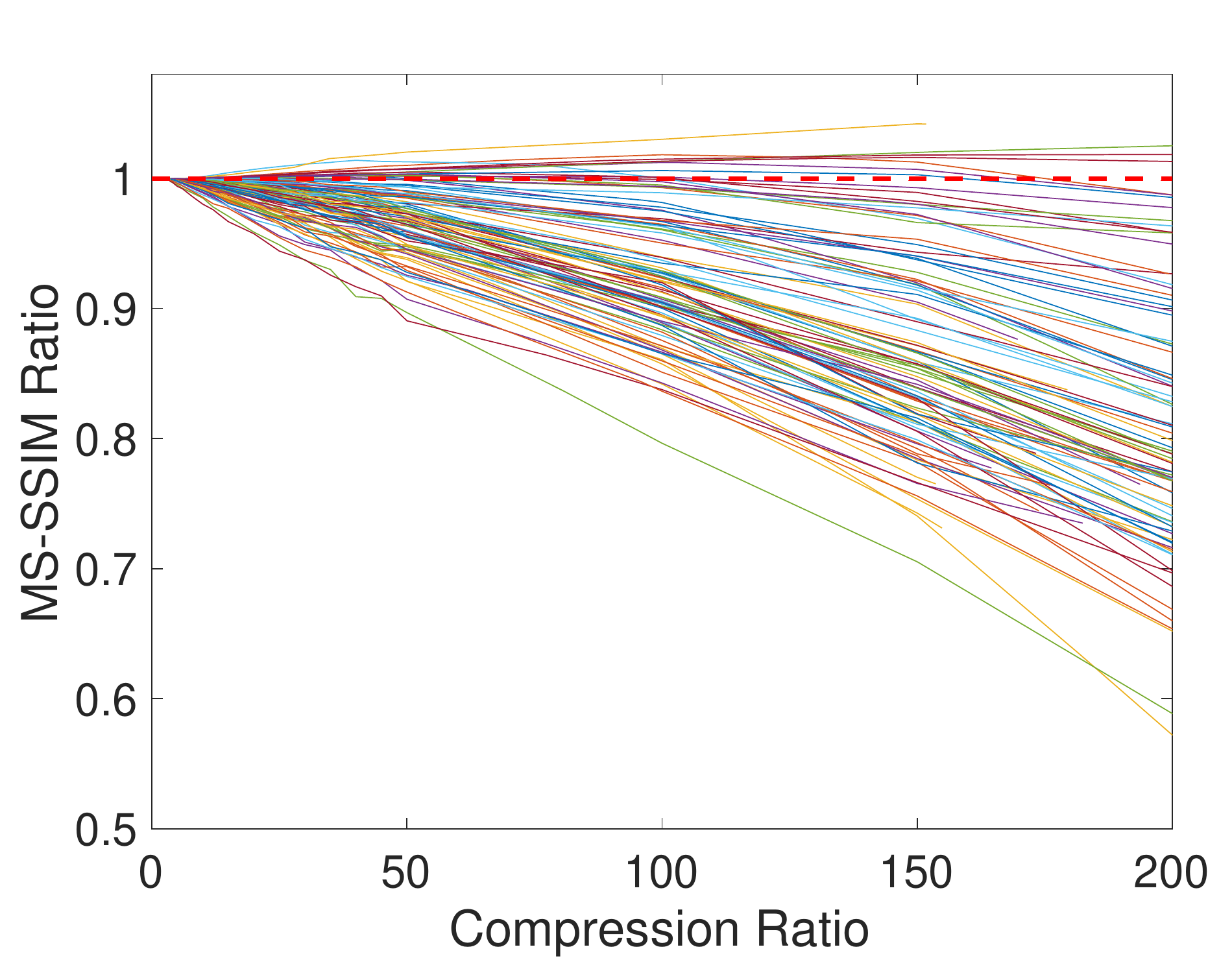}\\
		   {\small (g) PSNR ratio: MPEG-4/CARP} & {\small (h) MS-SSIM ratio: MPEG-4/CARP}\\
		  \includegraphics[width=7cm, height=3cm,trim={1.0cm 0.9cm 0 0},clip]{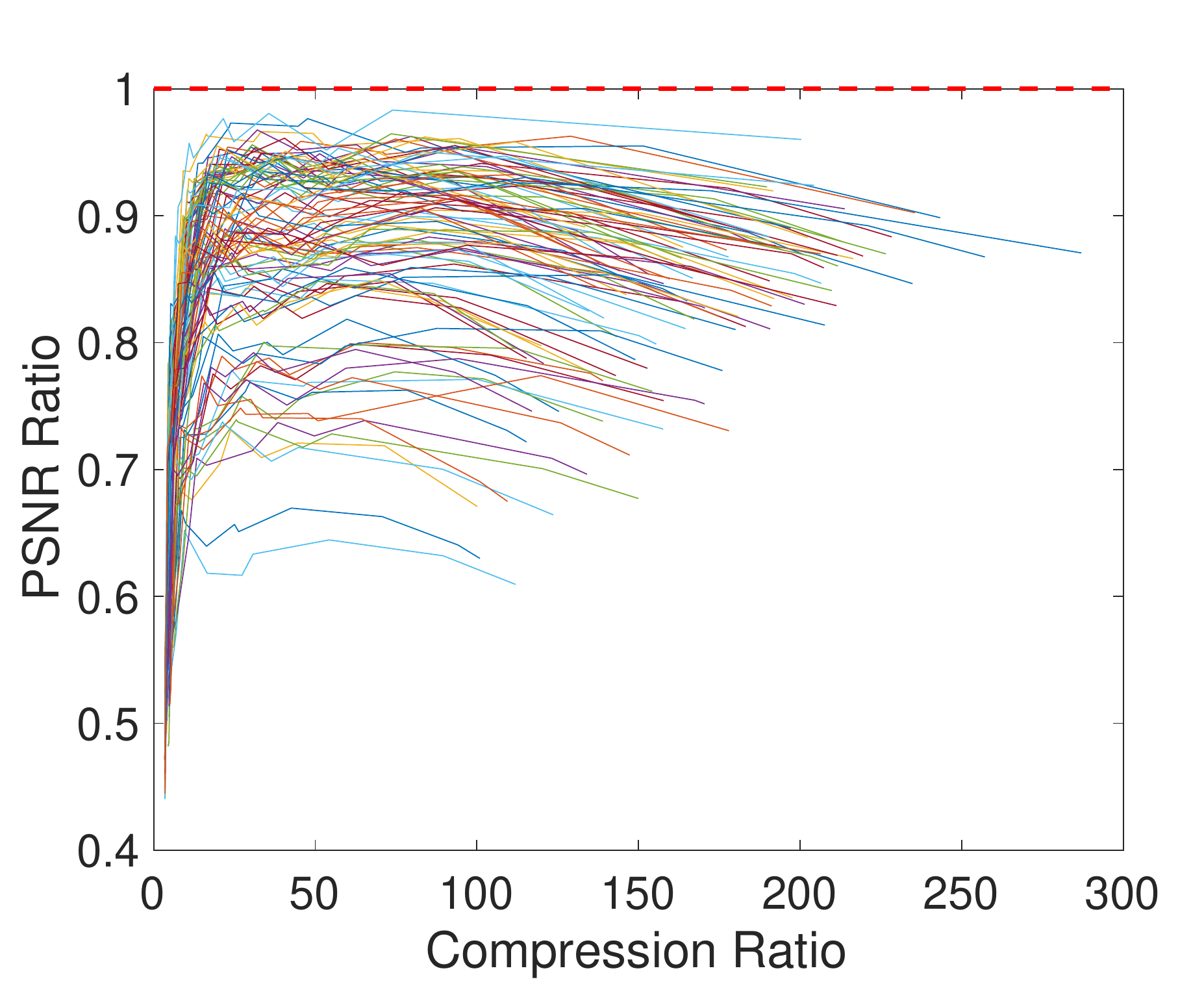} &
		  \includegraphics[width=7cm, height=3cm,trim={1.0cm 0.9cm 0 0},clip]{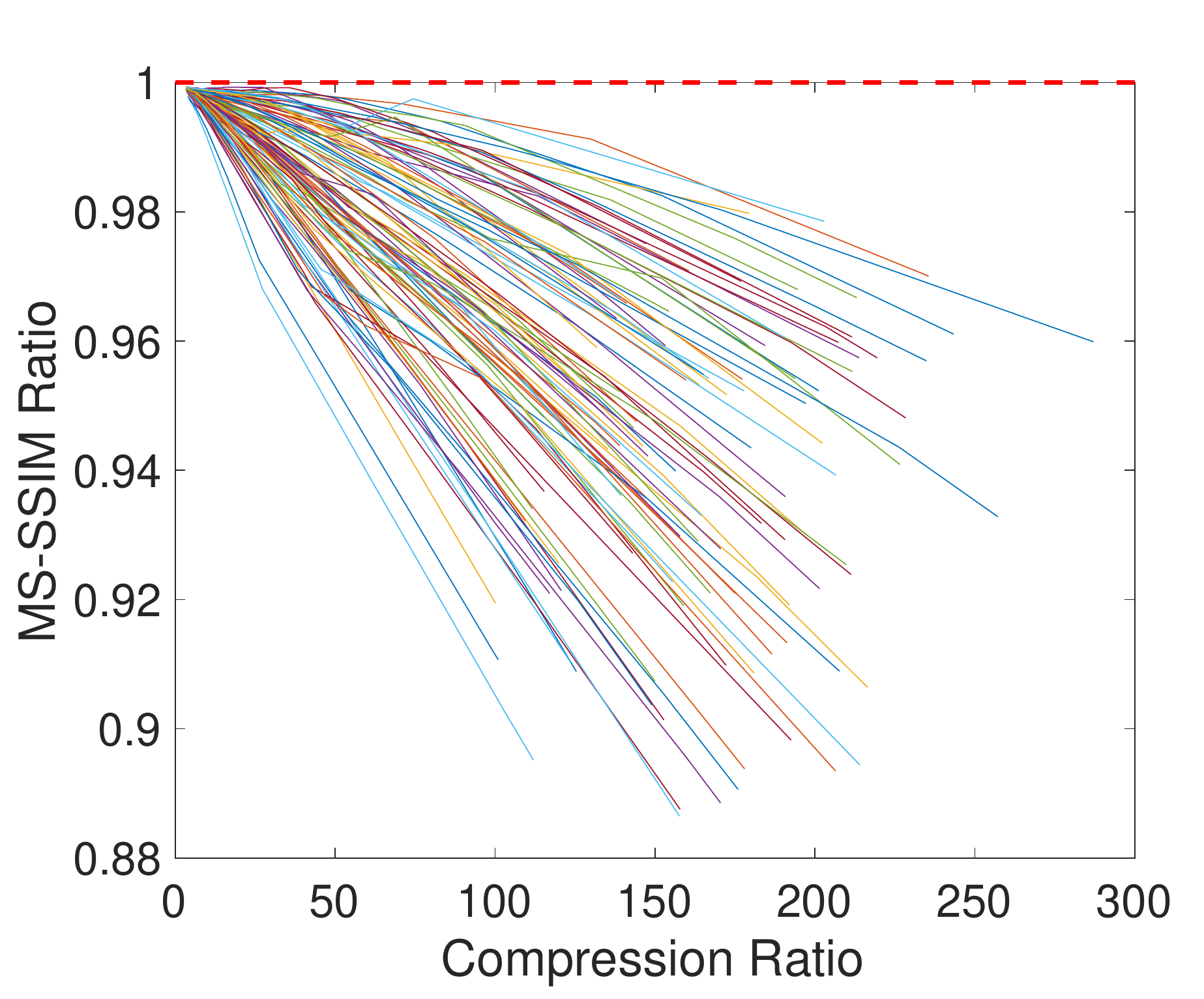} \\
		  {\small (i) PSNR ratio: BPG/CARP} & {\small (j) MS-SSIM ratio: BPG/CARP}\\
		  \includegraphics[width=7cm, height=3cm,trim={1.0cm 0.9cm 0 0},clip]{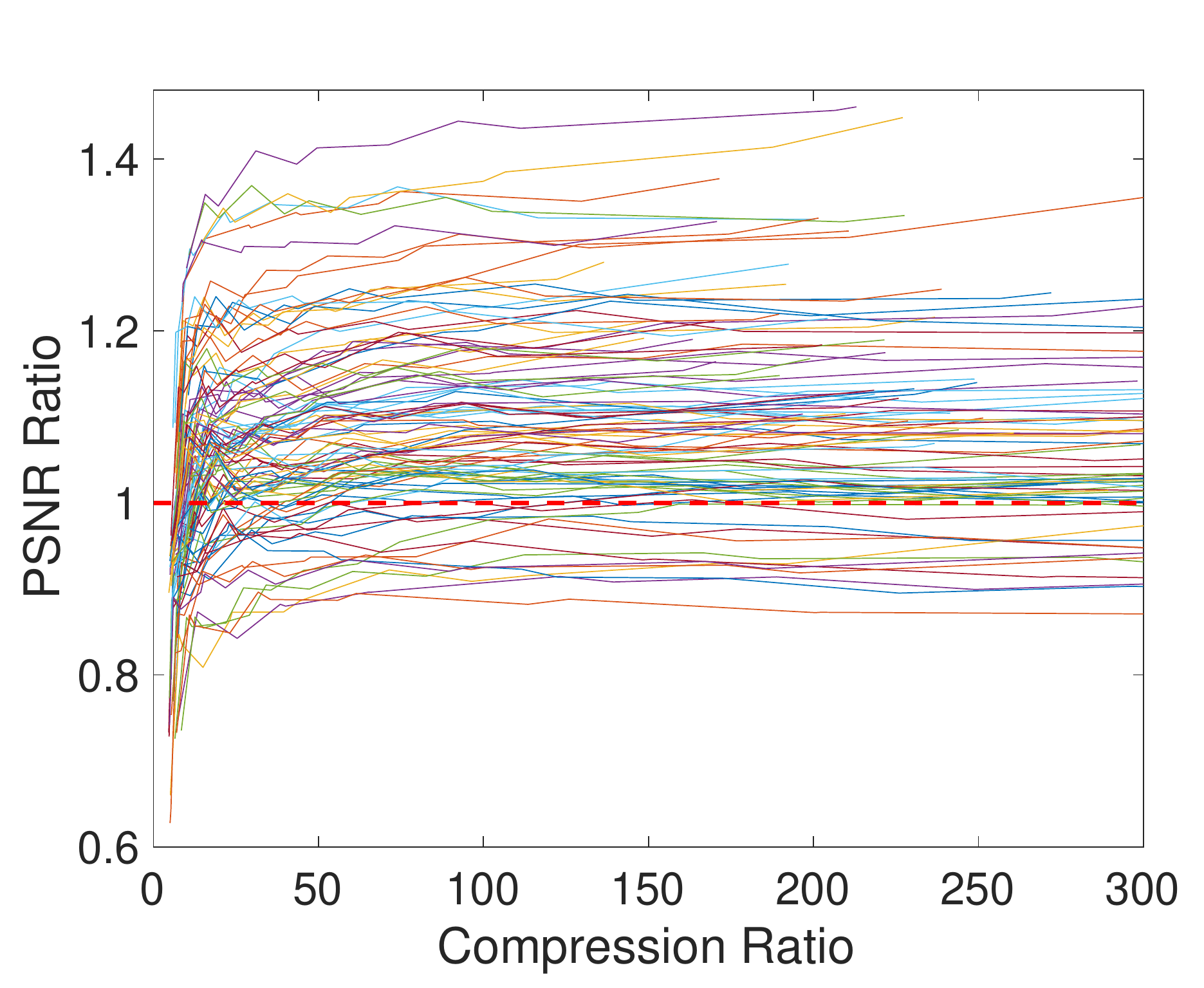}&
		  \includegraphics[width=7cm, height=3cm,trim={1.0cm 0.9cm 0 0},clip]{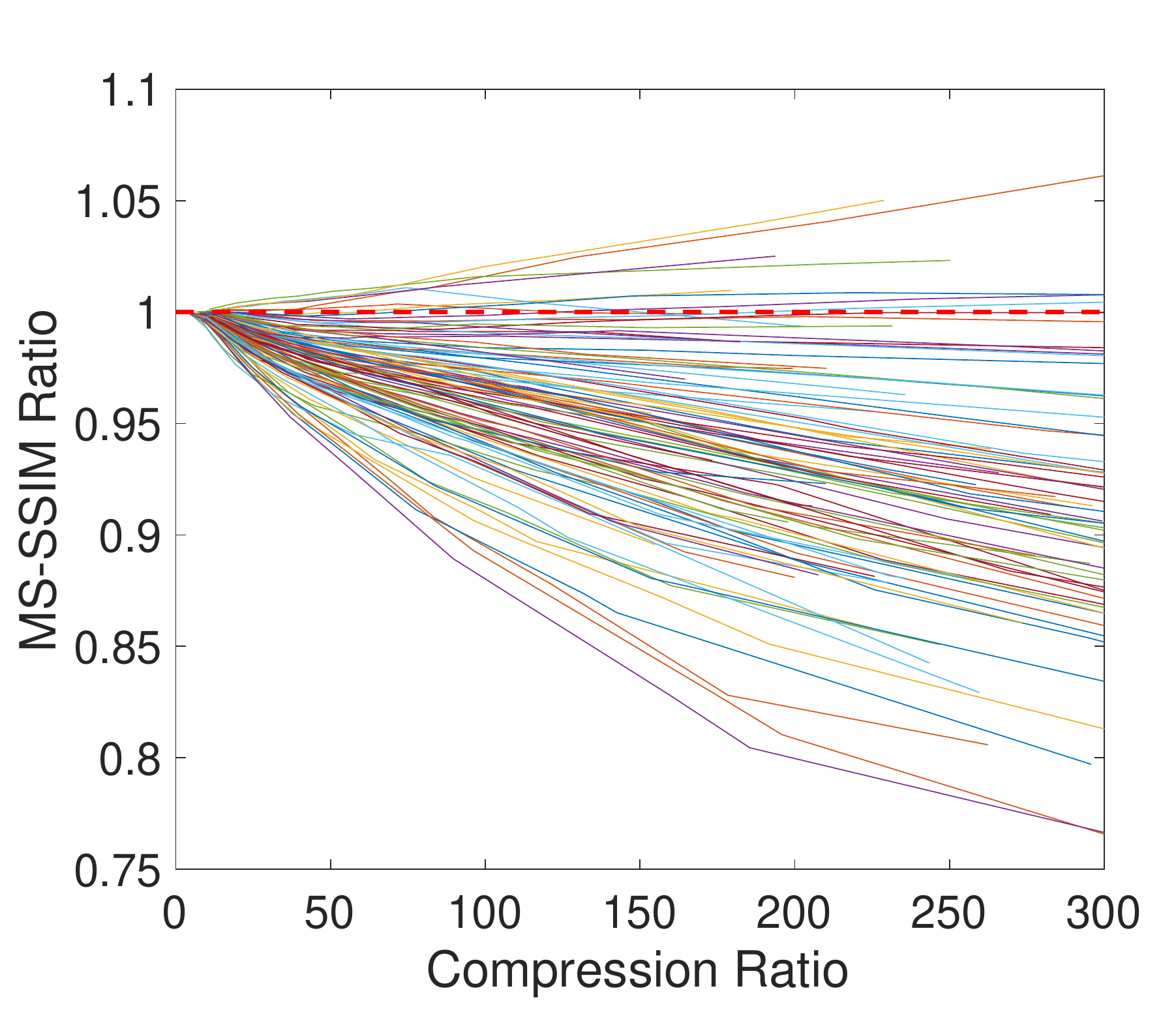} \\ 
		  {\small (k) PSNR ratio: HEVC/CARP} & {\small (l) MS-SSIM ratio: HEVC/CARP}
	\end{tabular}
\caption{YouTube videos: PSNR curves and MS-SSIM curves of CARP for 100 videos in (a) and (b); PSNR ratio curve and MS-SSIM ratio curves of JPEG, JPEG200, MPEG-4, BPG and HEVC relative to CARP in (c)--(l), respectively.}\label{3Dmean_old}
\end{figure}
Here we randomly select 20 videos from each task totaling 100 videos. Some frames of the sampled videos are displayed in the middle of Figure~\ref{bpg}. Note that these YouTube videos have a low resolution of 256 by 256, and thus they favor the MPEG-4 standard as MPEG-4 is optimized at low bit-rate video communications \cite{mpeg4}. 

CARP is applicable for streaming data by taking the entire video as input, treating time as an additional dimension, thus constituting a genuine video compression method like MPEG-4.
In addition to the popular MPEG-4 standard for video compression, we also consider using JPEG and JPEG2000 through a frame-by-frame compression while CARP is directly applicable to 3D images with no modification. Figure~\ref{3Dmean_old} presents the PSNR and MS-SSIM ratio curves (alternative methods over CARP) at various compression ratios for all the 100 videos, as well as the PSNR and MS-SSIM curves of CARP. CARP substantially outperforms methods of JPEG and JPEG2000 for nearly all individual videos at all compression ratios. CARP is better than MPEG-4 on all videos when the compression ratio is smaller than 10; for compression ratios between 10 and 20, MPEG-4 and CARP each performs better at a subset of the videos; for compression ratios above 20, CARP increasingly outperforms MPEG-4 at more videos. Moreover, CARP never underperforms MPEG-4 much on any individual image with the maximum PSNR ratio around 1.1. We note again that all the videos are at a low resolution that substantially favors MPEG-4. HEVC is better than CARP in PSNR on most videos when the compression ratio is larger than 10 as shown in Figure~\ref{3Dmean_old}(k), while CARP outperforms HEVC in MS-SSIM on almost all the individual videos as shown in Figure~\ref{3Dmean_old}(l). Overall, as shown in Figure~\ref{fig:summary}, CARP and HEVC give the best average PSNR and MS-SSIM than all other methods at all compression ratios from 0 to 50, and CARP tops HEVC under the MS-SSIM metric but loses to HEVC by a small margin under the PSNR metric.

\begin{figure}[!h]
 \centering
 \includegraphics[width=0.9\linewidth, height=12cm]{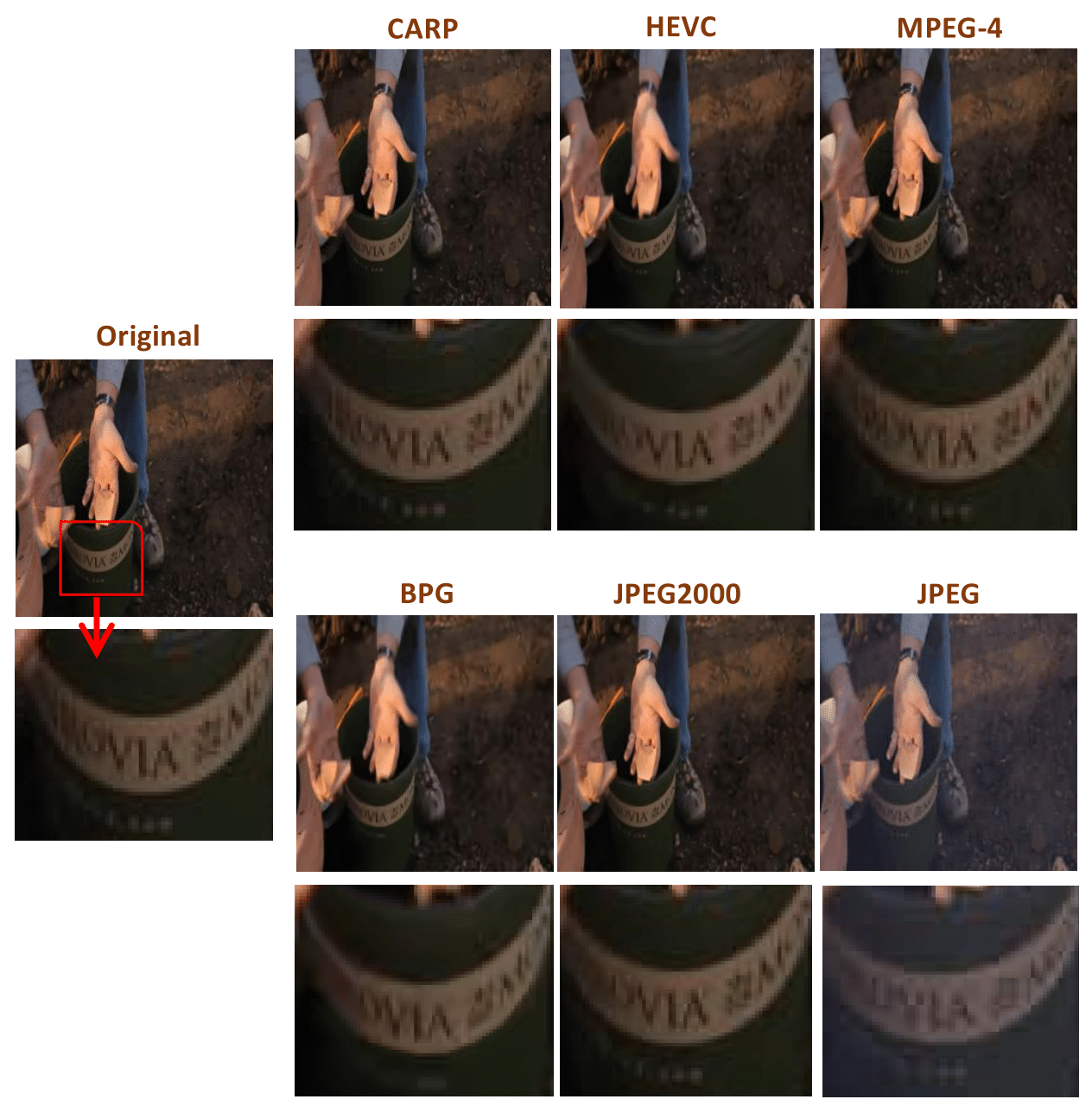}
 \caption{The comparison of reconstructed YouTube videos by six different methods when compression ratio is set to 30. Left to right and top to bottom:  CARP, HEVC, MPEG-4, BPG, JPEG2000, and JPEG.}\label{3Dold}
\end{figure}

For visual comparison, we select a video from the ``replotting a plant'' task and compare one frame of the reconstructed video to that of the original one in Figure \ref{3Dold}. The zoomed region shown in the bottom row shows that the reconstructed frame via CARP captures most details in the original frame (e.g., the words on the label), while the regions reconstructed via the other three methods in the bottom row of Figure \ref{3Dold} are more blurry.

\subsection{Surveillance video dataset}

We next investigate the performance of CARP on higher-resolution videos through a surveillance video dataset~\cite{vezzani2010video}, where each video has a resolution of 1024 by 1024. We randomly select one surveillance video for a parking lot, shown in the bottom of Figure \ref{bpg}. We divide the entire video into 180 segments of equal length to help assess the longitudinal variability of compression performances of each method and reduce the computational time of each method. 

\begin{figure}[H]
\centering 
	\begin{tabular}{cc}
		 \includegraphics[width=7cm, height=3cm,trim={1.0cm 0.8cm 0 0},clip]{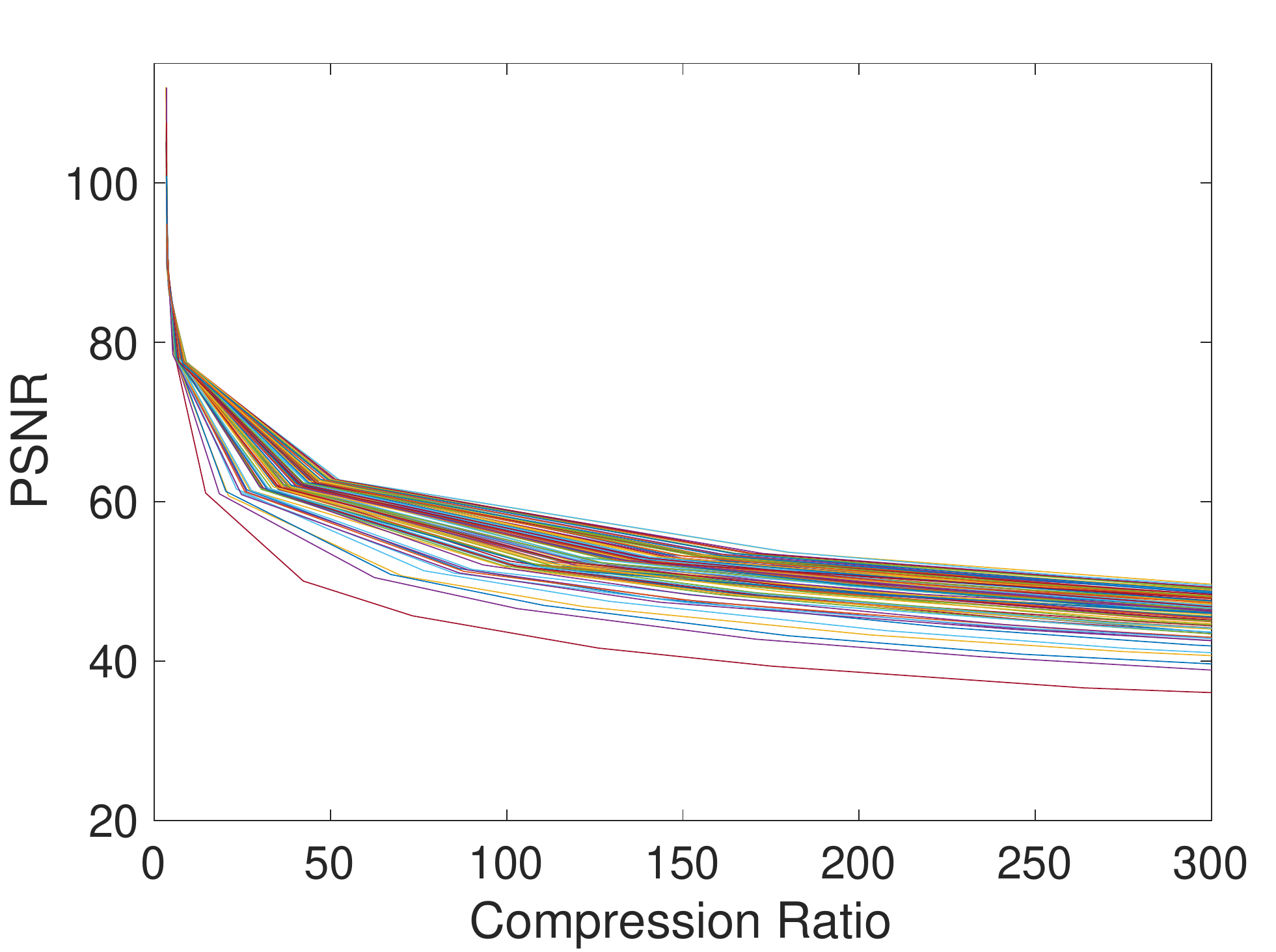} & 
		  \includegraphics[width=7cm, height=3cm,trim={1.2cm 0.8cm 0 0},clip]{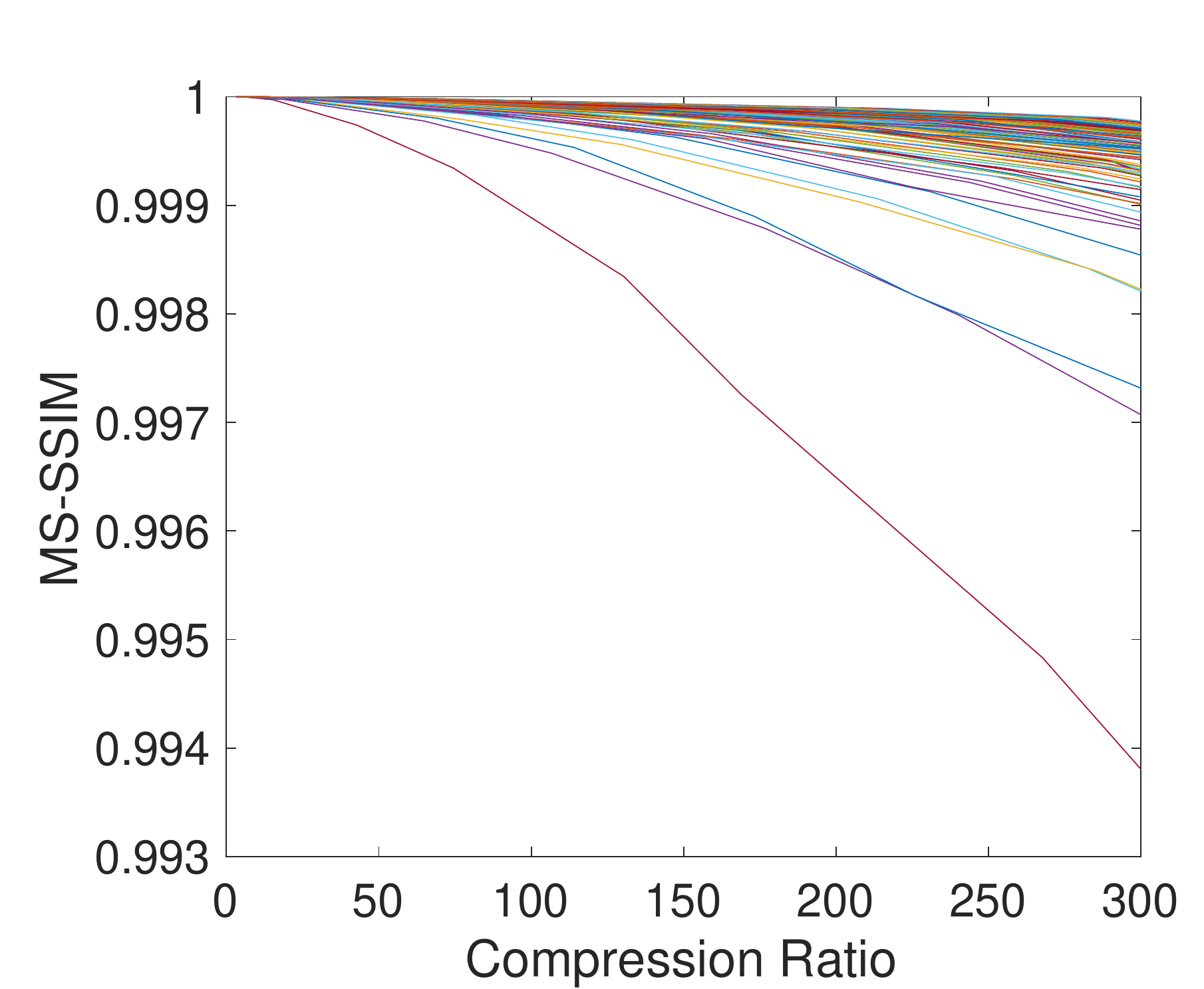}\\
		  {\small (a) PSNR of CARP for 180 videos} & {\small (b) 
		  MS-SSIM of CARP for 180 videos} \\
		  \includegraphics[width=7cm, height=3cm,trim={1.0cm 0.8cm 0 0},clip]{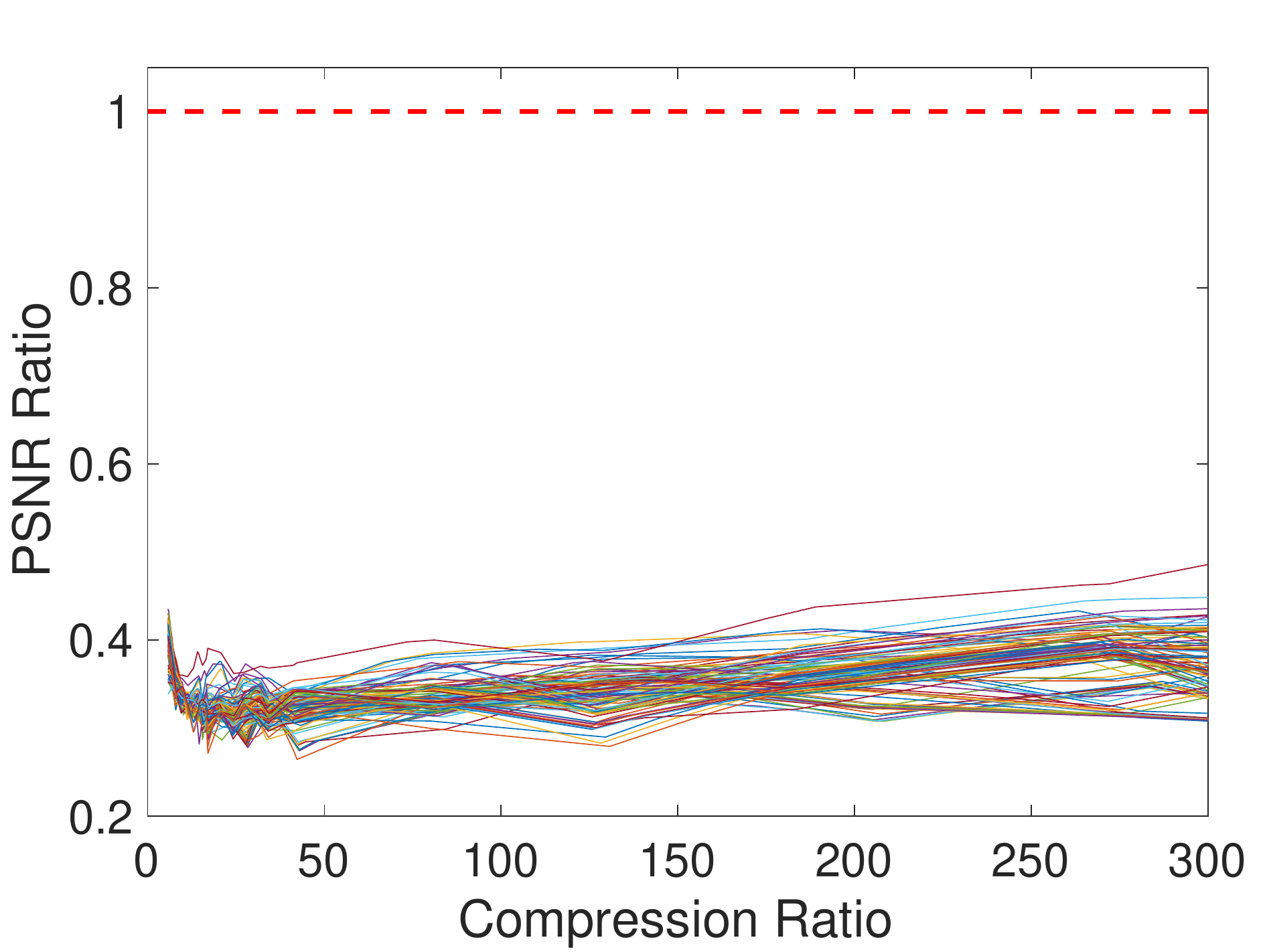} &
		  \includegraphics[width=7cm, height=3cm,trim={1.2cm 0.9cm 0 0},clip]{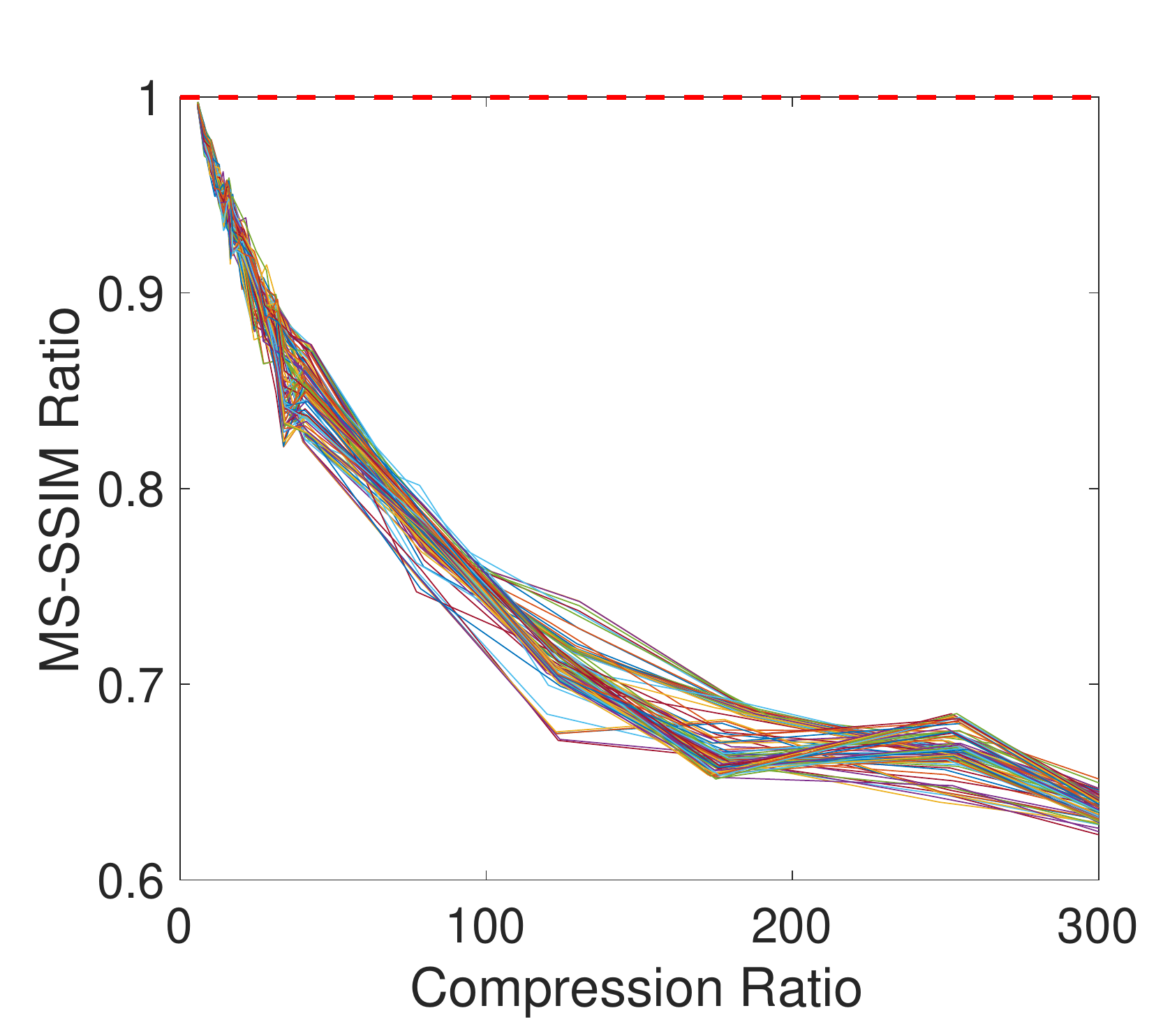}\\
		   {\small (c) PSNR ratio: JPEG/CARP} & {\small (d) MS-SSIM ratio: JPEG/CARP}\\
		  \includegraphics[width=7cm, height=3cm,trim={1.0cm 0.8cm 0 0},clip]{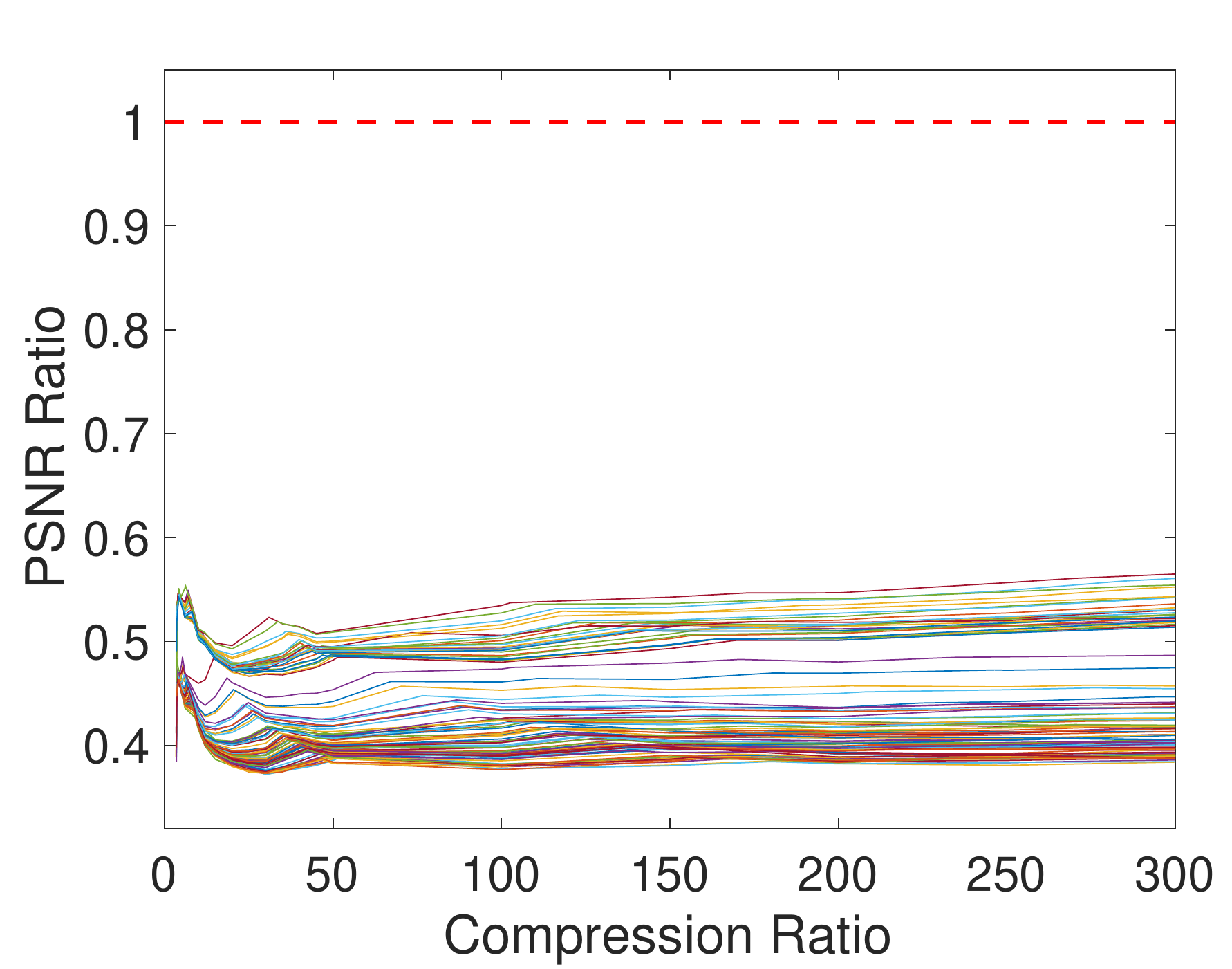}&
		  \includegraphics[width=7cm, height=3cm,trim={1.3cm 0.9cm 0 0},clip]{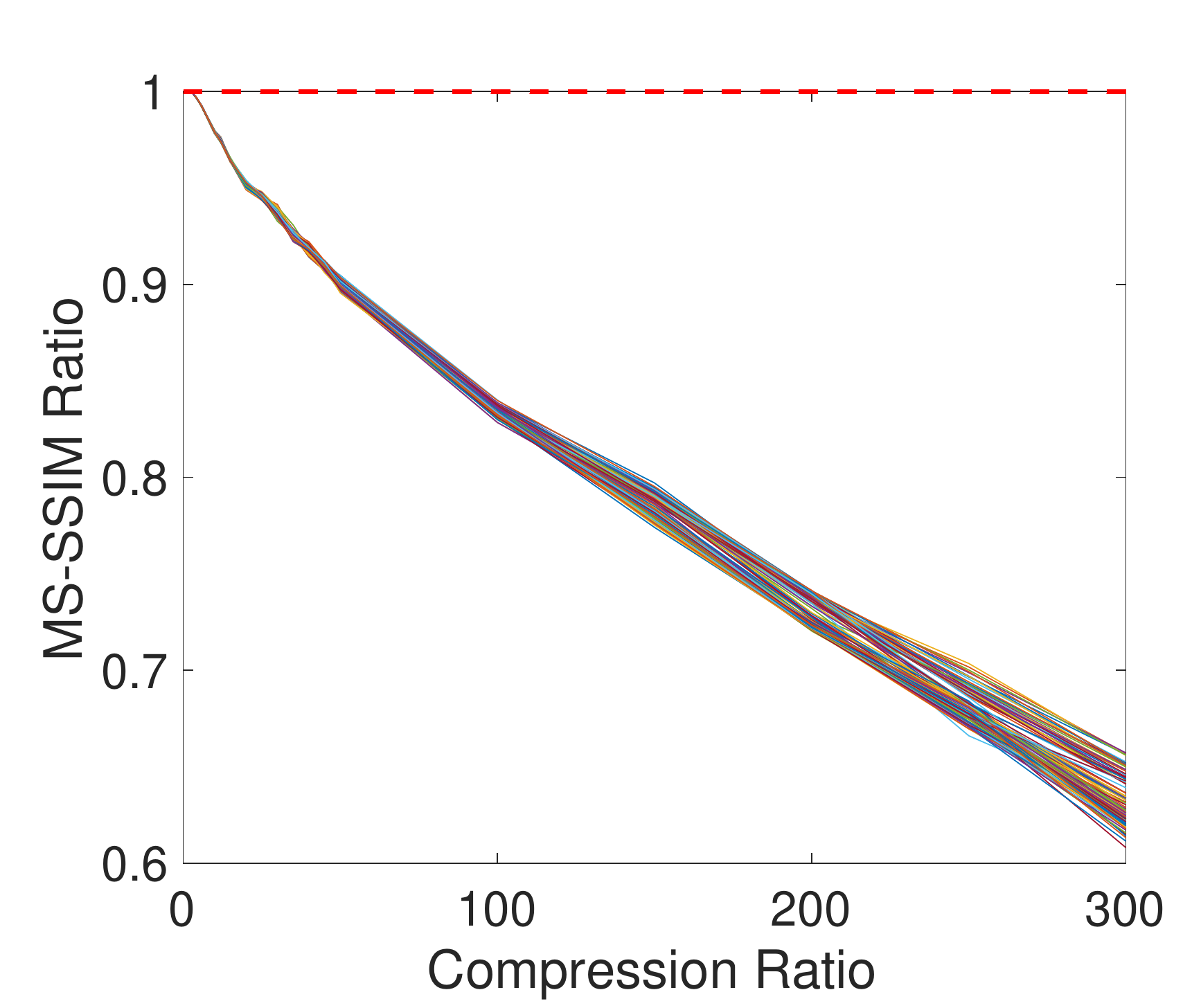} \\
		  {\small (e) PSNR ratio: JPEG2000/CARP} & {\small (f) MS-SSIM ratio: JPEG2000/CARP} \\
		  \includegraphics[width=7cm, height=3cm,trim={1.0cm 0.8cm 0 0},clip]{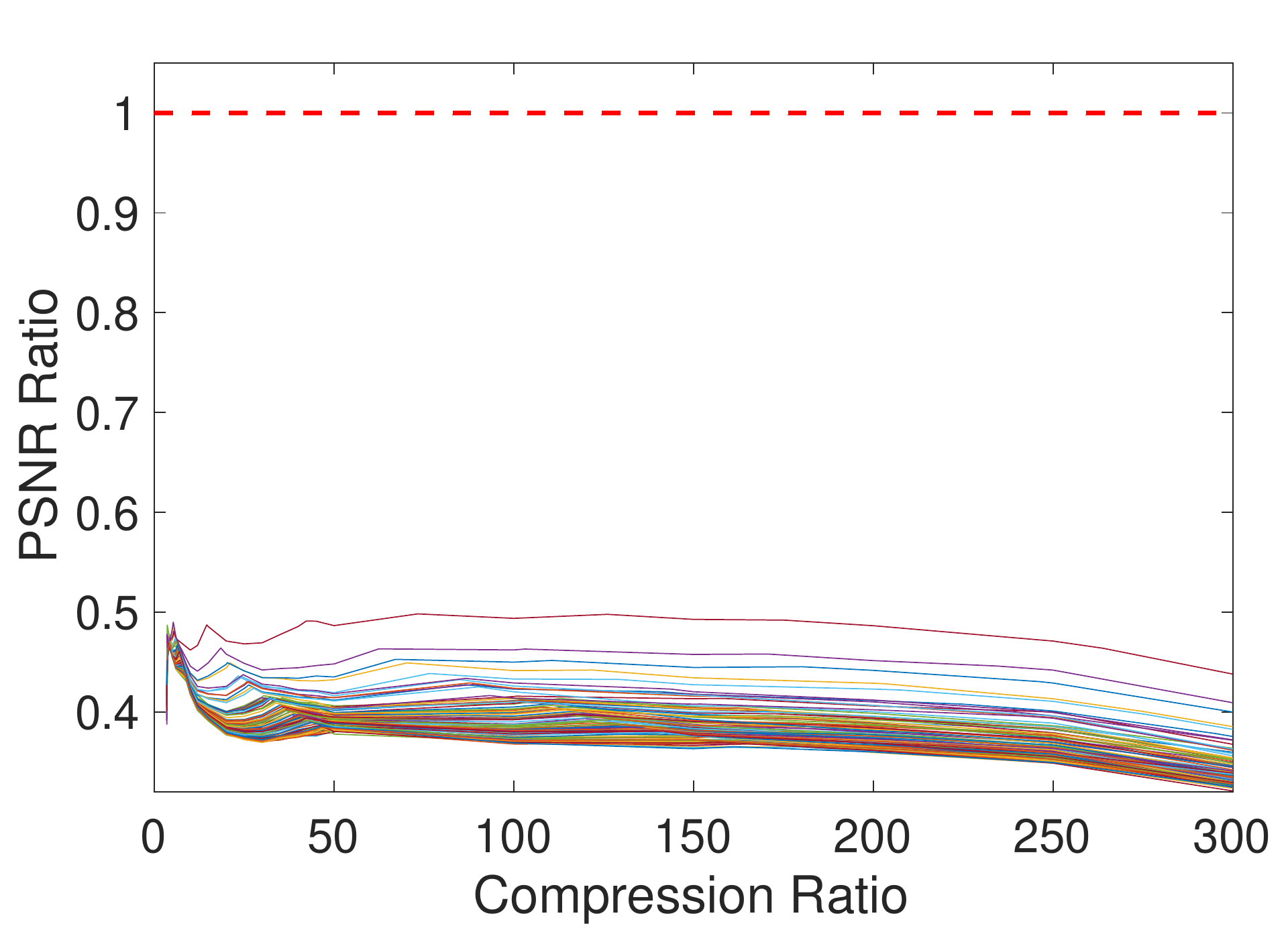} &
		  \includegraphics[width=7cm, height=3cm,trim={1.2cm 0.8cm 0 0},clip]{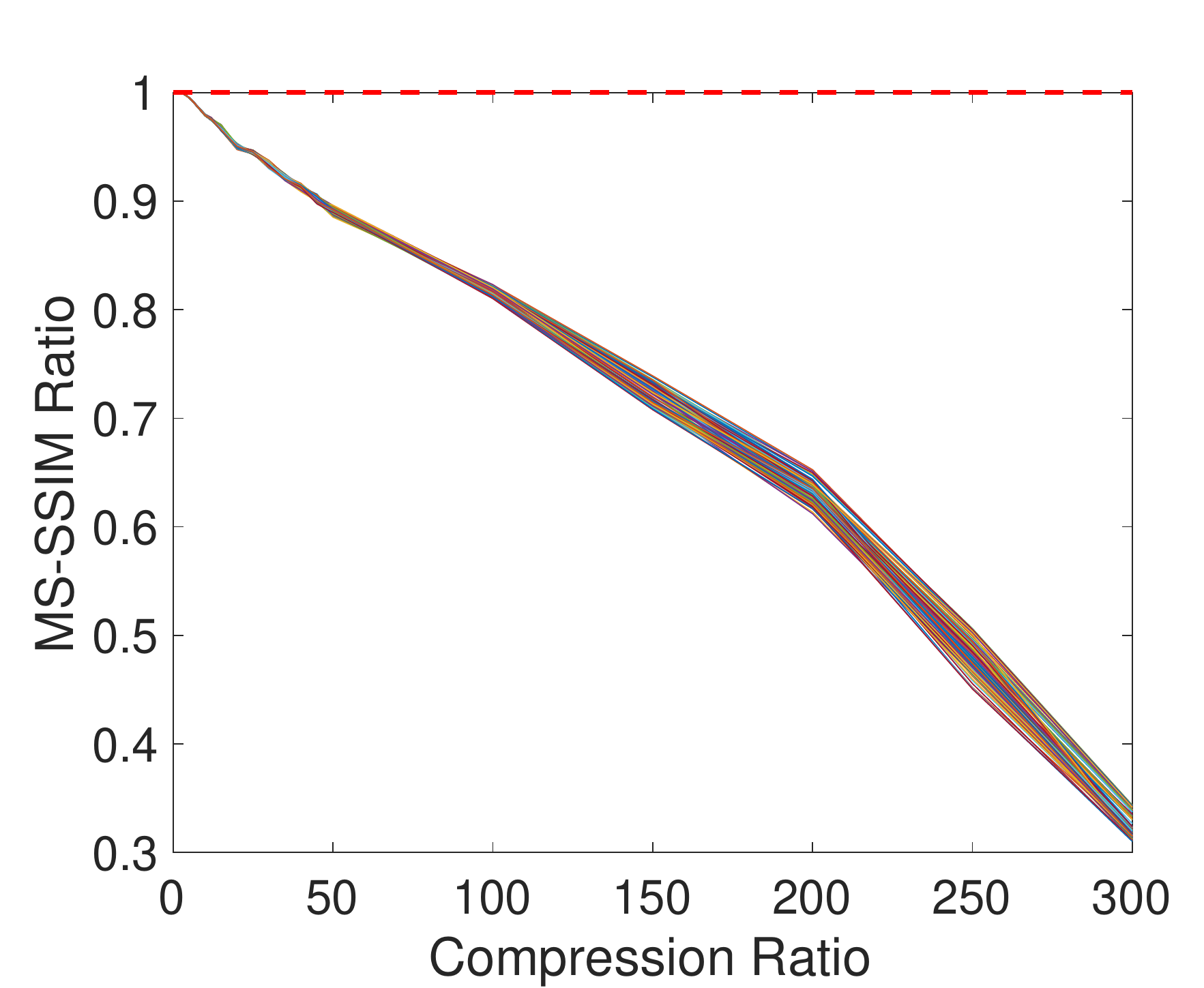}\\
		   {\small (g) PSNR ratio: MPEG-4/CARP} & {\small (h) MS-SSIM ratio: MPEG-4/CARP}\\
		  \includegraphics[width=7cm, height=3cm,trim={1.1cm 0.9cm 0 0},clip]{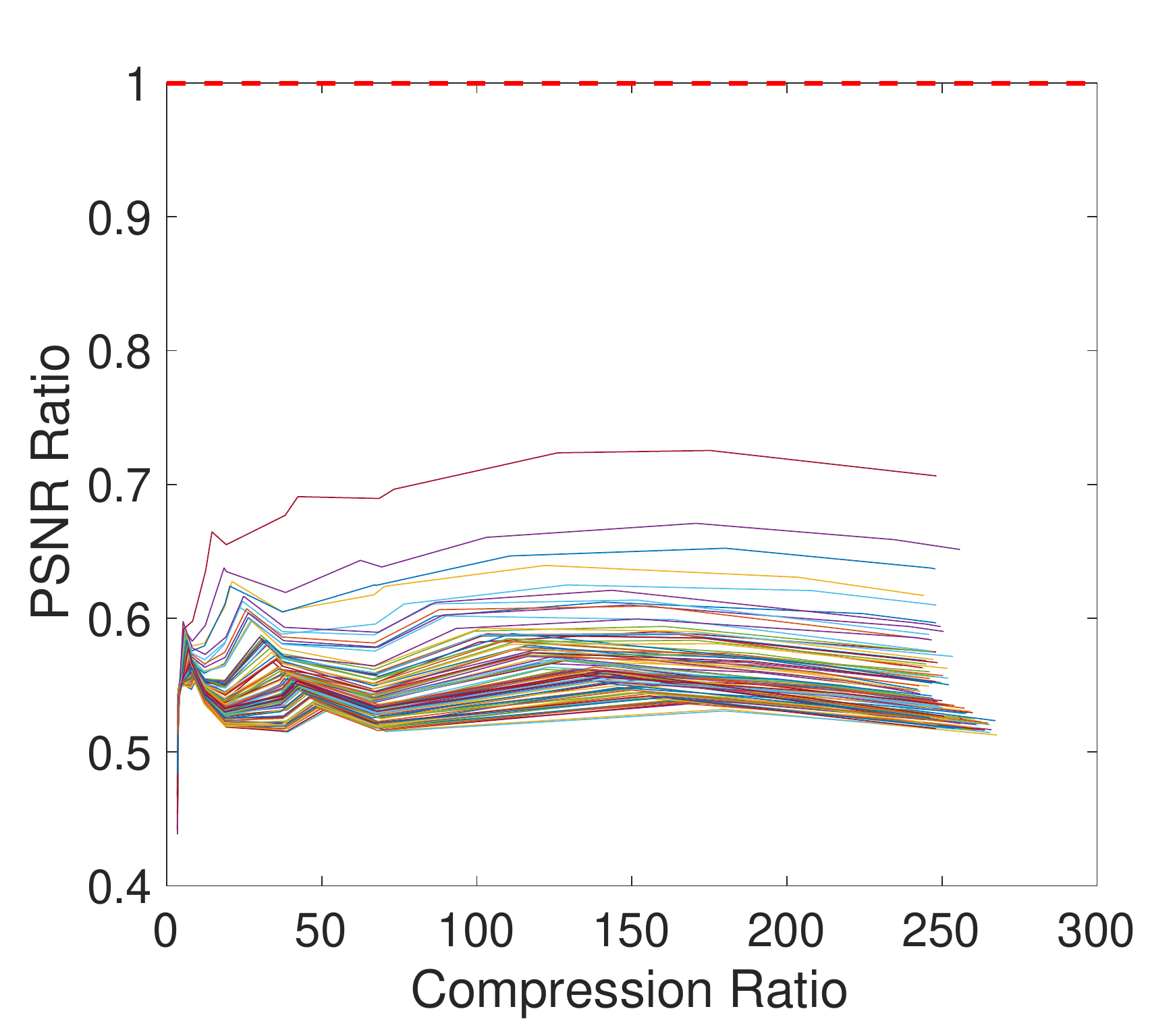} &
		  \includegraphics[width=7cm, height=3cm,trim={1.3cm 0.9cm 0 0},clip]{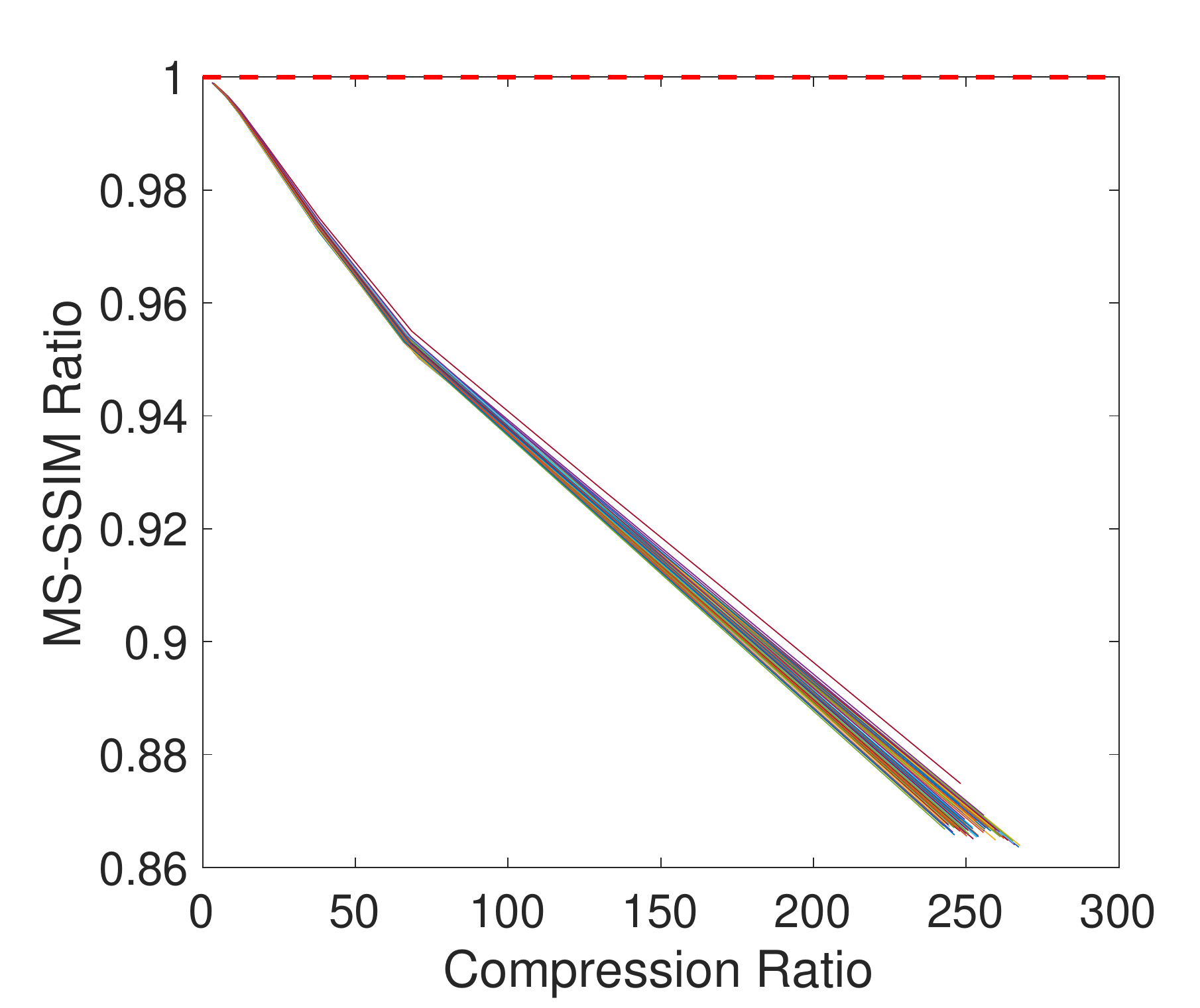} \\
		  {\small (i) PSNR ratio: BPG/CARP} & {\small (j) MS-SSIM ratio: BPG/CARP}\\
		  \includegraphics[width=7cm, height=3cm,trim={1.2cm 0.9cm 0 0},clip]{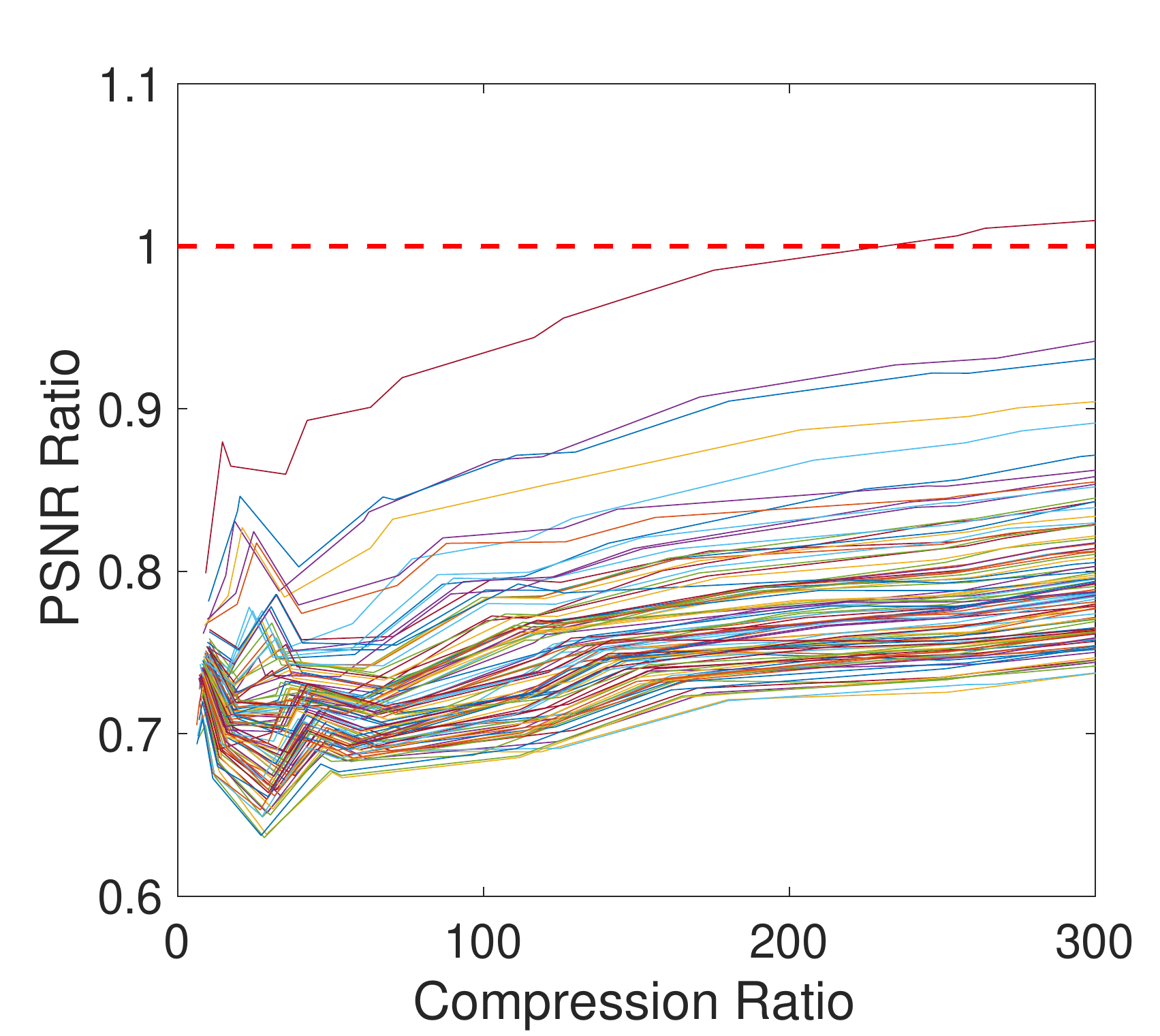}&
		  \includegraphics[width=7cm, height=3cm,trim={1.0cm 0.9cm 0 0},clip]{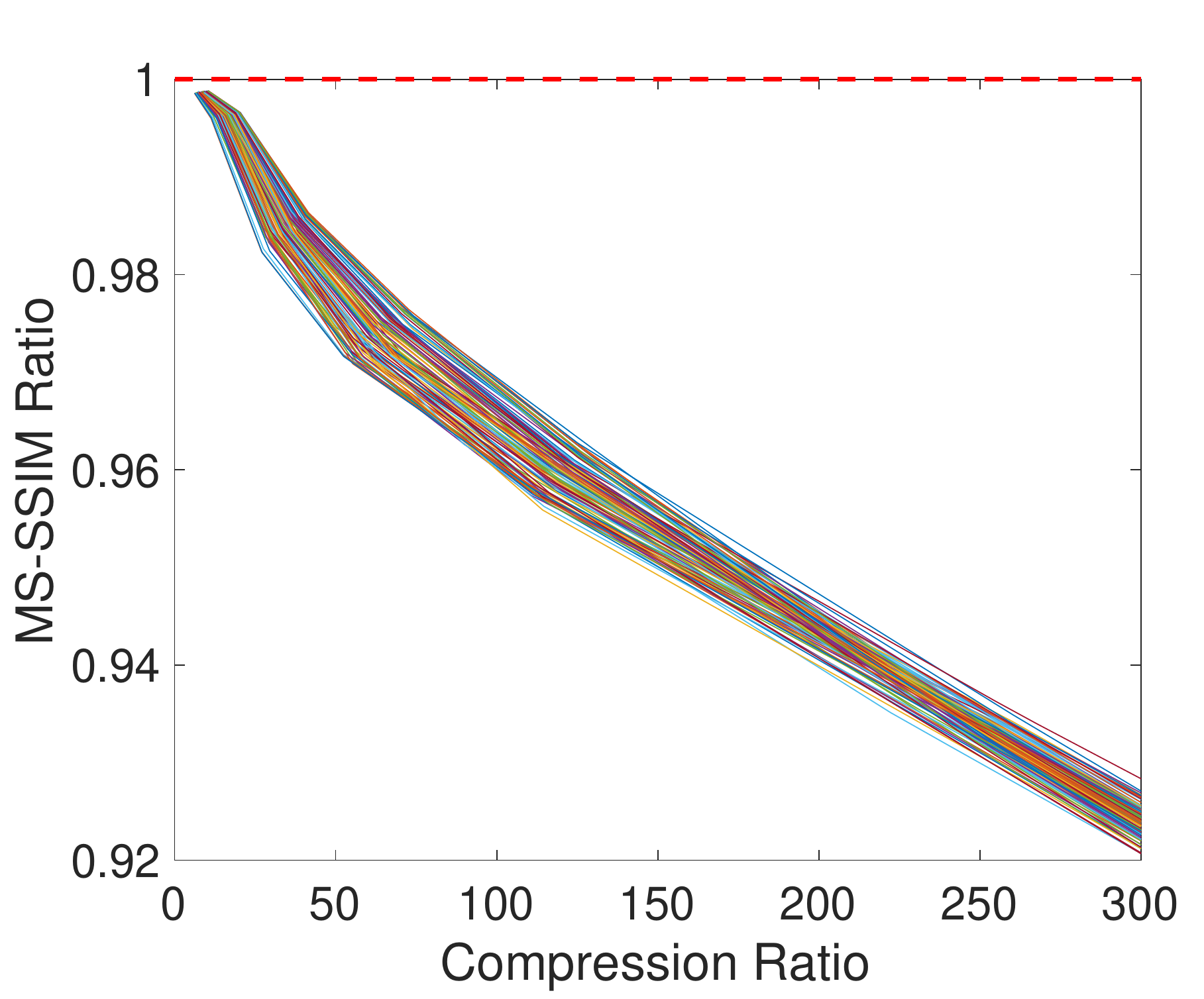} \\ 
		  {\small (k) PSNR ratio: HEVC/CARP} & {\small (l) MS-SSIM ratio: HEVC/CARP}
	\end{tabular}
\caption{Surveillance videos: PSNR curves and MS-SSIM curves of CARP for 180 videos in (a) and (b); PSNR ratio curve and MS-SSIM ratio curves of JPEG, JPEG200, MPEG-4, BPG and HEVC relative to CARP in (c)--(l), respectively.}\label{3Dmean_new}
\end{figure}

Figure \ref{3Dmean_new} plots the PSNR and MS-SSIM ratio curves (alternative method over CARP) among all the 180 videos as well as the PSNR and MS-SSIM curve for CARP at various compression ratios. We can see that CARP gives the best PSNRs and MS-SSIMs uniformly for all videos at all compression ratios (up to 300). Overall, CARP gives best average PSNR and MS-SSIM than all other methods at all compression ratios from 0 to 50, as shown in Figure~\ref{fig:summary}(e) and Figure~\ref{fig:summary}(f).

\begin{figure}[h!]
 \centering
 \includegraphics[width=0.9\linewidth]{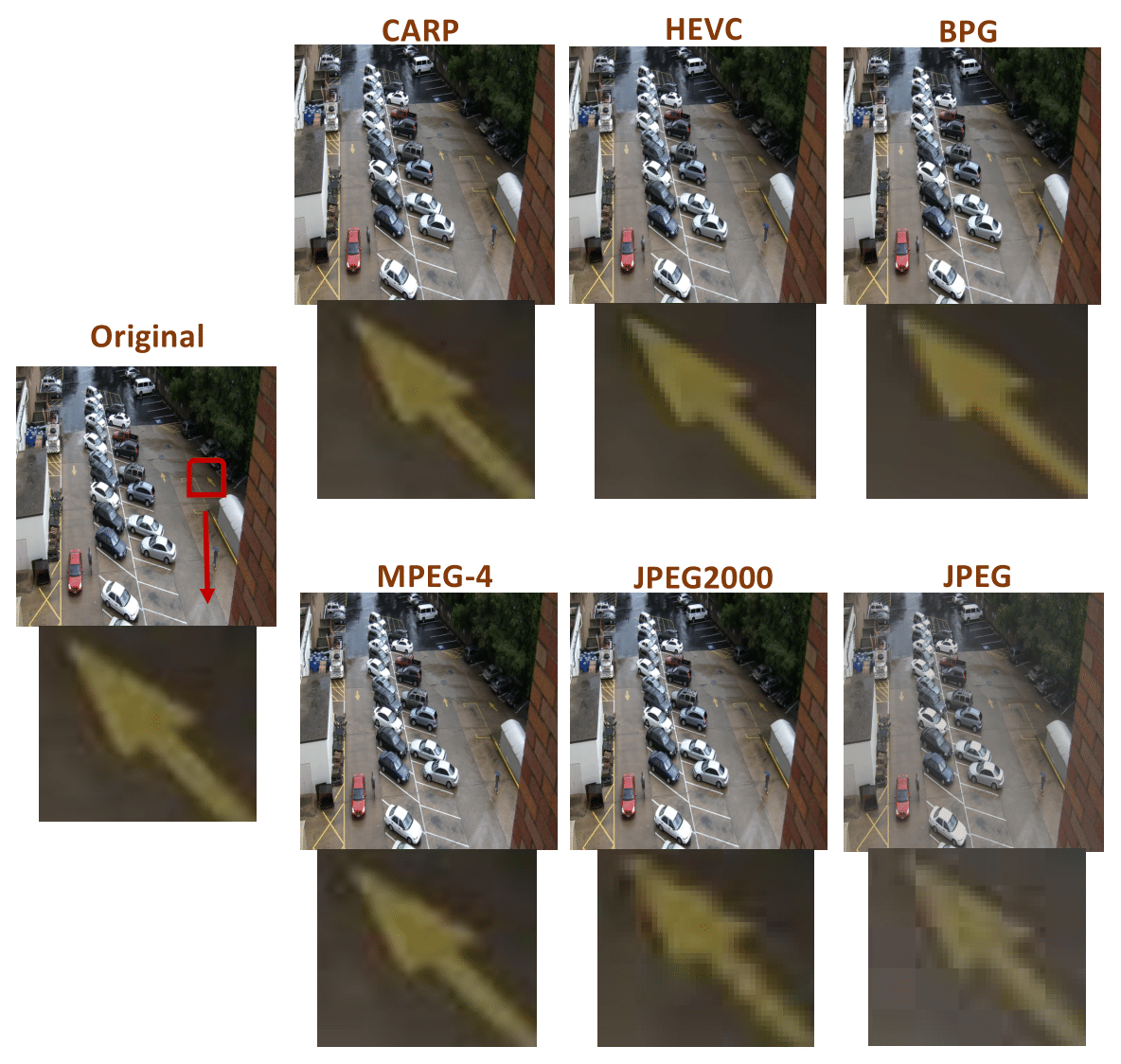}
 \caption{The comparison of reconstructed surveillance video among six different methods when compression ratio is set to 30. Left to right and top to bottom:  CARP, HEVC, MPEG-4, BPG, JPEG2000, and JPEG.}\label{3Dnew_2}
\end{figure}

For visual comparison, we randomly select one video and compare one frame of reconstructed videos. Figure \ref{3Dnew_2} shows the original frame and reconstructed frames by CARP, HEVC, BPG, MPEG-4, JPEG2000, and JPEG, when the compression ratio is set to 30. The zoomed region is the shadow area at the top-right corner in the original frame, shown in the bottom row. In comparison, the reconstructed frame via CARP captures most details of the region in the original frame, while the regions reconstructed via the other three methods in the bottom of Figure \ref{3Dnew_2} are more blurry (e.g., the edge of the yellow arrow). The arrows and their background are smoother in HEVC and BPG than that in CARP, but CARP gets much closer to the original pattern compared to HEVC and BPG.

\section{Conclusion} \label{sec:conclusion}
CARP uses a principled Bayesian hierarchical model to learn an optimal permutation on the image space, which effectively allows adaptive wavelet transforms on the image and achieves progressive transmission. CARP is directly applicable to 3D images with no modification. 
We conduct extensive experiments and show that CARP dominates the state-of-the-art image/video compression methods for all image types under consideration and often on nearly all of the individual images. CARP is computationally efficient in that it scales linearly with the number of pixels of an image. Currently for the 2D still images, the average computation time under our implementation of CARP, without any parallel computing, is around 3.17 second/image, while 0.82 second/image for JPEG, 0.40 second/image for JPEG2000, 88.75 second/image for E2E-DL, tested on a Macbook Pro with 2.2 GHz Intel Core i7 processor. BPG has a similar processing time to JPEG2000, but we cannot provide an exact time because the kernel of BPG is not accessible. The computing time of CARP can be further reduced with more optimized implementation. In particular, one main computational task in CARP is to compute the marginal likelihood of the wavelet regression model on each node in the partition tree, which can be parallelized over the nodes in the partition tree. We plan to implement a GPU parallelized version of CARP in the future to achieve substantial speedup.

\bibliographystyle{ieee_fullname}
\bibliography{refs}

\end{document}